\documentclass[traditabstract]{aa}
\usepackage{savesym} 
\usepackage{txfonts} 
\usepackage[english]{babel} 
\usepackage{t1enc} 
\usepackage{graphicx} 
\usepackage{epstopdf}
\usepackage{subfig} 
\usepackage{pdflscape}
\usepackage{fixltx2e} 
\usepackage{natbib} 
\usepackage{footmisc}
\usepackage{url} 
\usepackage{longtable}
\usepackage{multirow}
\usepackage{setspace}
\usepackage[table]{xcolor}
\usepackage{gensymb}
\usepackage{siunitx}

\begin{document} 

\pdfoutput=1

\title{Anomalous peculiar motions of high-mass young stars in the Scutum spiral arm}
\author{K. Immer\inst{1} \and J. Li\inst{2} \and L.~H. Quiroga-Nu\~nez\inst{3,1} \and M.~J. Reid\inst{4} 
\and B. Zhang\inst{5} \and L. Moscadelli\inst{6} \and K.~L.~J. Rygl\inst{7}}

\institute{
Joint Institute for VLBI ERIC, Oude Hoogeveensedijk 4, 7991 PD Dwingeloo, The Netherlands; email: {\tt immer@jive.eu} 
\and
Purple Mountain Observatory CAS, No.8 Yuanhua Road, Qixia District, Nanjing 210034, The People's Republic of China 
\and
Leiden Observatory, Leiden University, Niels Bohrweg 2, NL-2333CA, Leiden, The Netherlands
\and
Center for Astrophysics $\vert$ Harvard \& Smithsonian, 60 Garden Street, Cambridge, MA 02138, USA
\and
Shanghai Astronomical Observatory, 80 Nandan Road, Shanghai 200030, China
\and
Instituto Nationale di Astrofisica, Osservatorio Astrofisico di Arcetri, Largo Enrico Fermi 5, I-50125 Firenze, Italia
\and
Italian ALMA Regional Centre, INAF-Istituto di Radioastronomia, Via P. Gobetti 101, 40129 Bologna, Italy
}

\date{Received 07/09/2018; accepted 03/11/2019} 
 
\abstract{
We present trigonometric parallax and proper motion measurements toward 22 GHz water and 6.7 GHz methanol masers in 16 high-mass star-forming regions. These sources are all located in the Scutum spiral arm of the Milky Way. The observations were conducted as part of the Bar and Spiral Structure Legacy (BeSSeL) survey. A combination of 14 sources from a forthcoming study and 14 sources from the literature, we now have a sample of 44 sources in the Scutum spiral arm, covering a Galactic longitude range from 0$\degree$ to 33$\degree$.
A group of 16 sources shows large peculiar motions of which 13 are oriented toward the inner Galaxy. A likely explanation for these high peculiar motions is the combined gravitational potential of the spiral arm and the Galactic bar.
}
 
 
\maketitle 
 
\section{Introduction}  
\label{Intro}
 
Our Galaxy has been identified as a barred spiral galaxy \citep{deVaucouleurs1970}. However, due to our position in the Milky Way, many details concerning the exact shape of our Galaxy are still under debate \citep[e.g., the number and position of the spiral arms, the rotation curve or the shape of the bar;][]{Reid2014b}. 
To resolve some of these issues, reliable distance measurements to objects in the spiral arms are necessary. The high astrometric accuracies provided by very long baseline interferometry (VLBI) allow the determination of trigonometric parallaxes to maser sources in high-mass star-forming regions all over the Galaxy \citep{Honma2012, Reid2014b, Sanna2017}. Since these regions are located within the spiral arms \citep[e.g.,][]{Elmegreen2011}, this approach allows a precise mapping of the spiral structure of the Milky Way.
This technique also provides proper motion measurements for the maser sources; thus, we determine the positions of the sources in 6D space (coordinates, distance, line-of-sight velocity, and proper motion).

In recent years, the Bar and Spiral Structure Legacy \citep[BeSSeL\footnote{\url{http://bessel.vlbi-astrometry.org/}},][]{Brunthaler2011} survey, an NRAO\footnote{The National Radio Astronomy Observatory is a facility of the National Science Foundation operated under cooperative agreement by Associated Universities, Inc.} Very Long Baseline Array (VLBA) key science project, 
has measured the parallaxes and proper motion of over 100 star-forming regions across the Milky Way  \citep[e.g.,][]{Reid2014,Sato2014,Wu2014,Choi2014,Sanna2014,Hachisuka2015}.
In this paper we discuss BeSSeL survey results for maser sources in the Scutum (also called Scutum-Centaurus or Scutum-Crux) spiral arm.

The Scutum spiral arm is located in the inner Galaxy (from the Sun's point of view), between the Sagittarius and the Norma arms. The Scutum arm was identified as strong condensations in neutral hydrogen and CO observations \citep{Shane1972, Cohen1980, Dame1986, Dame2001} in longitude-velocity plots.
\citet{Sato2014} presented the first BeSSeL survey results for 16 Scutum arm sources, discussing the location of the arm and its pitch angle, as well as  the peculiar motions of the sources. We now have a larger sample of sources in the Scutum arm and can thus improve on pinpointing the spiral arm location, and we can   study the kinematics in the arm.

We present new distances and kinematic information for 16 high-mass star-forming regions in the Scutum arm. Another 14 sources of this arm will be published in Li et al. (in prep.). While Li et al. discuss the pitch angle of the Scutum spiral arm, we focus on the kinematics in the arm, especially a group of sources with anomalously large peculiar motion. In addition to  the sources from Li et al., we also included 14 sources from the literature \citep{Bartkiewicz2008, Xu2011, Immer2013, Sato2014, Zhang2014} in our peculiar motion analysis. 
We  first describe the observations (Sect. \ref{Observations}) and then present our results (Sect. \ref{Results}), including the parallax and proper motion fits for the 16 sources, the association of the sources with the Scutum spiral arm, and the measured peculiar motion of all 44 sources. In Sect. \ref{Discussion}, we  discuss the anomalously large peculiar motions of 16 sources and explore their possible origins.

\section{Observations and data reduction}
\label{Observations}

We conducted multi-epoch phase-referenced observations with the VLBA of 13 6.7 GHz methanol masers and 
three 22 GHz water masers, located within high-mass star-forming regions.
The coordinates of the sources, program codes, and 
observing epochs are given in Table \ref{Sources-Coord}. For the methanol masers, four epochs spread over one year are scheduled 
to optimally sample the extent of the Earth's orbit around the Sun as viewed by the targets
(corresponding to the peaks of the sinusoidal parallax signature) in right ascension and to minimize the correlation between the parallax and proper motion contributions \citep{Xu2006,Menten2007}. Since water maser spots can be shorter-lived than one year \citep{Tarter1986}, 
two additional epochs are observed at 22 GHz. Compact background sources, used as position references for the maser targets, were selected 
from the ICRF2 catalog \citep{Fey2015} and a dedicated VLBA survey of extragalactic sources \citep{Immer2011}. The background
quasars that were observed with each target are listed in Table \ref{Sources-Coord}.

The general observation setup and the data reduction procedures are detailed in \citet{Reid2009}. Except for G028.86+00.06, four adjacent frequency bands with 16 MHz width (8 MHz for G028.86+00.06) were observed in right and left circular polarization. The maser emission
was centered in the third (second for G028.86+00.06) frequency band, at the LSR velocity of the target (see Table \ref{Sources-Coord}).
The data were correlated in two steps, one for the continuum data and one for the maser line data. For the maser line data, the frequency 
resolution was 8 kHz (BR198), 16 kHz (BR149), or 31.25 kHz (G028.86+00.06). This corresponds to a velocity resolution of 0.36 km s$^{-1}$ and 0.72 km s$^{-1}$, respectively, for the 6.7 GHz masers 
and 0.11 km s$^{-1}$ for the 22 GHz  masers (BR198; 0.42 km s$^{-1}$ for the water maser G028.86+00.06), assuming rest frequencies of 6668.56 MHz for the 
CH$_{3}$OH(5$_{1}-$6$_{0}$) and 22235.0 MHz for the H$_{2}$O(6$_{16}$-5$_{23}$) maser transitions. The spectra of the masers are shown in Fig. \ref{Spectra}.

The calibration and analysis of the BeSSeL survey data was conducted using ParselTongue \citep{Kettenis2006} scripts for the NRAO Astronomical
Image Processing System \citep[AIPS,][]{Greisen2003}. The different calibration steps are described in \citet{Reid2009}. After calibration,
the background sources and the target masers were imaged with the AIPS task IMAGR. Then their relative positions were determined by fitting the brightness distribution of the feature with an elliptical Gaussian model, using the AIPS task JMFIT. 

Different numbers of maser spots were selected for parallax fitting in each target maser (see Table \ref{Sources-Info}). The positional differences between the 
maser spots and each of the quasars were calculated and then modeled with a combination of a sinusoidal parallax signature and linear
proper motion in each coordinate. Systematic uncertainties (typically from unmodeled atmospheric and ionospheric delays) are normally 
much larger than the formal position uncertainties from the Gaussian fitting \citep{Menten2007}, leading to high reduced $\chi^2$ in the 
parallax and proper motion modeling. To account for these uncertainties, we added additional error values in quadrature to the formal position uncertainties on both
coordinates and adjusted them until the reduced $\chi^2$ for each coordinate was near unity \citep[see][]{Reid2009}.

If several maser spots are combined in the final parallax fit, we increased the formal uncertainty on the parallax fit by a 
factor of $\mathrm{\sqrt{N_{S\!pot}}}$, where N$_{S\!pot}$ is the number of combined maser spots. We make the conservative assumption that position variations of the maser spots are 100\% correlated, which can be caused by small changes in the background sources or residual unmodeled atmospheric 
delays affecting equally all maser spots per epoch. Data from the different background quasars are assumed to be uncorrelated.

In the combined parallax fitting, we allowed for different proper motion values for different maser spots due to possible internal 
motions in the star-forming regions. However, if the position differences of a maser spot were measured to more than one background source, then the proper motion of this maser was constrained to be the same for all the background sources used in the parallax fitting of 
this spot since we assume that the background sources do not have any measurable proper motion.

For the 6.7 GHz masers, in sources where we had background sources on both sides of the target, we detected gradients of the parallax values
on the sky for the parallaxes determined from different background sources \citep{Reid2017}. This is not an issue for the 22 GHz masers,
which indicates unmodeled ionospheric delays as the root of this problem since ionospheric dispersive delays scale with $\nu^{-2}$, and thus 
residual dispersive delays are much larger at 6.7 GHz. By simultaneously 
fitting for parallax parameters as well as these ionospheric effects, we 
improved the modeling of the position offset data (maser spot minus quasar position), yielding a more reliable parallax estimate. More details are given in \citet{Reid2017} and \citet{Zhang2019}.

\section{Results}
\label{Results}

\subsection{Parallax fits}

\begin{figure*}[ht]
        \centering
        \caption{Parallax and proper motion data and fits for a 6.7 GHz methanol and a 22 GHz water maser. The black symbols show the average of all quasar positions for 6.7 GHz masers. In each panel, the data of different maser spots are offset for clarity.
    First panel: Measured position of the maser on the sky. The positions  expected from the fit are shown as open circles.
    Second panel: East (upper part, solid lines) and north (lower part, dashed lines) position offsets vs. time.
    Third panel: As  second panel, but with proper motion removed, showing only the parallax effect for the maser underlined in the legend.}
       \subfloat[6.7 GHz methanol maser $-$ G023.25$-$00.24]{\includegraphics[width=0.8\textwidth]{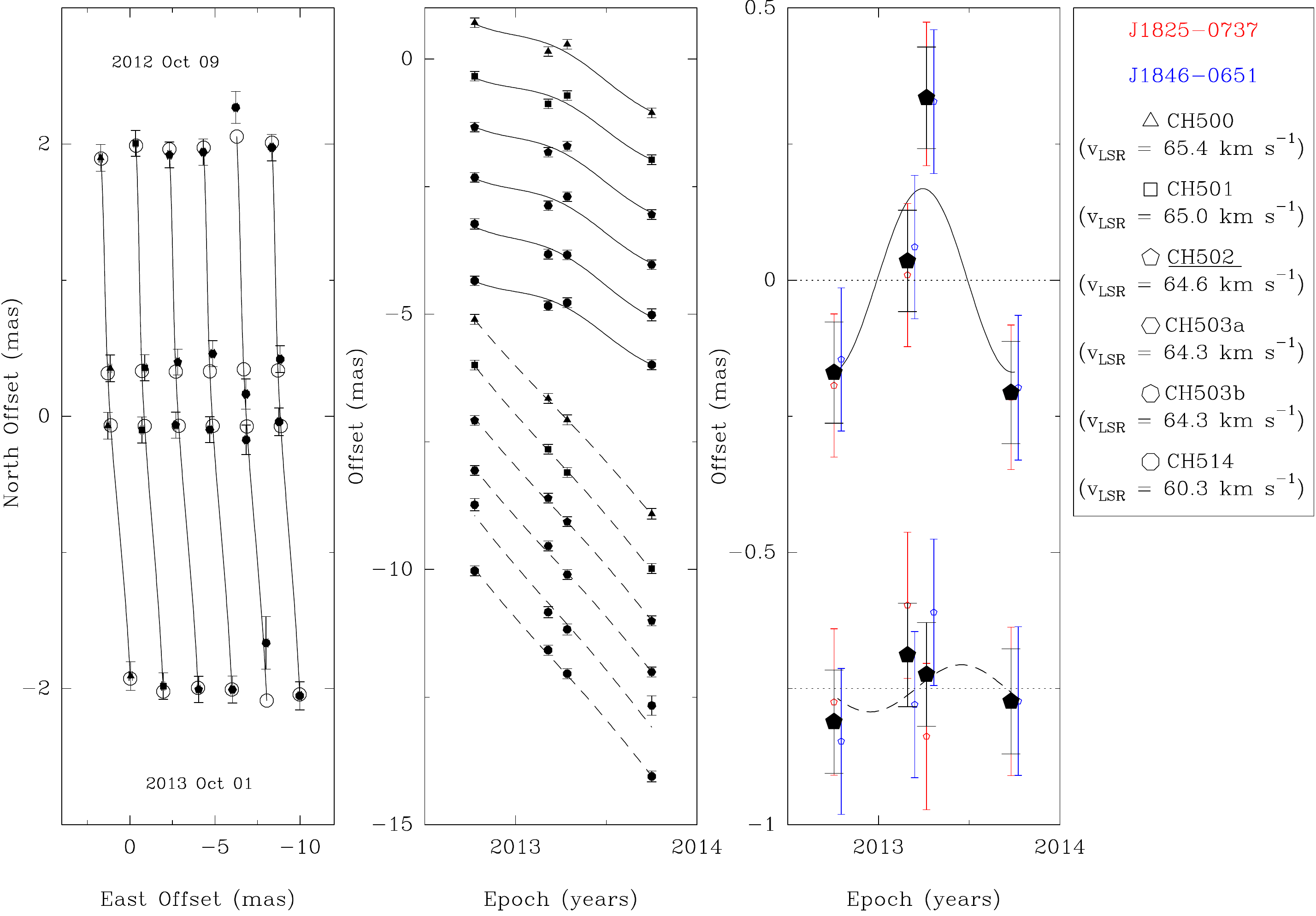}\label{G02325-Parallax}}\\
       \subfloat[22 GHz water maser $-$ G028.86$+$00.06]{\includegraphics[width=0.8\textwidth]{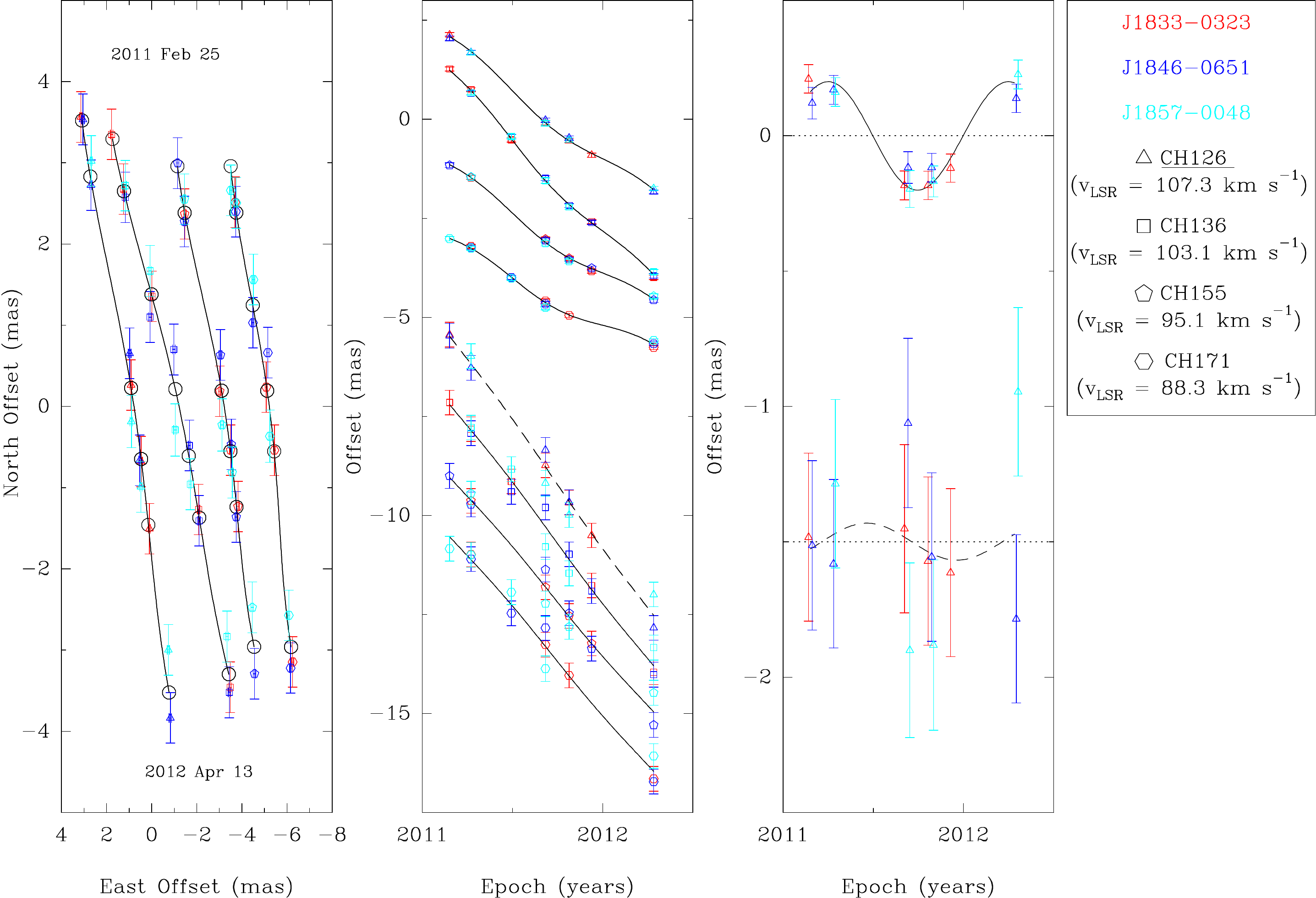}\label{G02886-Parallax}}\\
      \label{Parallaxes}
\end{figure*}

The parallax and proper motion results are listed in Table \ref{Source-Results}. 
The parallax fits are shown in Figs. \ref{Parallaxes} and \ref{Parallax-Appendix}. The internal motion
of eight maser sources are plotted in Fig. \ref{ProperMotion} in the online Appendix.

\begin{table*}[ht]
\centering
\caption{Parallax and proper motion results\tablefootmark{a}.}
\renewcommand{\arraystretch}{1.5}
\begin{tabular}{
  l
  >{$}c<{$}
  c
  >{$}c<{$}
  >{$}c<{$}
  >{$}c<{$}
}
\hline \hline
Source  &       {\pi}                   &       {Distance}                                      &       {\mu_x}\tablefootmark{b}                        &       {\mu_y}\tablefootmark{b}                        &       {\text{v}_{lsr}}                         \\[4pt]
        &       \text{[mas]}                    &       {[kpc]}                                 &       {[\text{mas yr}^{-1}]}                      &       {[\text{mas yr}^{-1}]}                  &       {[\text{km s}^{-1}]}                        \\[4pt]\hline
G023.25$-$00.24 &       0.169   \pm     0.051   &       5.92    $^{+2.56        }_{     -1.37   }$ &       -1.72   \pm     0.14    &       -4.06   \pm     0.20    &       63      \pm     3\\
G028.14$-$00.00 &       0.158   \pm     0.023   &       6.33    $^{+1.08        }_{     -0.80   }$ &       -2.11   \pm     0.03    &       -4.85   \pm     0.03    &       100     \pm     5\\
G028.86+00.06$^{*}$     &       0.201   \pm     0.019   &       4.98    $^{+0.52        }_{     -0.43   }$ &       -3.19   \pm     0.05    &       -5.72   \pm     0.24    &       100     \pm     10\\
G029.86$-$00.04$^{*}$\tablefootmark{c}  &       0.221   \pm     0.015   &       4.52    $^{+0.33        }_{     -0.28   }$ &       -2.30   \pm     0.01    &       -5.33   \pm     0.02    &       100     \pm     4\\
G029.95$-$00.01$^{*}$\tablefootmark{d}  &       0.222   \pm     0.018   &       4.50    $^{+0.40        }_{     -0.33   }$ &       -2.32   \pm     0.02    &       -5.38   \pm     0.02    &       99      \pm     5\\
G029.98+00.10$^{*}$     &       0.156   \pm     0.010   &       6.41    $^{+0.44        }_{     -0.39   }$ &       -1.39   \pm     0.03    &       -3.49   \pm     0.20    &       109     \pm     10\\
G030.22$-$00.18 &       0.284   \pm     0.032   &       3.52    $^{+0.45        }_{     -0.36   }$ &       -0.87   \pm     0.08    &       -4.70   \pm     0.15    &       113     \pm     3\\
G030.41$-$00.23$^{*}$   &       0.253   \pm     0.021   &       3.95    $^{+0.36        }_{     -0.30   }$ &       -1.98   \pm     0.09    &       -4.07   \pm     0.12    &       103     \pm     3\\
G030.70$-$00.06$^{*}$   &       0.153   \pm     0.020   &       6.54    $^{+0.98        }_{     -0.76   }$ &       -2.33   \pm     0.08    &       -5.13   \pm     0.09    &       89      \pm     3\\
G030.74$-$00.04$^{*}$   &       0.326   \pm     0.055   &       3.07    $^{+0.62        }_{     -0.44   }$ &       -2.38   \pm     0.16    &       -4.46   \pm     0.16    &       88      \pm     3\\
G030.78+00.20   &       0.140   \pm     0.032   &       7.14    $^{+2.12        }_{     -1.33   }$ &       -1.85   \pm     0.05    &       -3.91 \pm       0.12    &       82      \pm     5       \\
G030.81$-$00.05 &       0.321   \pm     0.037   &       3.12    $^{+0.41        }_{     -0.32   }$ &       -2.34   \pm     0.08    &       -5.66   \pm     0.16    &       105     \pm     5\\
G030.97$-$00.14 &       0.294   \pm     0.022   &       3.40    $^{+0.28        }_{     -0.24   }$ &       -2.14   \pm     0.02    &       -3.82   \pm     0.13    &       77      \pm     3       \\
G031.41+00.30$^{*}$     &       0.267   \pm     0.029   &       3.75    $^{+0.46        }_{     -0.37   }$ &       -1.38   \pm     0.07    &       -5.70   \pm     0.08    &       97      \pm     15\\
G032.04+00.05\tablefootmark{e}  &       0.195   \pm     0.008   &       5.13    $^{+0.22        }_{     -0.20   }$ &       -1.71   \pm     0.11    &       -4.51   \pm     0.13    &       98      \pm     5\\
G033.09$-$00.07 &       0.134   \pm     0.016   &       7.46    $^{+1.01        }_{     -0.80   }$ &       -2.46 \pm       0.02    &       -5.32   \pm     0.03    &       100     \pm     5\\ \hline \hline
\end{tabular}
\tablefoot{\\
\tablefoottext{a}{Additional 14 Scutum arm sources will be published in Li et al. (in prep.).}\\
\tablefoottext{b}{The proper motion values in the table correspond to the proper motion of the maser spot underlined in Figs. \ref{Parallaxes} and \ref{Parallax-Appendix}, except for sources marked with a star for which the proper motion corresponds to the proper motion estimate of the central object (see online Appendix \ref{PropMot-Measurements} for more details).}\\
\tablefoottext{c}{Weighted combined result of this work for the 6.7 GHz methanol maser ($\pi$=(0.307$\pm$0.024) mas, $\mu_x$ = ($-$2.36$\pm$0.07) mas yr$^{-1}$, $\mu_y$ = ($-$5.60$\pm$0.15) mas yr$^{-1}$, v$_{lsr}$=(101$\pm$3) km s$^{-1}$) and the results of \citet{Zhang2014} for the 12 GHz methanol maser ($\pi$=(0.161$\pm$0.020) mas, $\mu_x$ = ($-$2.30$\pm$0.01) mas yr$^{-1}$, $\mu_y$ = ($-$5.32$\pm$0.02) mas yr$^{-1}$, v$_{lsr}$=(100$\pm$3) km s$^{-1}$).}\\
\tablefoottext{d}{Weighted combined result of this work for the 6.7 GHz methanol maser ($\pi$=(0.352$\pm$0.040) mas, $\mu_x$ = ($-$2.38$\pm$0.12) mas yr$^{-1}$, $\mu_y$ = ($-$5.37$\pm$0.23) mas yr$^{-1}$, v$_{lsr}$=(99$\pm$5) km s$^{-1}$) and results of \citet{Zhang2014} for the 12 GHz methanol maser ($\pi$=(0.190$\pm$0.020) mas, $\mu_x$ = ($-$2.32$\pm$0.02) mas yr$^{-1}$, $\mu_y$ = ($-$5.38$\pm$0.02) mas yr$^{-1}$, v$_{lsr}$=(98$\pm$3) km s$^{-1}$).}\\
\tablefoottext{e}{Weighted combined result of this work for the 6.7 GHz methanol maser ($\pi$=(0.299$\pm$0.053) mas, $\mu_x$ = ($-$1.67$\pm$0.11) mas yr$^{-1}$, $\mu_y$ = ($-$4.47$\pm$0.14) mas yr$^{-1}$, v$_{lsr}$=(98$\pm$5) km s$^{-1}$) and results of \citet{Sato2014} for the 12 GHz methanol maser ($\pi$=(0.193$\pm$0.008) mas, $\mu_x$ = ($-$2.21$\pm$0.40) mas yr$^{-1}$, $\mu_y$ = ($-$4.80$\pm$0.40) mas yr$^{-1}$, v$_{lsr}$=(97$\pm$5) km s$^{-1}$).}\\
}
\label{Source-Results} 
\end{table*}

\subsection{Spiral arm association}

The masers of the BeSSeL survey are assigned to the different spiral arms 
by associating the high-mass star-forming regions, which harbor the masers, with
molecular clouds in the CO Galactic plane survey of \citet{Dame2001} and then 
associating the clouds with a spiral structure in l-v space. These assignments 
can be tested by locating the masers in plan view plots of the Milky Way, where 
they tend to trace out continuous arcs that can be identified as portions of 
spiral arms. Our sources are selected to be part of the Scutum spiral arm (see Li et al., in prep.).

\subsection{Peculiar motion}
\label{PeculiarMotion}

\begin{table*}
\centering
\caption{Peculiar motions calculated based on the Galactic parameters of \citet{Reid2019}.}
\renewcommand{\arraystretch}{1.2}
\begin{tabular}{
  l
  >{$}c<{$}
  >{$}c<{$}
  >{$}c<{$}
  c
}
\hline \hline
Source  &       {\text{U}_{s}}\tablefootmark{a}                 &       {\text{V}_{s}}\tablefootmark{b}                 &       {\text{W}_{s}}\tablefootmark{c} & Reference\\
        &       {[\text{km s}^{-1}]}                    &       {[\text{km s}^{-1}]}                       &       {[\text{km s}^{-1}]}
        & \\ \hline
G000.31$-$00.20   &  18  \pm  3 &    -6 \pm  6 &    -8 \pm 6 & (1)  \\
G005.88$-$00.39   &  -3  \pm  3 &   -13 \pm  5 &   -11 \pm 5 & (2)\\
G009.21$-$00.20   &  21  \pm 38 &    -5 \pm 41 &     4 \pm 7 & (1)  \\
G010.32$-$00.15   &   3  \pm  5 &   -72 \pm  5 &    16 \pm 1 & (1) \\
G010.34$-$00.14   &  -5  \pm  5 &   -20 \pm  4 &     8 \pm 2 & (1) \\
G011.10$-$00.11   &  -7  \pm  6 &    -9 \pm  6 &    -7 \pm 5 & (1) \\
G011.91$-$00.61   &   3  \pm  8 &     9 \pm  6 &   -12 \pm 5 & (2)\\
G012.68$-$00.18   &  41  \pm 10 &    -8 \pm  5 &     2 \pm 3 & (3) \\
G012.81$-$00.19   &   5  \pm  7 &     4 \pm  2 &     8 \pm 2 & (3) \\
G012.88$+$00.48   &  12  \pm  7 &    -9 \pm  3 &    -9 \pm 2 & Combined: (3), (4) \\
G012.90$-$00.24   &  16  \pm 10 &    -6 \pm  5 &    -8 \pm 2 & (3) \\
G012.90$-$00.26   &  17  \pm  5 &    -6 \pm  2 &    -1 \pm 1 & (3)\\
G013.87$+$00.28   &   8  \pm 21 &    -8 \pm 34 &   -10 \pm 38& (2)\\
G016.58$-$00.05   &  21  \pm  7 &    -8 \pm  6 &     4 \pm 6 & (2)\\
G018.87$+$00.05   &  -7  \pm 17 &     1 \pm  6 &    -4 \pm 6 & (1) \\
G019.36$-$00.03   &  -3  \pm 13 &   -16 \pm  8 &     5 \pm 6 & (1) \\
G022.03$+$00.22   &   1  \pm 15 &    13 \pm  6 &    -2 \pm 6 & (1) \\
G023.25$-$00.24   &     -21  \pm 43     &       -34 \pm 31 &    -2 \pm 7 & This work     \\
G023.65$-$00.12   &  33  \pm  9 &     9 \pm  3 &     5 \pm 1 & (5)\\
G024.63$-$00.32   & -25  \pm 26 &   -12 \pm 14 &   -12 \pm 7 & (1) \\
G025.70$+$00.04   & -34  \pm 34 &    -7 \pm 90 &    -6 \pm 11& (2)\\
G027.36$-$00.16   &     -83      \pm 48 &       -34     \pm     36 &    -3 \pm 8 & (4)     \\
G028.14$-$00.00   &  -7  \pm 28 &    -6 \pm  9 &    -3 \pm 2 & This work \\ 
G028.30$-$00.38   &  12  \pm 13 &     0 \pm  6 &    -1 \pm 5 & (1) \\ 
G028.39$+$00.08   & -21  \pm 10 &    14 \pm  5 &    -9 \pm 2 & (1) \\
G028.83$-$00.25   &  12  \pm 30 &    -8 \pm 11 &    -2 \pm 6 & (1)\\
G028.86$+$00.06   &  53  \pm 10 &   -15 \pm  9 &    13 \pm 3 & This work \\
G029.86$-$00.04   &  43  \pm  6 &    -4 \pm  4 &    -1 \pm 1 & Combined: This work, (6)\\
G029.95$-$00.01   &  43  \pm  7 &    -5 \pm  4 &    -1 \pm 1 & Combined: This work, (6)\\
G029.98$+$00.10   & -58  \pm 17 &    13 \pm 12 &    -3 \pm 3 & This work \\
G030.19$-$00.16   &  14  \pm  7 &    20 \pm  3 &    10 \pm 1  & (1)\\
G030.22$-$00.18   &  50  \pm 10 &    25 \pm  3 &   -16 \pm 3 & This work \\
G030.41$-$00.23   &  33  \pm  8 &    17 \pm  3 &     6 \pm 2 & This work\\
G030.70$-$00.06   &  -9  \pm 20 &   -15 \pm  5 &    -1 \pm 3 & This work\\
G030.74$-$00.04   &  45  \pm 10 &     6 \pm  5 &     9 \pm 2 & This work\\
G030.78$+$00.20   &     -67      \pm 36 &       -26     \pm     24 &     3 \pm 3 & This work     \\
G030.81$-$00.05   &  67  \pm  6 &     7 \pm  5 &     0 \pm 2 & This work \\
G030.97$-$00.14   &  25  \pm  5 &     2 \pm  3 &    10 \pm 1 & This work \\
G031.28$+$00.06   &  23  \pm 19 &    16 \pm  4 &     3 \pm 1 & Combined: (1), (6)\\
G031.41$+$00.30   &  49  \pm 11 &     2 \pm 12 &   -16 \pm 3 & This work \\
G031.58$+$00.07   &  -4  \pm 29 &     2 \pm 10 &     1 \pm 2 & (6)\\
G032.04$+$00.05   &   3  \pm  6 &     7 \pm  5 &    -5 \pm 3 & Combined: This work, (2) \\
G033.09$-$00.07   & -26  \pm 19 &    -3 \pm  5 &    -1 \pm 1 & This work \\ 
G033.39$+$00.00   & -52  \pm 32 &    -3 \pm 32 &    -2 \pm 7 & (1)\\ \hline \hline
\end{tabular}
\label{Source-Peculiar-Motion} 
\tablefoot{\\
\tablefoottext{a}{Increases toward the Galactic center.}\\
\tablefoottext{b}{Increases in the direction of Galactic rotation.}\\
\tablefoottext{c}{Increases toward the north Galactic pole.}\\
\tablefoottext{}{References. (1) Li et al. (in prep.); (2) \citet{Sato2014}; (3) \citet{Immer2013}; (4) \citet{Xu2011}; (5) \citet{Bartkiewicz2008}; (6) \citet{Zhang2014}}
}
\end{table*}

In the peculiar motion analysis we included four additional 12 GHz methanol masers and ten 22 GHz water masers from the literature (see Col. 7 of Table \ref{Source-Peculiar-Motion}), as well as 14 maser sources that will be published and discussed in more detail in Li et al. (in prep.), yielding a total of 44 maser sources tracing the Scutum spiral arm. 

\begin{figure*}
        \centering
        \caption{Positions and peculiar motion vectors of the Scutum arm masers (blue: this paper, white: Li et al. (in prep.), gray: literature) for sources with uncertainties on the vector components of less than 20 km s$^{-1}$. A representative velocity vector of 20 km s$^{-1}$ is shown in the upper right corner. The red shaded area around this vector displays the median uncertainty on the velocity vectors ($\Delta$U$_{S}$ = 10 km s$^{-1}$, 
        $\Delta$V$_{S}$ = 5 km s$^{-1}$). The 22 GHz water, 6.7 GHz methanol, and 12 GHz methanol masers are indicated with circles, squares, and diamonds, respectively. The positions of the Sun and the Galactic center are indicated. The dotted lines represent the distance uncertainty on each source. The black and gray dashed lines represent the position and width of the Scutum spiral arm, as described in \citet{Sato2014}.}
        \includegraphics[width=0.8\textwidth]{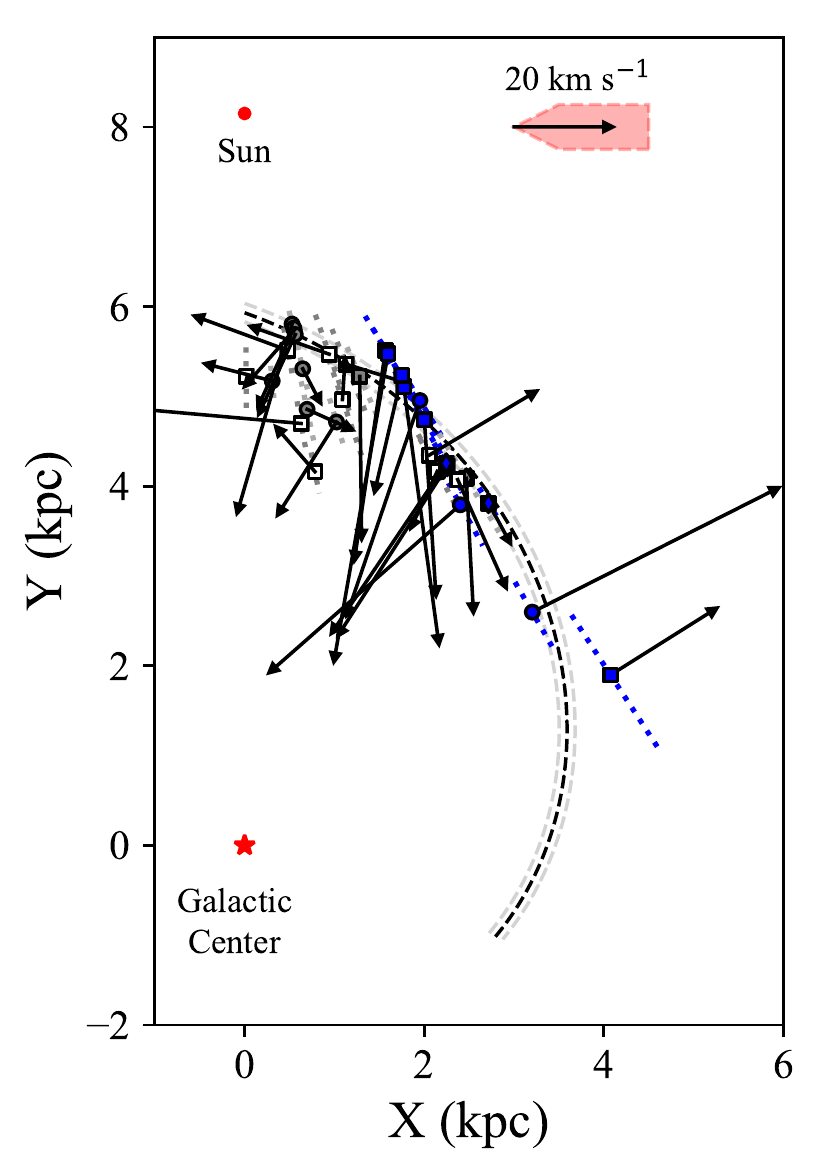}
        \label{Scutum-Arm-PeculiarMotion}
\end{figure*}

Transforming the measured three-dimensional motions of the 44 Scutum arm masers (proper motion plus line-of-sight 
velocity) to a reference frame that rotates with the Galactic disk, we yield their peculiar 
motion (U$_{s}$, V$_{s}$, W$_{s}$), where U$_{s}$ increases toward the Galactic 
center, V$_{s}$ in the direction of Galactic rotation, and W$_{s}$ toward the north Galactic pole.
As the rotation model, we used a Universal rotation curve \citep{Persic1996} with the \textit{A5} fit values of \citet[][Table 3]{Reid2019} and a distance to the Galactic center of 8.15 kpc, giving a rotation speed of 236 km s$^{-1}$. The motion of the Sun 
in this reference frame is U$_{\odot}$ = 10.6 km s$^{-1}$, V$_{\odot}$ = 10.7 km s$^{-1}$, and 
W$_{\odot}$ = 7.6 km s$^{-1}$. The calculated peculiar motion components of the Scutum arm sources are listed in Table 
\ref{Source-Peculiar-Motion}. The medians of the uncertainties on the U$_{S}$, V$_{S}$, and W$_{S}$ components are 10 km s$^{-1}$, 5 km s$^{-1}$, and 3 km s$^{-1}$, respectively.
Figure \ref{Scutum-Arm-PeculiarMotion} shows the peculiar motion of the Scutum maser sources with uncertainties of less than 20 km s$^{-1}$ in the plane of the Galaxy. 

\section{Discussion}
\label{Discussion}

\begin{figure*}
        \centering
     \caption{Distribution of the three peculiar motion vector components U$_{S}$ (upper panels), V$_{S}$ (middle panels), and W$_{S}$ (lower panels) for the Scutum arm sources (left) and the combined distribution of the Sagittarius, Local, and Perseus arms sources (right) with uncertainties on the peculiar motion components of less than 20 km s$^{-1}$.}
     \subfloat{\includegraphics[width=0.45\textwidth]{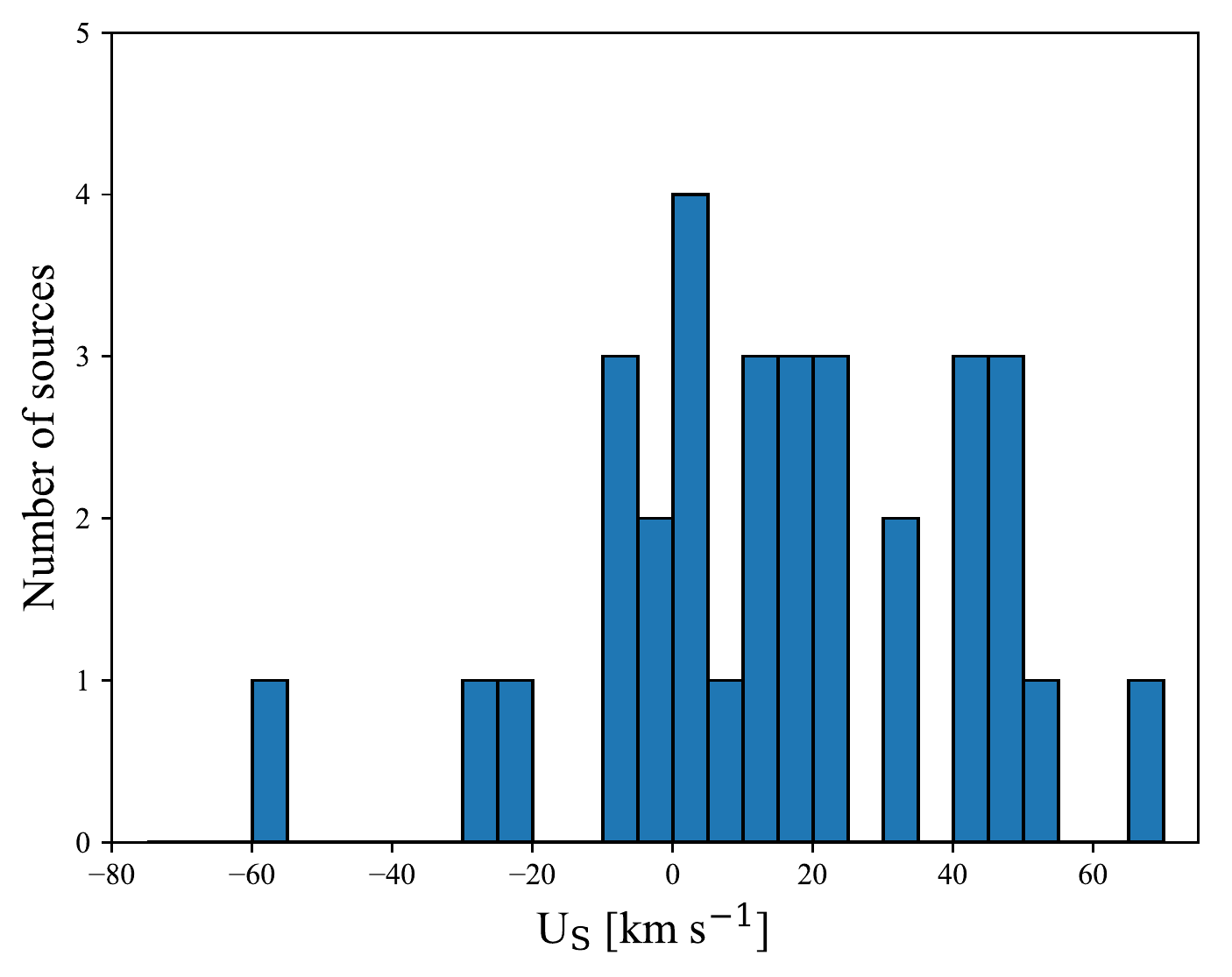}}
    \subfloat{\includegraphics[width=0.45\textwidth]{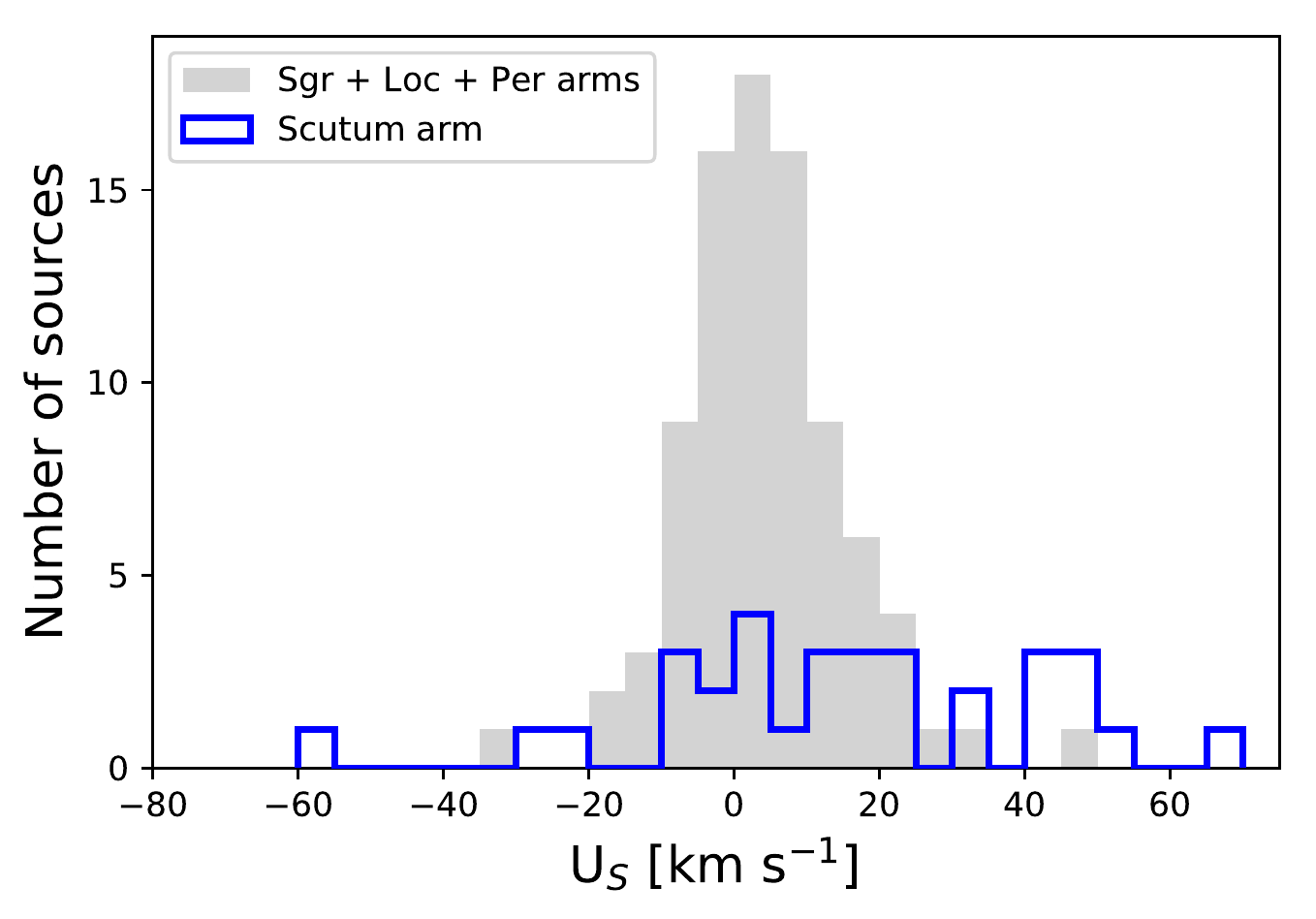}}\\
    \subfloat{\includegraphics[width=0.45\textwidth]{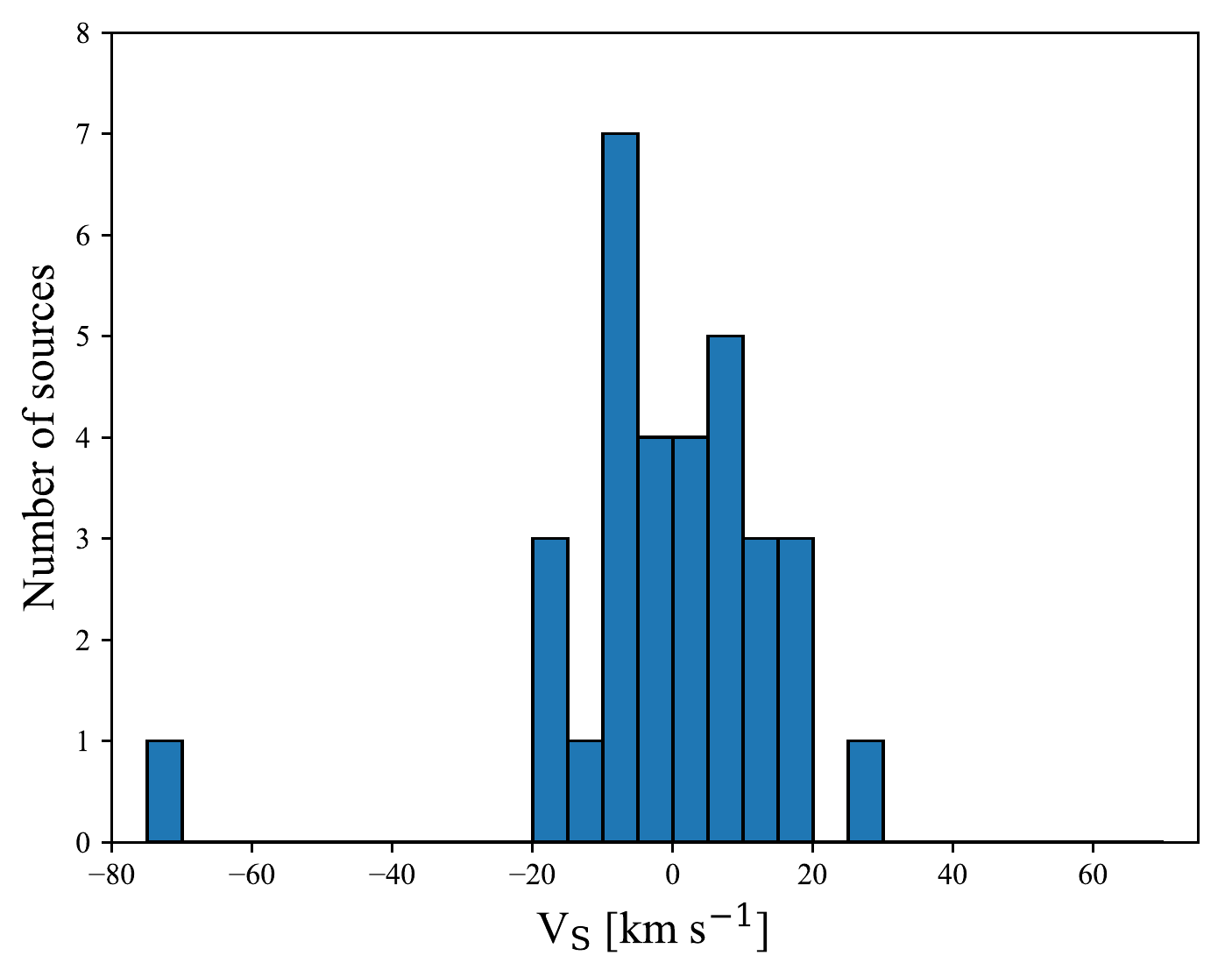}}
    \subfloat{\includegraphics[width=0.45\textwidth]{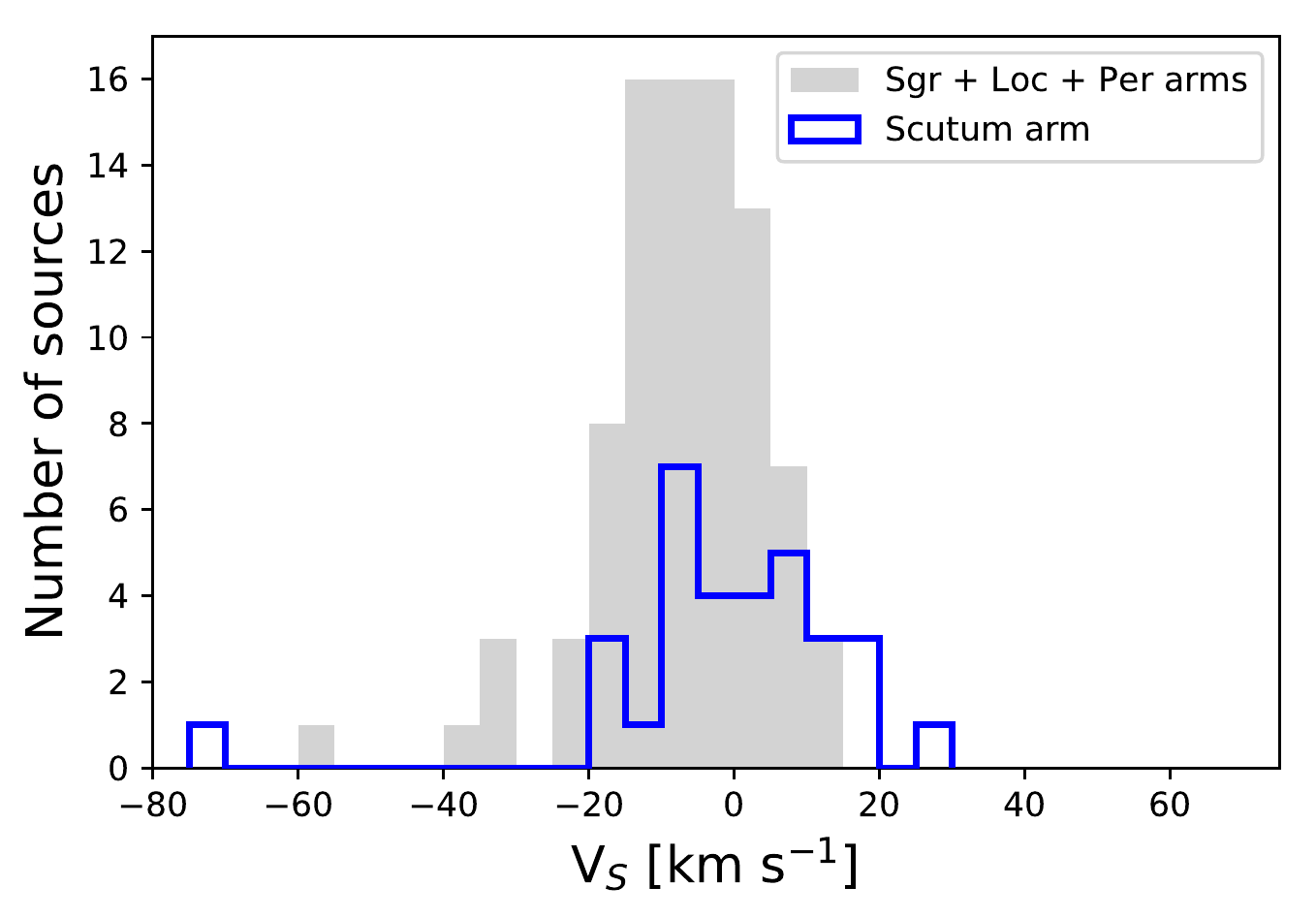}}\\
    \subfloat{\includegraphics[width=0.45\textwidth]{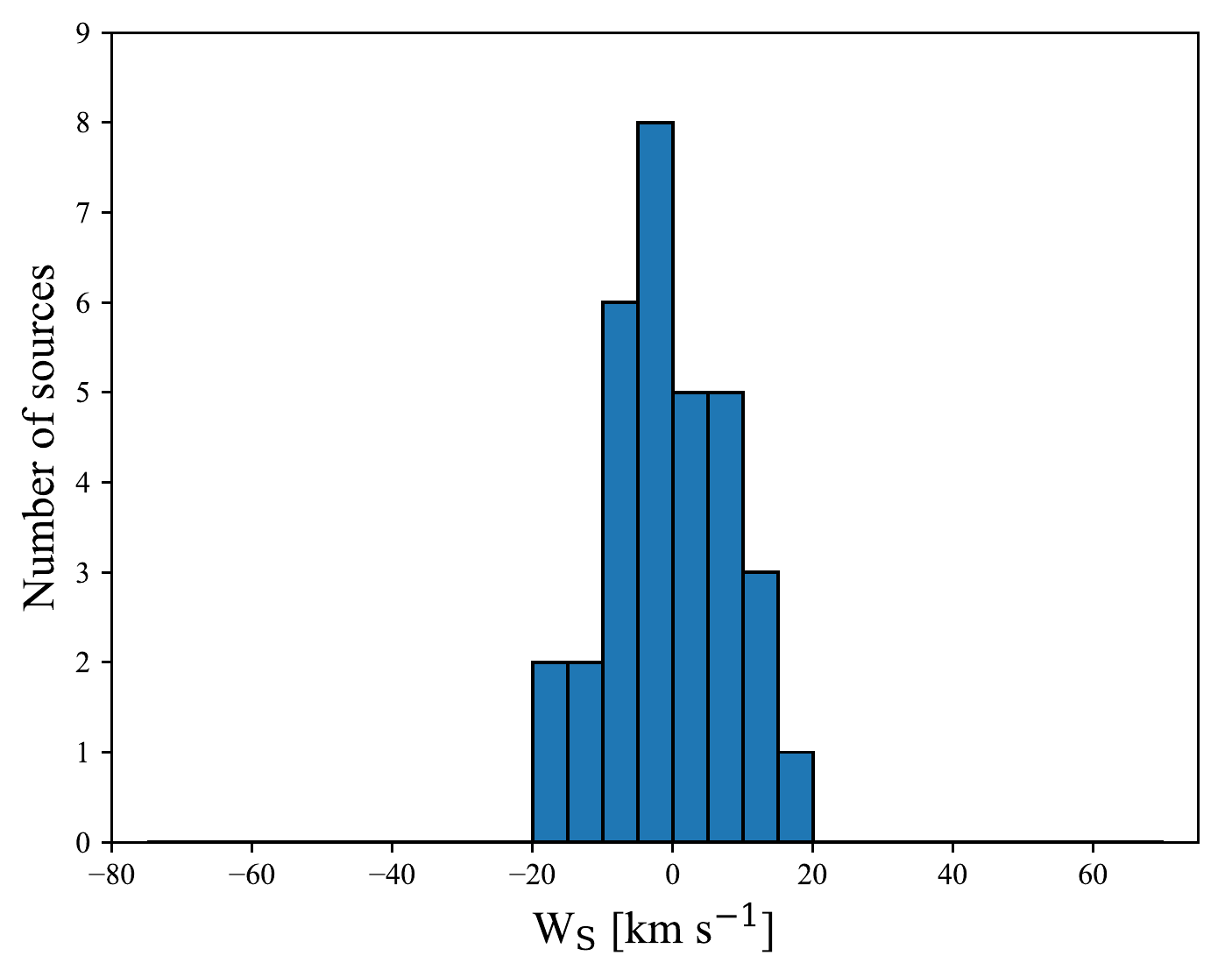}}
        \subfloat{\includegraphics[width=0.45\textwidth]{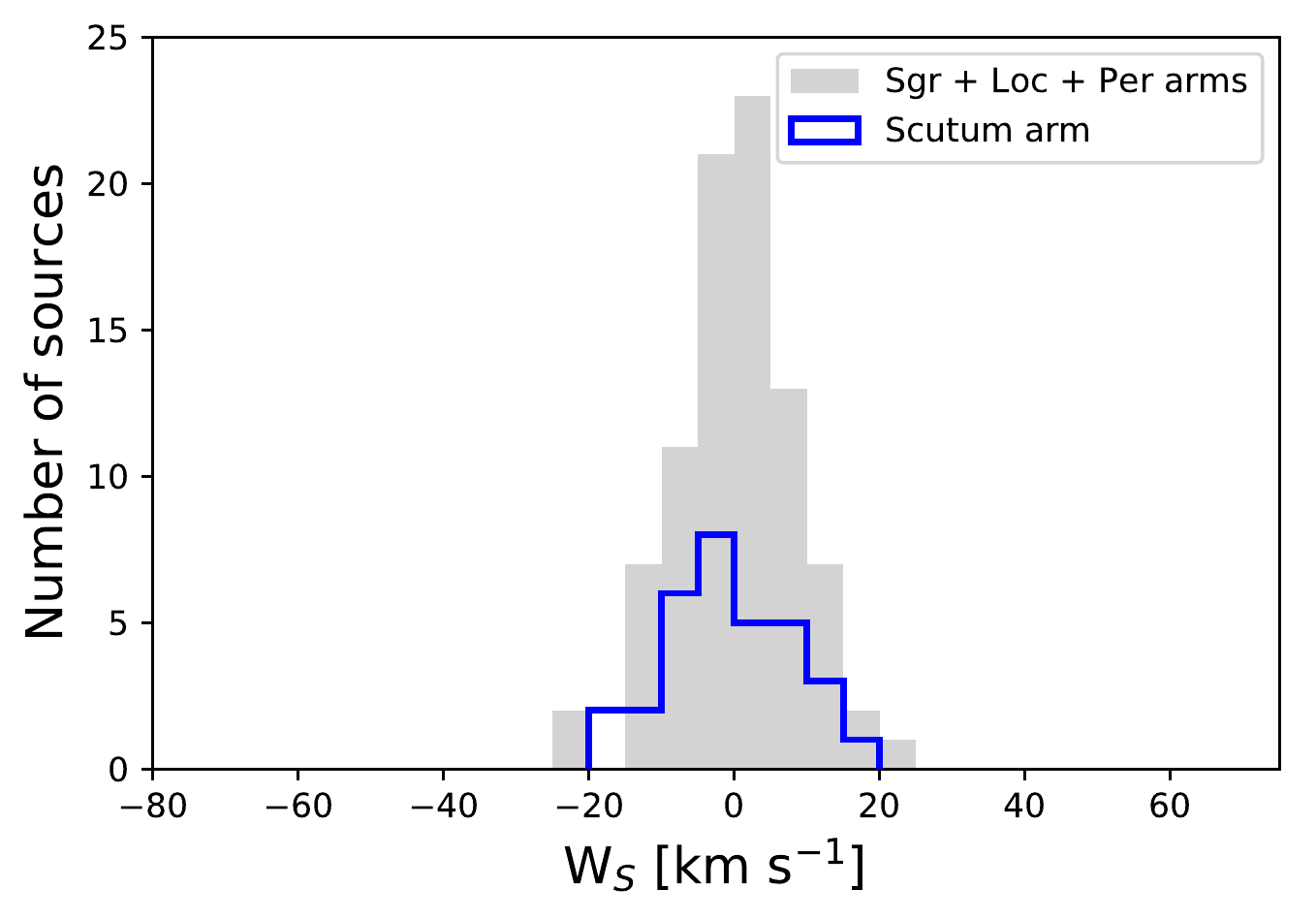}}
        \label{Scutum-Arm-PeculiarMotion-Histograms}
\end{figure*}

The left panels in Fig. \ref{Scutum-Arm-PeculiarMotion-Histograms} show the distribution of the peculiar motion vector components U$_{S}$, V$_{S}$, and W$_{S}$ for sources with an uncertainty on the components of less than 20 km s$^{-1}$ (32 maser sources). For the V$_{S}$ and W$_{S}$ components, the mean and median of the distributions are close to zero and most of the sources have values for these two components between $-$20 and 20 km s$^{-1}$. For U$_{S}$, however, the mean and median are 15.7 and 15.3 km s$^{-1}$, respectively, and half of the sources have velocity magnitudes > 20 km s$^{-1}$, indicating that the U$_{S}$ component in these sources dominates the peculiar motion. In 13 of these sources, their motion is pointed toward  the inner Galaxy. In the following section, we define the high peculiar motion sources as sources with |U$_{S}$| > 20 km s$^{-1}$. We also included G010.32$-$00.15 in this group since this maser shows large peculiar motions as well, but in the V$_{S}$ component.

Class II methanol masers (6.7, 12 GHz) are mostly located close to the protostar, while 22 GHz water masers are often detected in the outflow regions \citep[e.g.,][]{Hofner1996,Norris1998,Minier2000,Goddi2011}. Since these high peculiar motions are detected in sources with all three types of masers, it is unlikely that they are caused by internal motions in the star-forming regions due to outflows for example.

To understand if there is a fundamental difference in the characteristics of the host clouds of the high and the low peculiar motion masers, we compared the cloud masses, dust temperatures, bolometric luminosities, and H$_{2}$ column densities of our sources from the APEX Telescope Large Area Survey of the Galaxy (ATLASGAL) clump catalog of \citet{Urquhart2018} (Fig. \ref{Scutum-Arm-Characteristica}). The masses were corrected for our measured parallax distances. For each of the four characteristics, we conducted a Kolmogorov-Smirnov two-sample test to determine the probability that the low and high peculiar motion maser samples are drawn from the same distribution. For the cloud masses, dust temperatures, bolometric luminosities, and H$_{2}$ column densities, the test yielded p-values of 0.82, 0.26, 0.10, and 0.96, respectively. Thus, for the bolometric luminosity the probability is only 10\% that the two samples come from the same distribution.

To test whether these high peculiar motion sources are unique in the Galaxy or also detected in other spiral arms, we compared the distributions of the peculiar motion components in the Scutum arm with maser sources in the Sagittarius, Local, and Perseus arms. The source information was taken from \citet{Reid2014}, applying the same rotation model as for the Scutum arm sources. The distribution of the V$_{S}$ and W$_{S}$ components are similar. Although some sources in the other spiral arms show high velocities in the U$_{S}$ component as well, their occurrence is much lower than in the Scutum spiral arm (eight sources with |U$_{S}$| > 20 km s$^{-1}$ in the three arms combined compared to 16 sources in the Scutum arm). With a Kolmogorov-Smirnov two-sample test, we tested the hypothesis that the samples of the peculiar motion components of the Scutum arm and the other three spiral arms are drawn from the same distribution, yielding p-values of 0.0005, 0.02, and 0.78 for U$_{S}$, V$_{S}$, and W$_{S}$, respectively. Thus, for the U$_{S}$ and V$_{S}$ components, the probability is less than 5\% that the Scutum arm sources are drawn from the same distribution as the other spiral arm sources.
These results suggest that a local phenomenon led to the high peculiar motions of these clouds.

\begin{figure*}
        \centering
        \caption{Same as Fig. \ref{Scutum-Arm-PeculiarMotion}, but only the peculiar motion vectors of the high peculiar motion sources are shown (cyan). The black and gray dashed lines represent the position and width of the Scutum spiral arm as described in \citet{Sato2014}. The dotted blue, dot-dashed brown, and dashed magenta lines show the locations of the bars of \citet{Bissantz2002}, \citet{Wegg2015}, and \citet{Lopez-Corredoira2007}, respectively. A minor axis of 1.2 kpc is assumed for all three bar structures.}
    \includegraphics[width=0.8\textwidth]{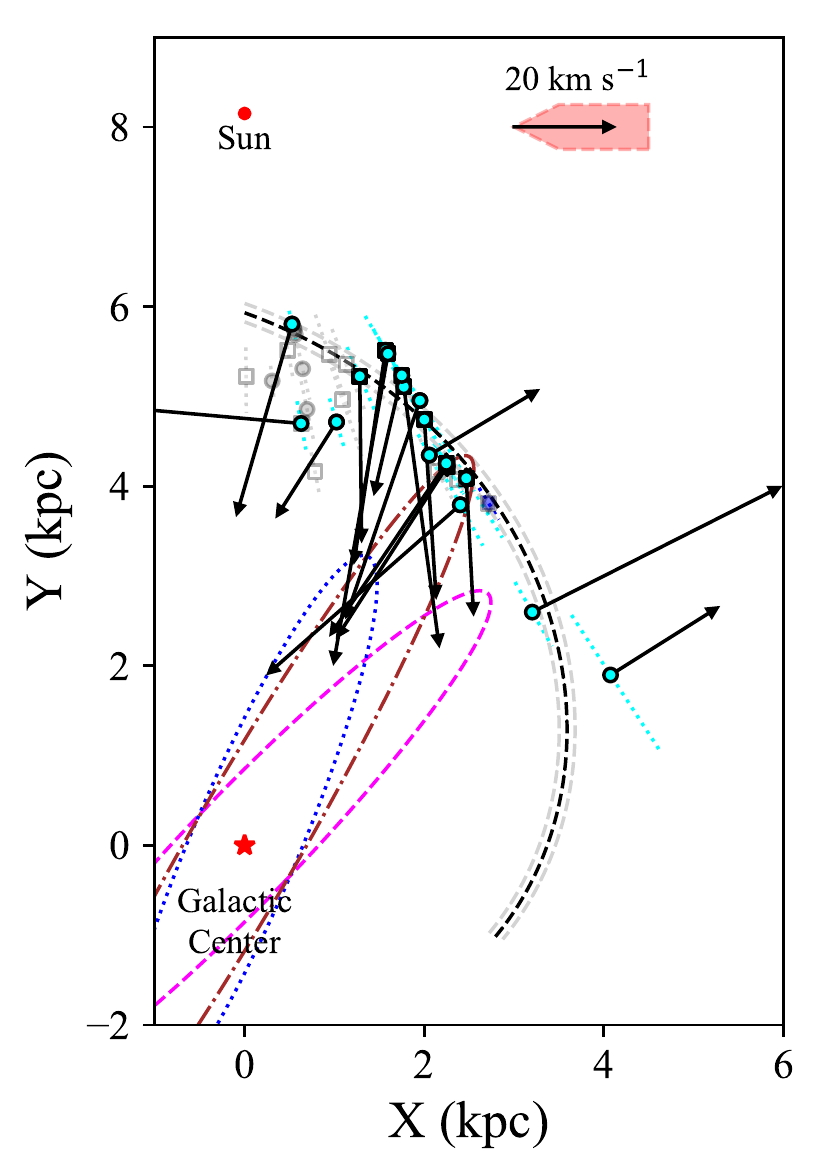}
        \label{Scutum-Arm-PeculiarMotion-LargeVel}
\end{figure*}

\paragraph{Gravitational attraction of the spiral arm and the Galactic bar}
\hspace{-2pt}

The formation of spiral arms in the Milky Way has been historically described via the spiral density-wave theory \citep[for a recent review, see][]{Lin1964,Shu2016}. Gas within the co-rotation radius of the Galaxy \citep[$\sim$8.5 kpc,][]{Dias2019} moves faster than the wave itself and is  accelerated radially outward when entering the interior side of the arm. Thus, this theory cannot explain the motion of our high peculiar motion sources, which is mostly directed radially inward. Including large-scale shock waves \citep[e.g.,][]{Roberts1969} or magnetic fields \citep[e.g.,][]{Roberts1970} can give the correct sign for the peculiar motions, but their measured magnitudes might be too large to be easily explained by these models. Since we do not see many high peculiar motion sources in most spiral arms, this cannot be the full answer.

\citet{deVaucouleurs1970-cap} first suggested that the Milky Way is a barred spiral galaxy. There is still a dispute about the exact shape and orientation of the bar. 
While some authors favor a shorter structure with a length of 2.5$-$3.5 kpc and a position angle with respect to the Sun--Galactic center direction of $\sim$23$\degree$ \citep[e.g.,][]{Bissantz2002,Gerhard2002,Babusiaux2005}, others identified a long bar with a length of 4 kpc and a position angle of 45$\degree$ \citep[e.g.,][]{Hammersley2000,Benjamin2005, Lopez-Corredoira2007, Anders2019}.
 Recently, \citet{Wegg2015} identified a long bar with a half-length of 5 kpc and a position angle of 28$-$33$\degree$. It is not clear if these structures are fundamentally different or how they are related.
Figure \ref{Scutum-Arm-PeculiarMotion-LargeVel} shows the bar structures in relation to the position of the Scutum arm sources (for all three bars, a minor axis of 1.2 kpc is assumed, as in \citealt{Lopez-Corredoira2007}). 

\citet{Lopez-Corredoira2007} estimated the mass of the Galactic bar as 6$\cdot$10$^{9}$ M$_{\sun}$. Using this mass for the bar, we determined its gravitational potential and subsequently the terminal velocity caused solely by the bar at the distance of the high peculiar motion sources which yields 95$-$130 km s$^{-1}$ for the 3.5 kpc bar, 95$-$165 km s$^{-1}$ for the 4 kpc bar, and 105$-$145 km s$^{-1}$ for the 5 kpc bar. Since the terminal velocity is larger than the peculiar motion of the high peculiar motion sources, the gravitational attraction of the bar is a possible explanation for these anomalous values. We do not expect the sources to move with the terminal velocity since they are not in free-fall and there is friction within the spiral arm. The Galactic models of \citet{Roberts1979} showed that large noncircular motions with radial velocities as high as 100 km s$^{-1}$ are expected near the ends of the bar.

The peculiar motion vectors of eight high peculiar motion sources are oriented toward the tip of the short bar, hinting toward the gravitational potential of this structure being the prominent attractor. The long bar of \citet{Wegg2015} crosses the spiral arm at l$\sim$31$\degree$ at a distance from the Sun of $\sim$5 kpc where several high peculiar motion sources are located.

Although the clustering of the high peculiar motion sources in the Scutum arm hints at the potential of the bar as the dominating potential, we cannot exclude the importance of the spiral arm when in interplay with the potential of the bar. Thus, we conclude that the most likely explanation for the high peculiar motions is the combined gravitational potential of the spiral arm and the Galactic bar.

\citet{Baba2009} have shown in their N-body+hydrodynamical simulations of a barred spiral galaxy that peculiar motions of more than 25 km s$^{-1}$ can be reproduced. However, their simulated peculiar motions appear randomly distributed over the Galaxy, and are thus not consistent with the results of the BeSSeL survey. 
The simulations of \citet{Khoperskov2019} show large radial velocities toward the Galactic center of ~20 km s$^{-1}$ within 0 kpc < X < 3 kpc and 3 kpc < Y < 6 kpc and a change in the direction of the peculiar motion from inward to outward near X = 4 kpc and Y = 2.5 kpc, which is consistent with most of our peculiar motions.

\section{Conclusion}

In this publication, we present new parallax and proper motion measurements of 16 maser sources in the Scutum spiral arm. Together with 14 maser sources from the literature and another 14 maser sources, which will be published in Li et al. (in prep.), we determined the peculiar motions in the Scutum arm. The main conclusions of our paper can be summarized as follows:

\begin{enumerate}
    \item Sixteen sources in the Scutum spiral arm show large peculiar motions. Thirteen of these motion vectors are oriented toward the inner Galaxy.
    \item These high peculiar motions are measured in all three maser types of our sources: 6.7 GHz methanol masers, 12 GHz methanol masers, and 22 GHz water masers.
    \item There is a hint that the masers with high peculiar motions are hosted in clouds that are on average more luminous than those of the low peculiar motion masers.
    \item This large number of sources with high peculiar motions is unique in our Galaxy; other spiral arms only harbor a small number of sources with large peculiar motions. This suggests that  a local phenomenon is the cause for these larger peculiar motions.
    \item The most likely explanation for the large peculiar motions is the combined gravitational attraction of the spiral arm and Galactic bar potential.
\end{enumerate}

\begin{acknowledgements}
We thank the anonymous referee for many helpful comments. K.L.J.R. is funded by the Advanced European Network of E-infrastructures for Astronomy with the SKA (AENEAS) project, supported by the European Commission Framework Programme Horizon 2020 Research and Innovation action under grant agreement No.~731016.
\end{acknowledgements}


\appendix
\label{Appendix}

\section{Observations}


\section{Maser spectra}

\begin{figure*}[ht]
    \caption{Spectra of the 16 maser sources listed in Table \ref{Source-Results}. Each spectrum is an amplitude scalar average of one-second integrations of all data on all baselines.}
        \centering
       \subfloat[6.7 GHz methanol maser $-$ G023.25$-$00.24 (third epoch)]{\includegraphics[width=0.5\textwidth]{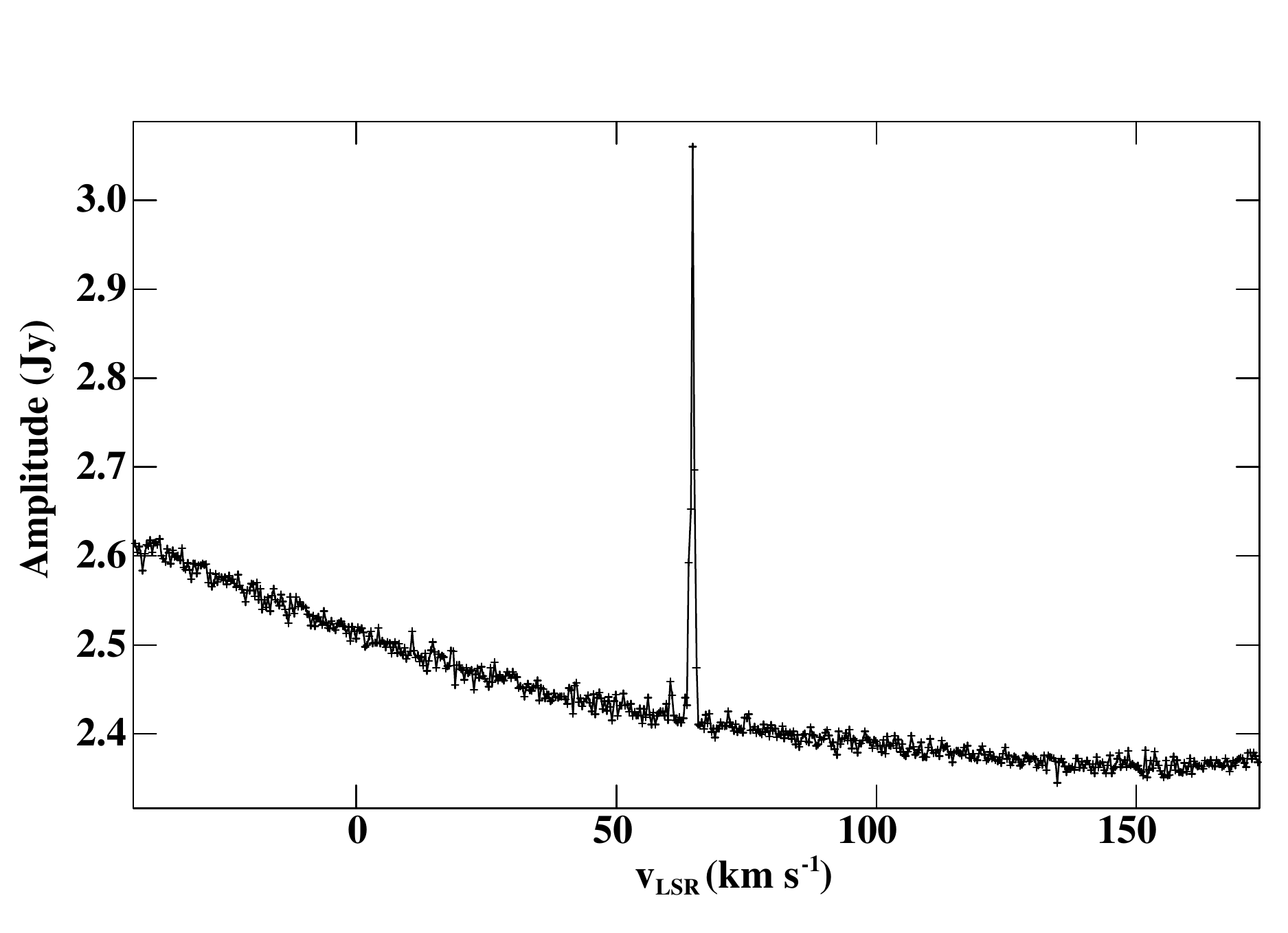}\label{G02325-Spectrum}}
       \subfloat[6.7 GHz methanol maser $-$ G028.14$-$00.00 (first epoch)]{\includegraphics[width=0.5\textwidth]{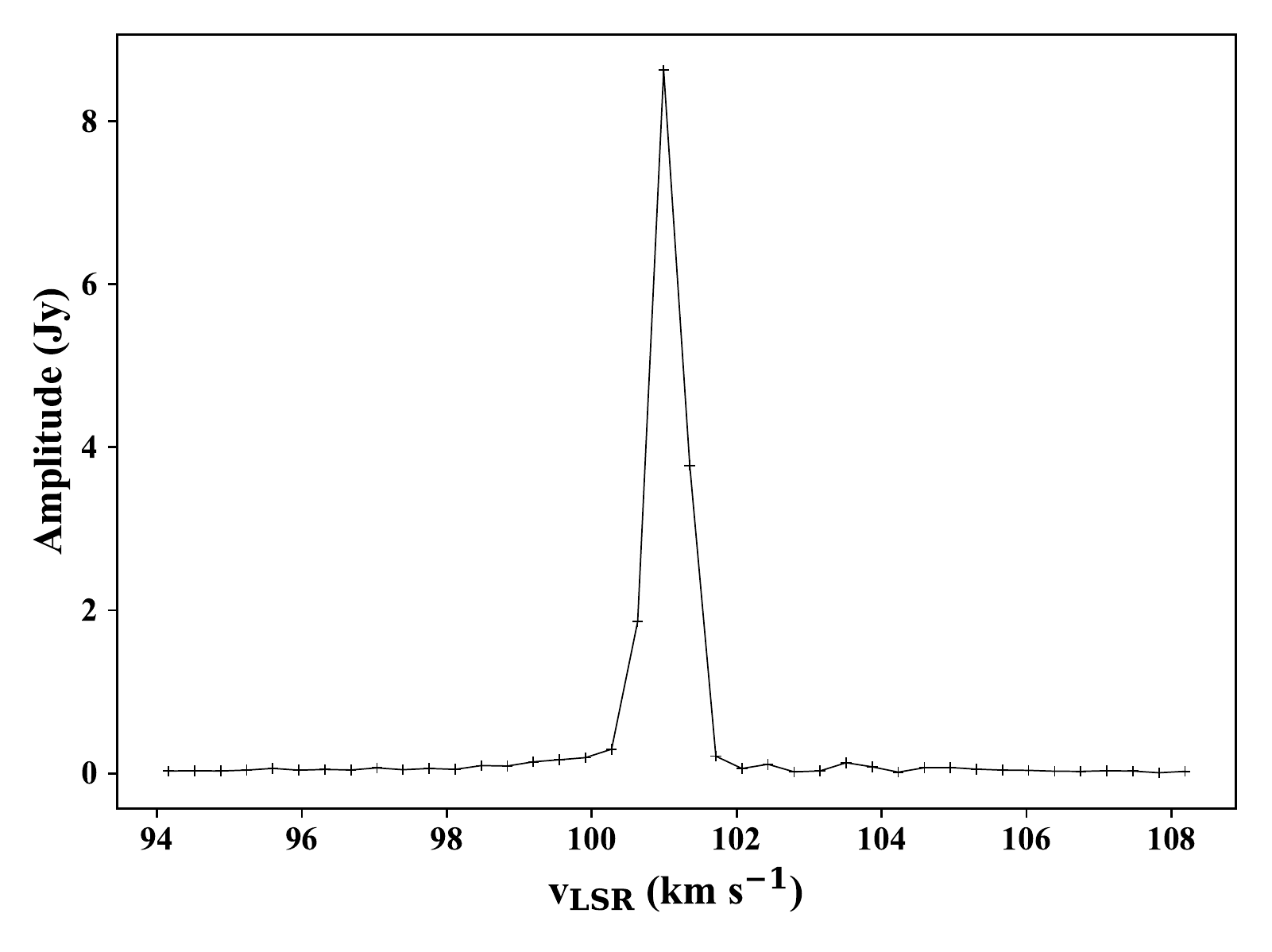}\label{G02814-Spectrum}}

       \subfloat[22 GHz water maser $-$ G028.86$+$00.06 (second epoch)]{\includegraphics[width=0.5\textwidth]{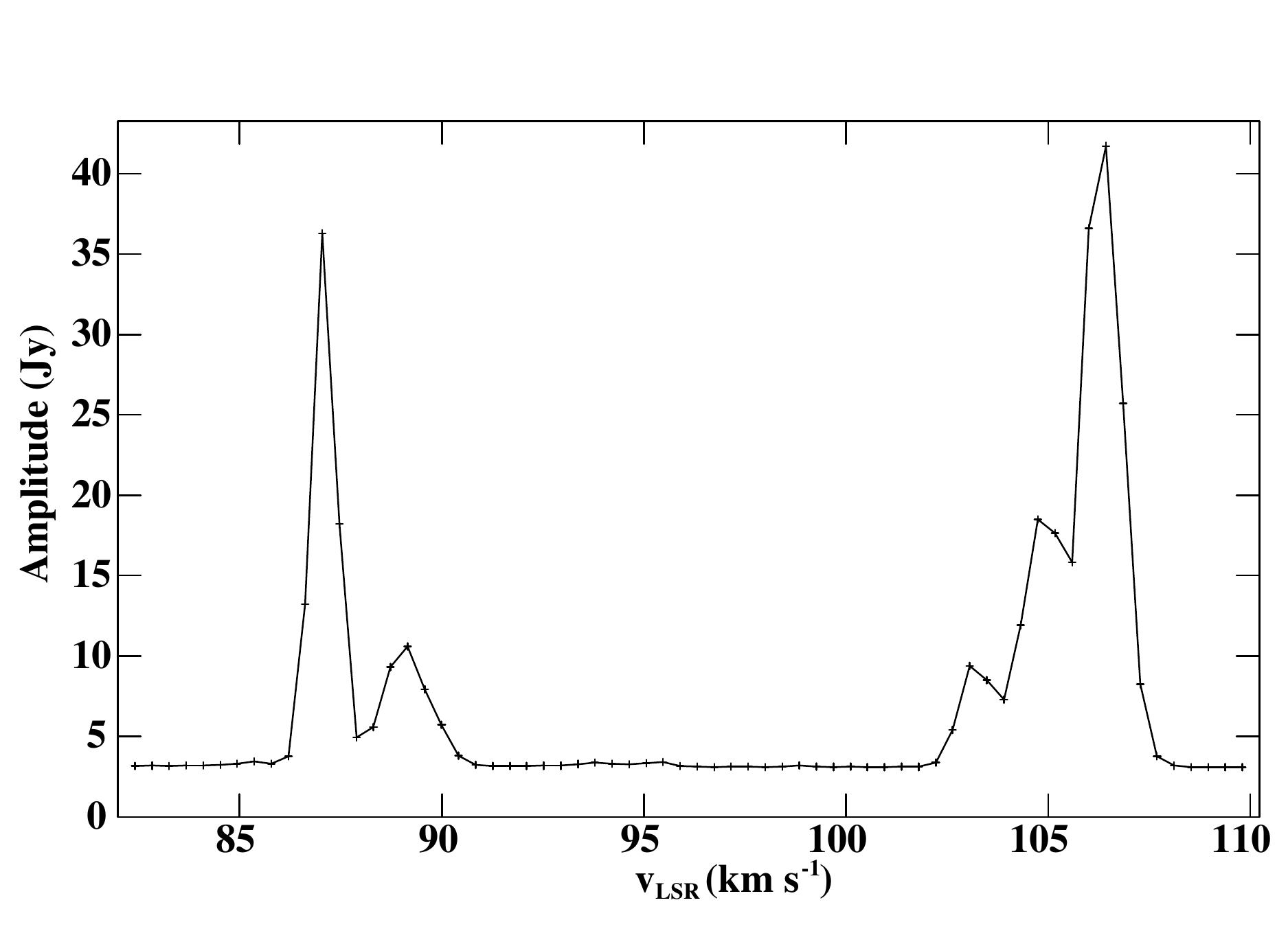}\label{G02886-Spectrum}}
       \subfloat[6.7 GHz methanol maser $-$ G029.86$-$00.04 (second epoch)]{\includegraphics[width=0.5\textwidth]{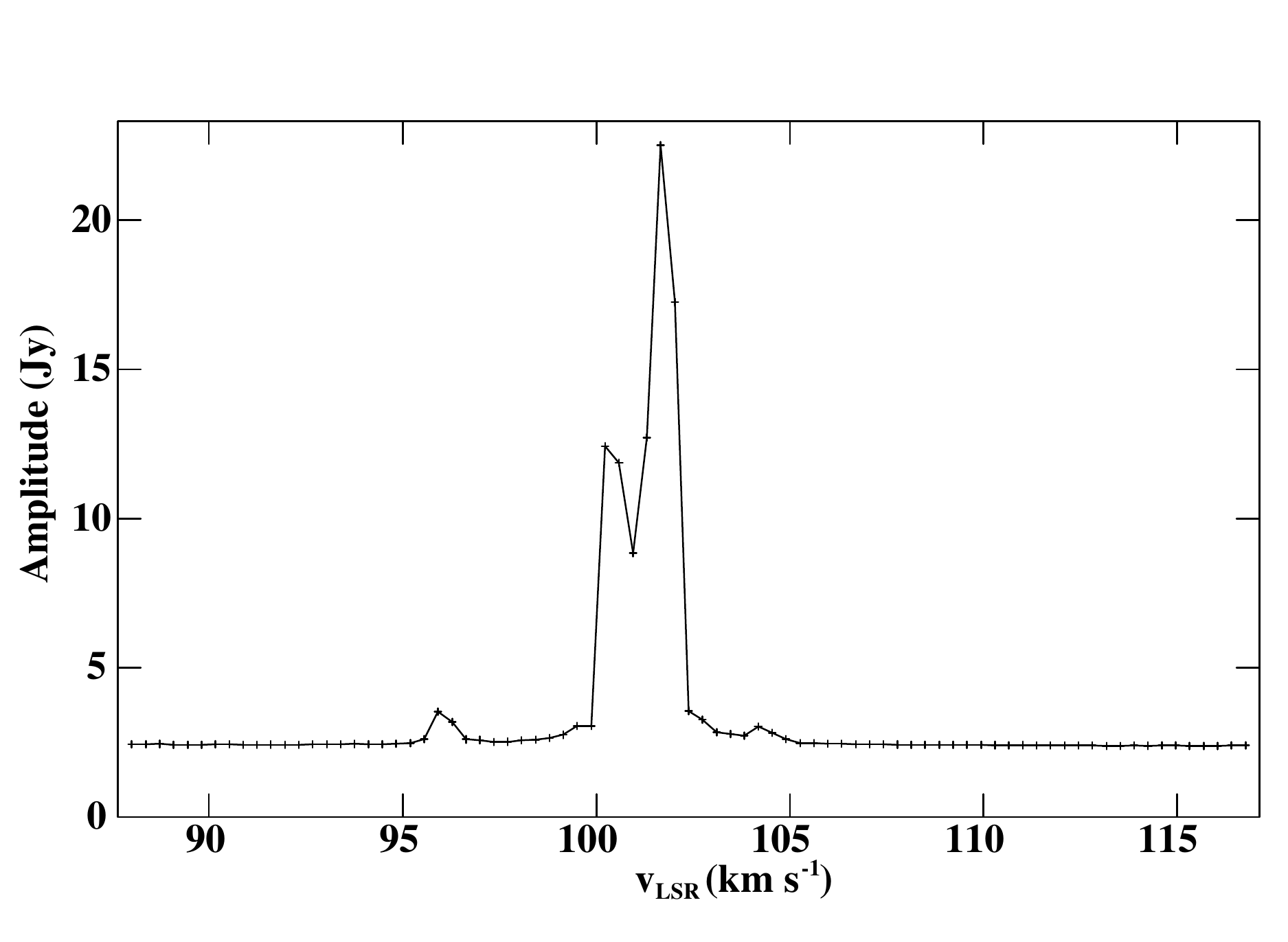}\label{G02986-Spectrum}}

       \subfloat[6.7 GHz methanol maser $-$ G029.95$-$00.01 (second epoch)]{\includegraphics[width=0.5\textwidth]{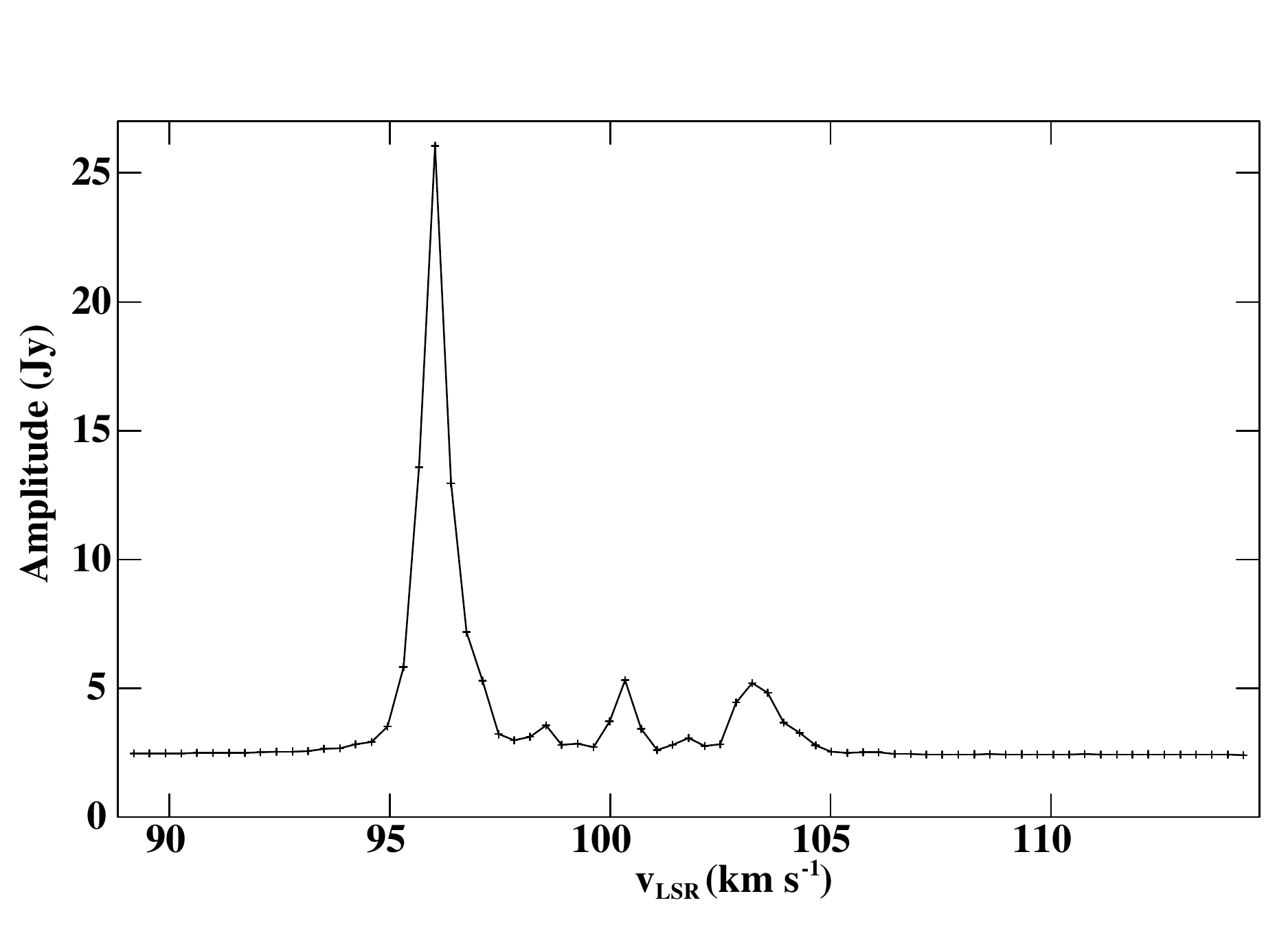}\label{G02995-Spectrum}}
       \subfloat[22 GHz water maser $-$ G029.98+00.10 (second epoch)]{\includegraphics[width=0.5\textwidth]{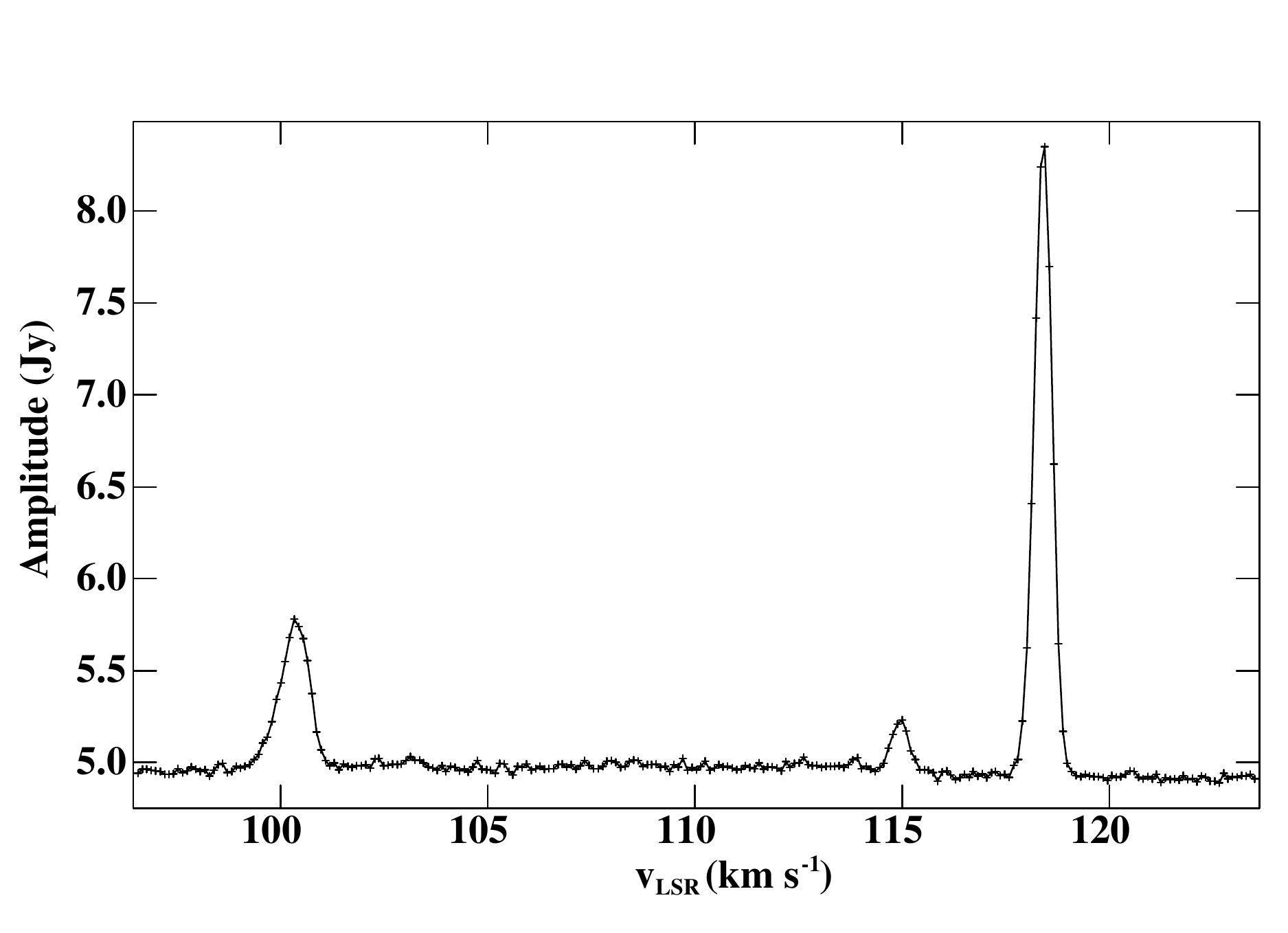}\label{G02998-Spectrum}}
        \label{Spectra}
\end{figure*}

\addtocounter{figure}{-1}

\begin{figure*}[ht]
        \caption{Continued.}
        \centering
                \addtocounter{subfigure}{6}
       \subfloat[6.7 GHz methanol maser $-$ G030.22$-$00.18 (second epoch)]{\includegraphics[width=0.5\textwidth]{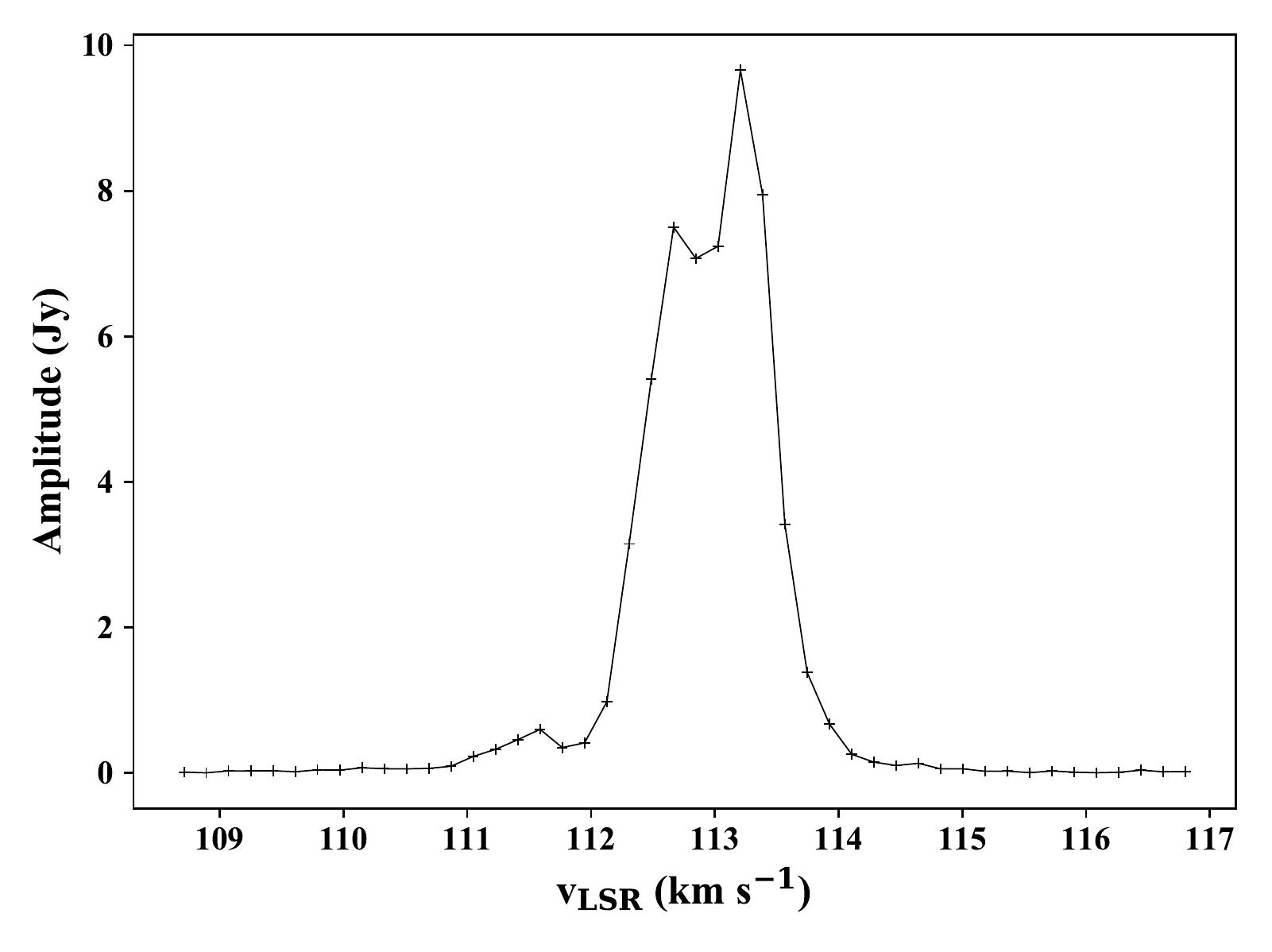}\label{G03022-Spectrum}}
       \subfloat[6.7 GHz methanol maser $-$ G030.41$-$00.23 (first epoch)]{\includegraphics[width=0.5\textwidth]{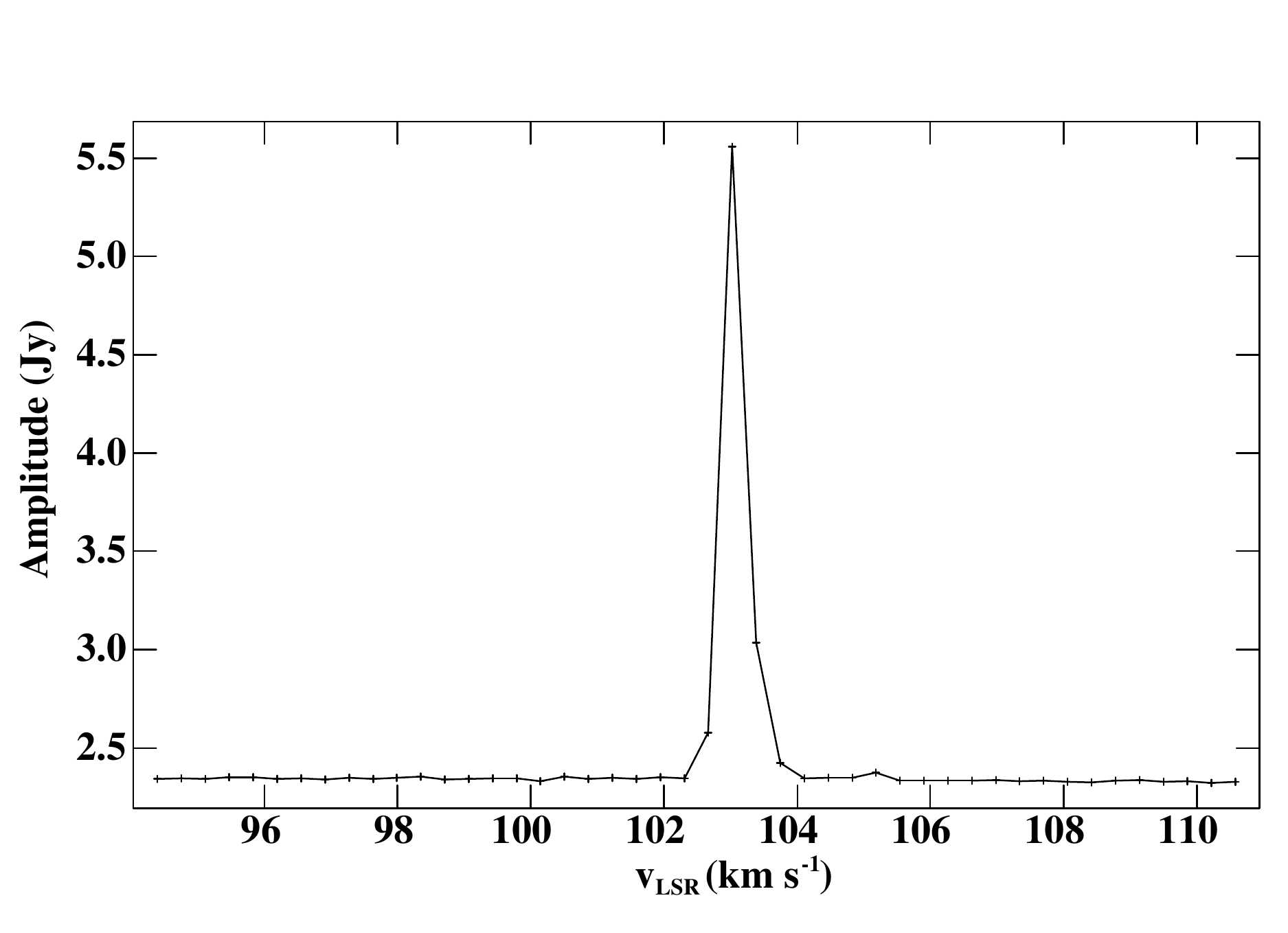}\label{G03041-Spectrum}}

       \subfloat[6.7 GHz methanol maser $-$ G030.70$-$00.06 (second epoch)]{\includegraphics[width=0.5\textwidth]{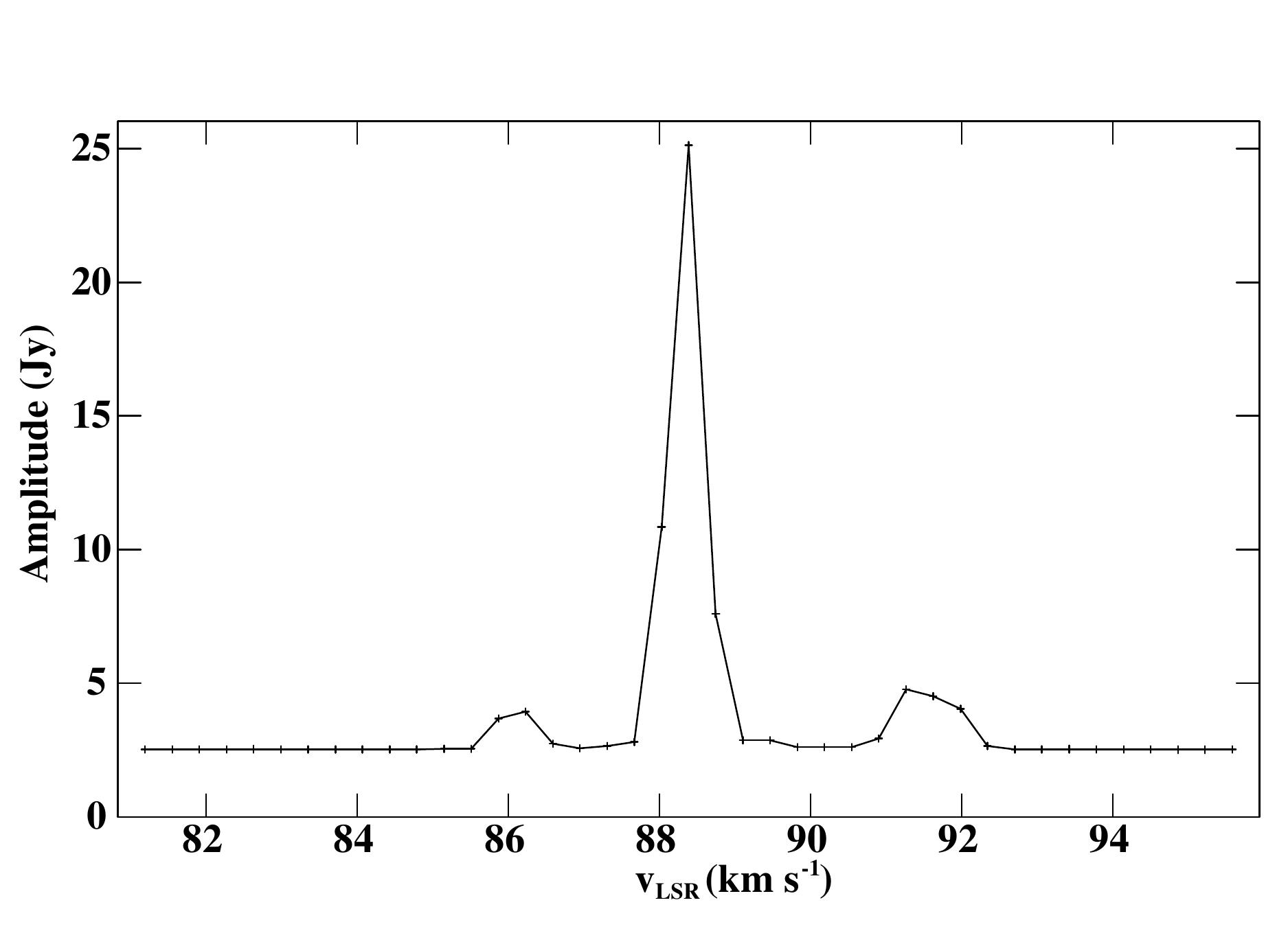}\label{G03070-Spectrum}}
       \subfloat[6.7 GHz methanol maser $-$ G030.74$-$00.04 (second epoch)]{\includegraphics[width=0.5\textwidth]{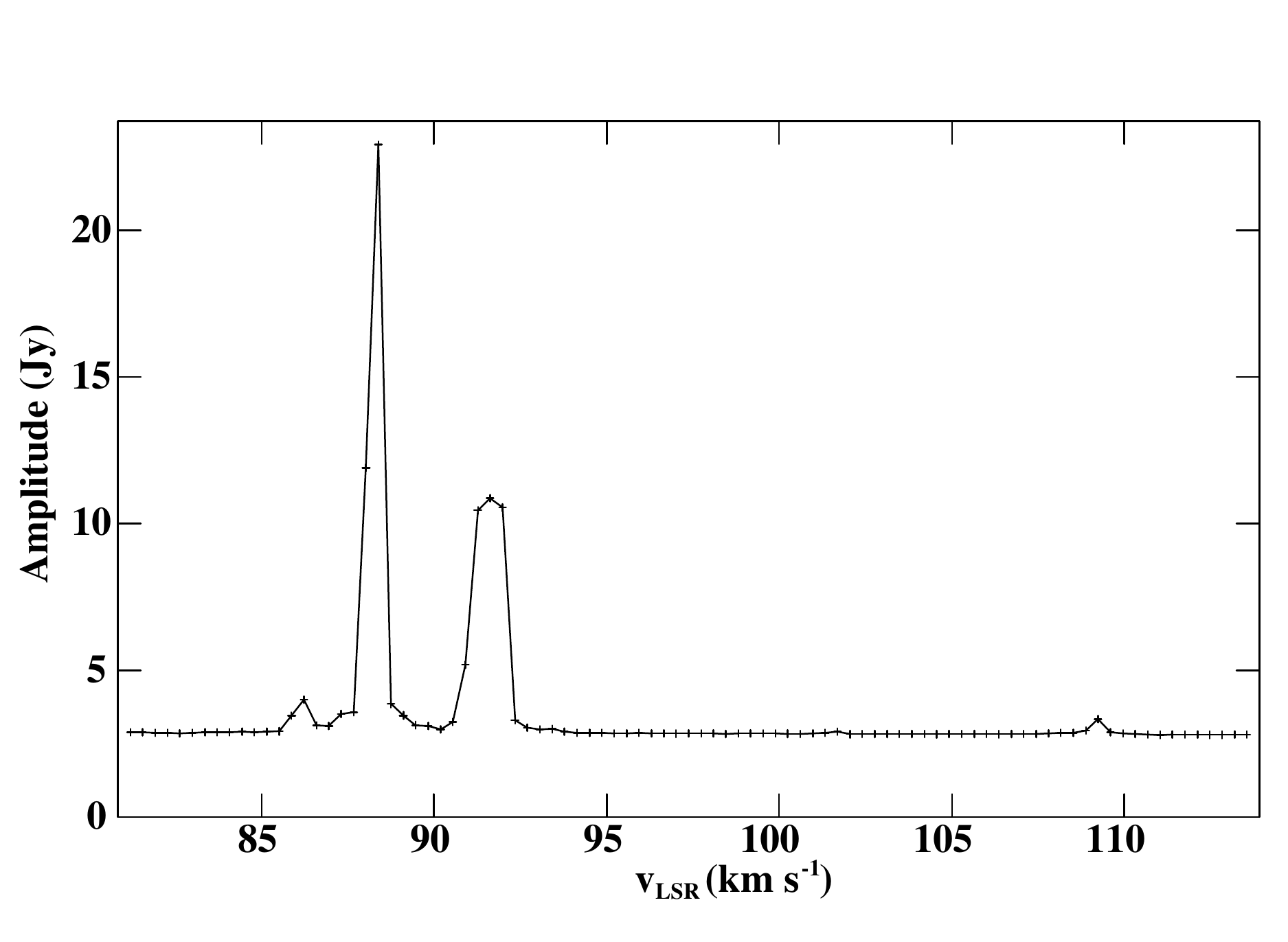}\label{G03074-Spectrum}}

       \subfloat[6.7 GHz methanol maser $-$ G030.78+00.20 (second epoch)]{\includegraphics[width=0.5\textwidth]{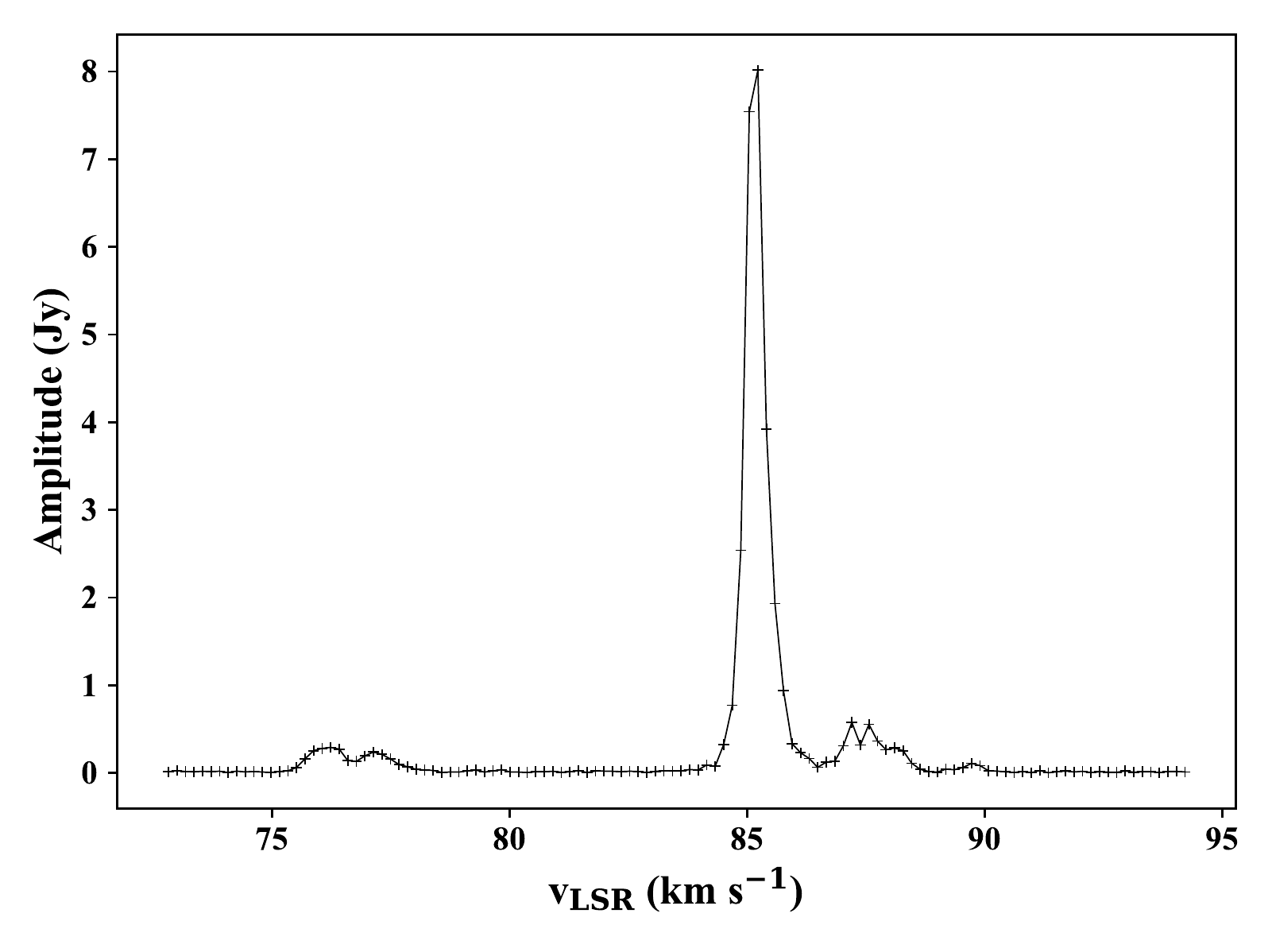}\label{G03078-Spectrum}}
       \subfloat[6.7 GHz methanol maser $-$ G030.81$-$00.05 (second epoch)]{\includegraphics[width=0.5\textwidth]{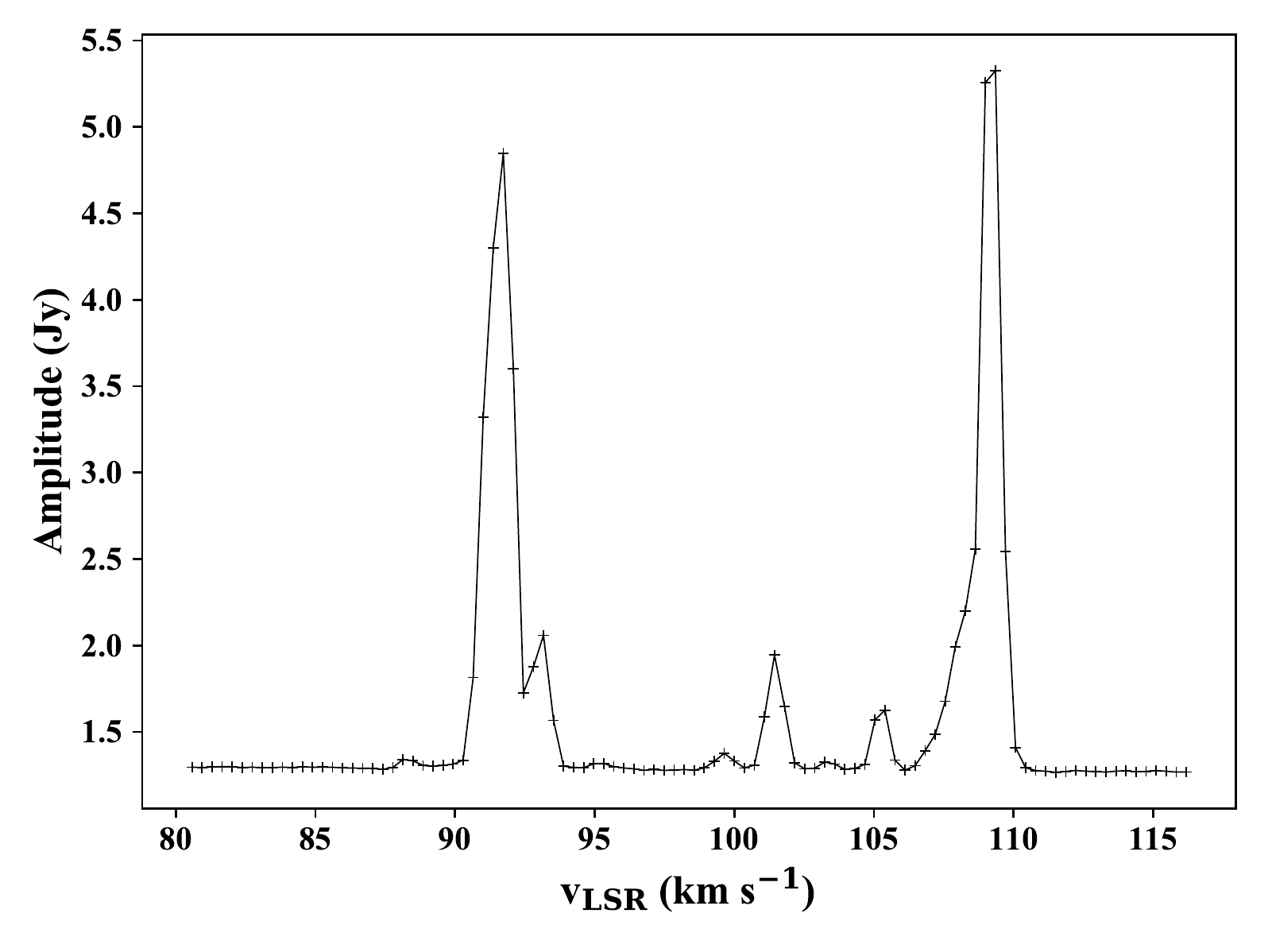}\label{G03081-Spectrum}}
\end{figure*}

\addtocounter{figure}{-1}

\begin{figure*}[ht]
    \caption{Continued.}
    \centering      
                \addtocounter{subfigure}{12}
       \subfloat[6.7 GHz methanol maser $-$ G030.97$-$00.14 (second epoch)]{\includegraphics[width=0.5\textwidth]{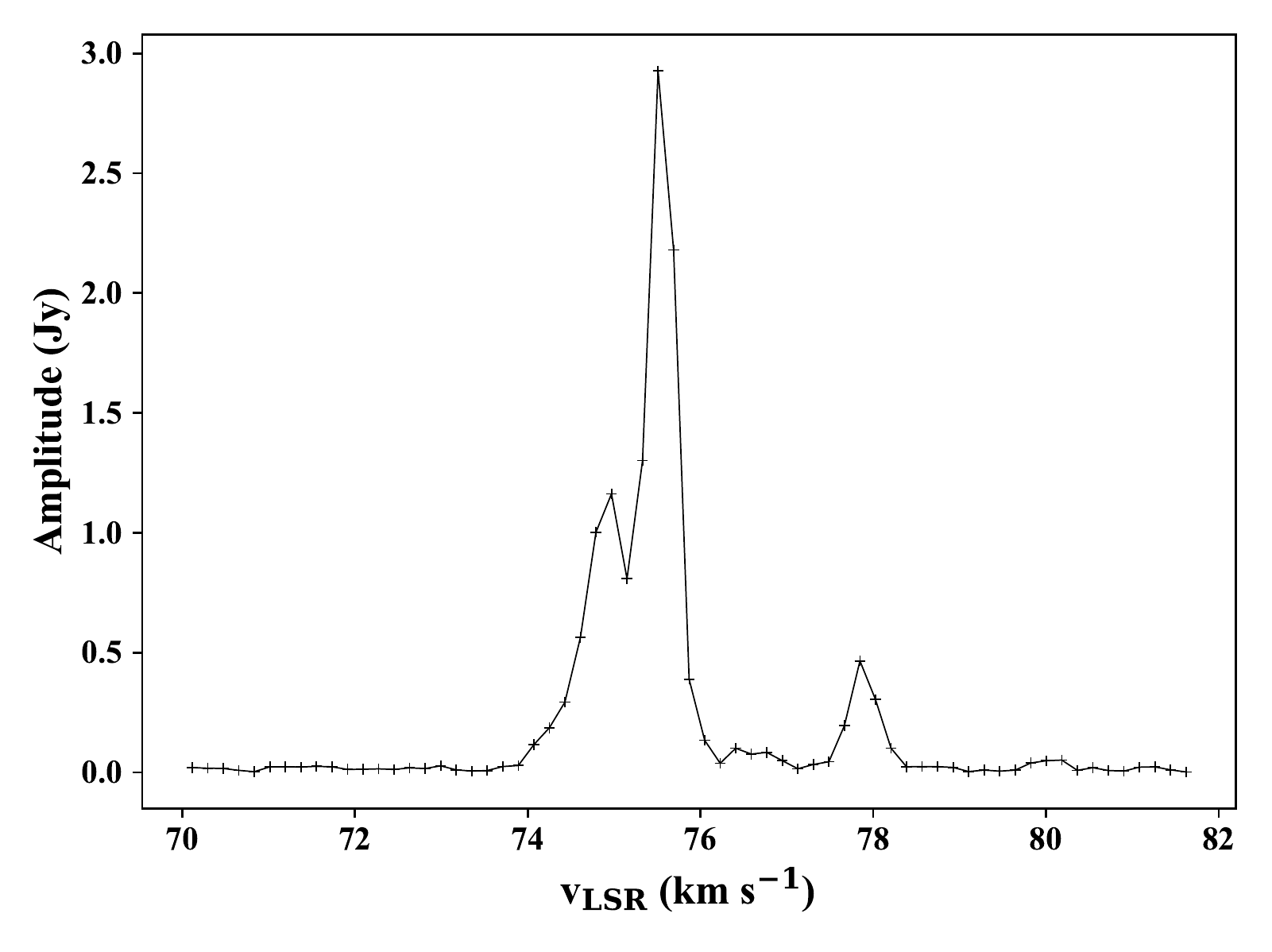}\label{G03097-Spectrum}}
       \subfloat[22 GHz water maser $-$ G031.41+00.30 (second epoch)]{\includegraphics[width=0.5\textwidth]{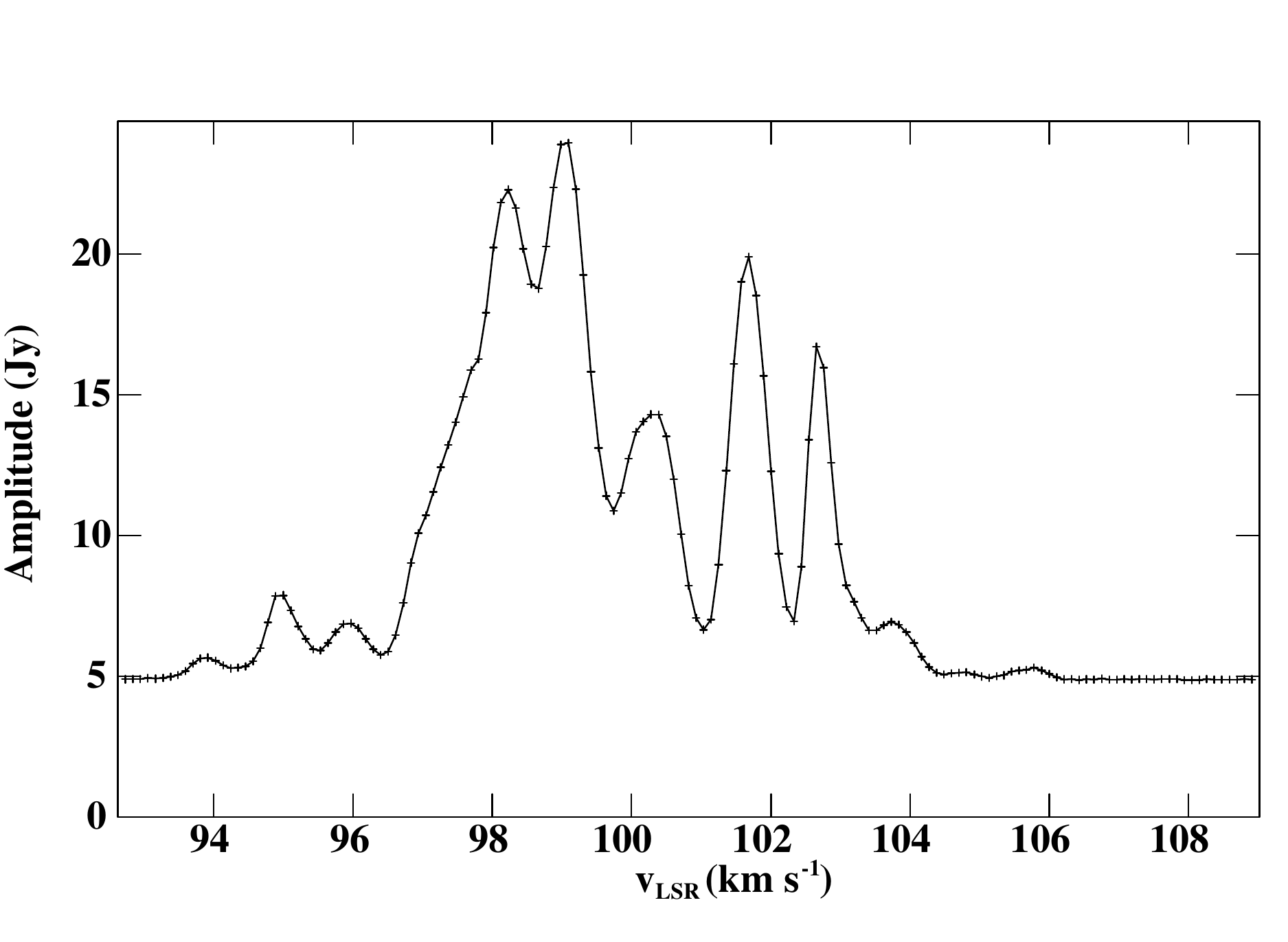}\label{G03141-Spectrum}}\\

       \subfloat[6.7 GHz methanol maser $-$ G032.04+00.05 (second epoch)]{\includegraphics[width=0.5\textwidth]{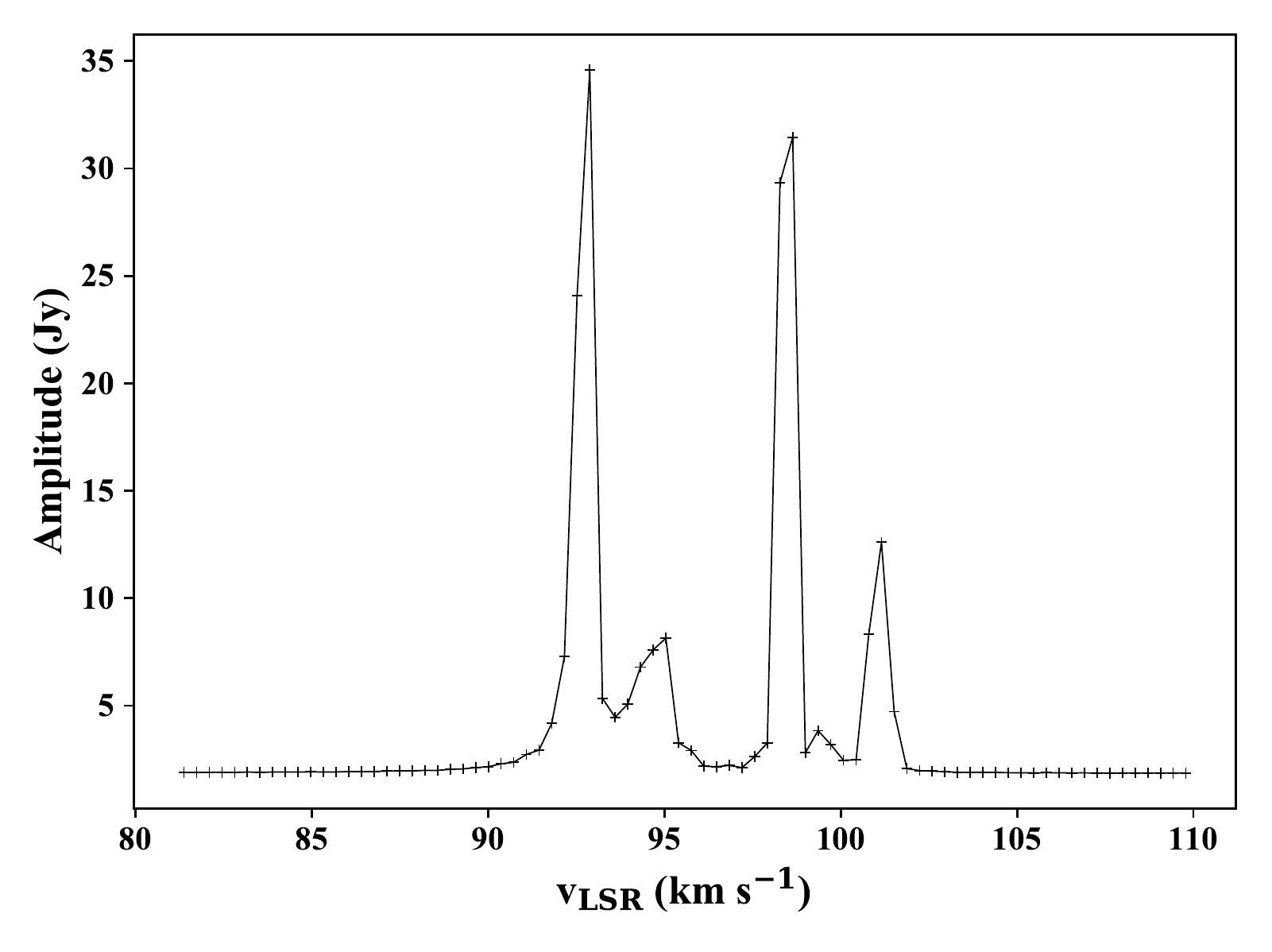}\label{G03204-Spectrum}}
       \subfloat[6.7 GHz methanol maser $-$ G033.09$-$00.07 (second epoch)]{\includegraphics[width=0.5\textwidth]{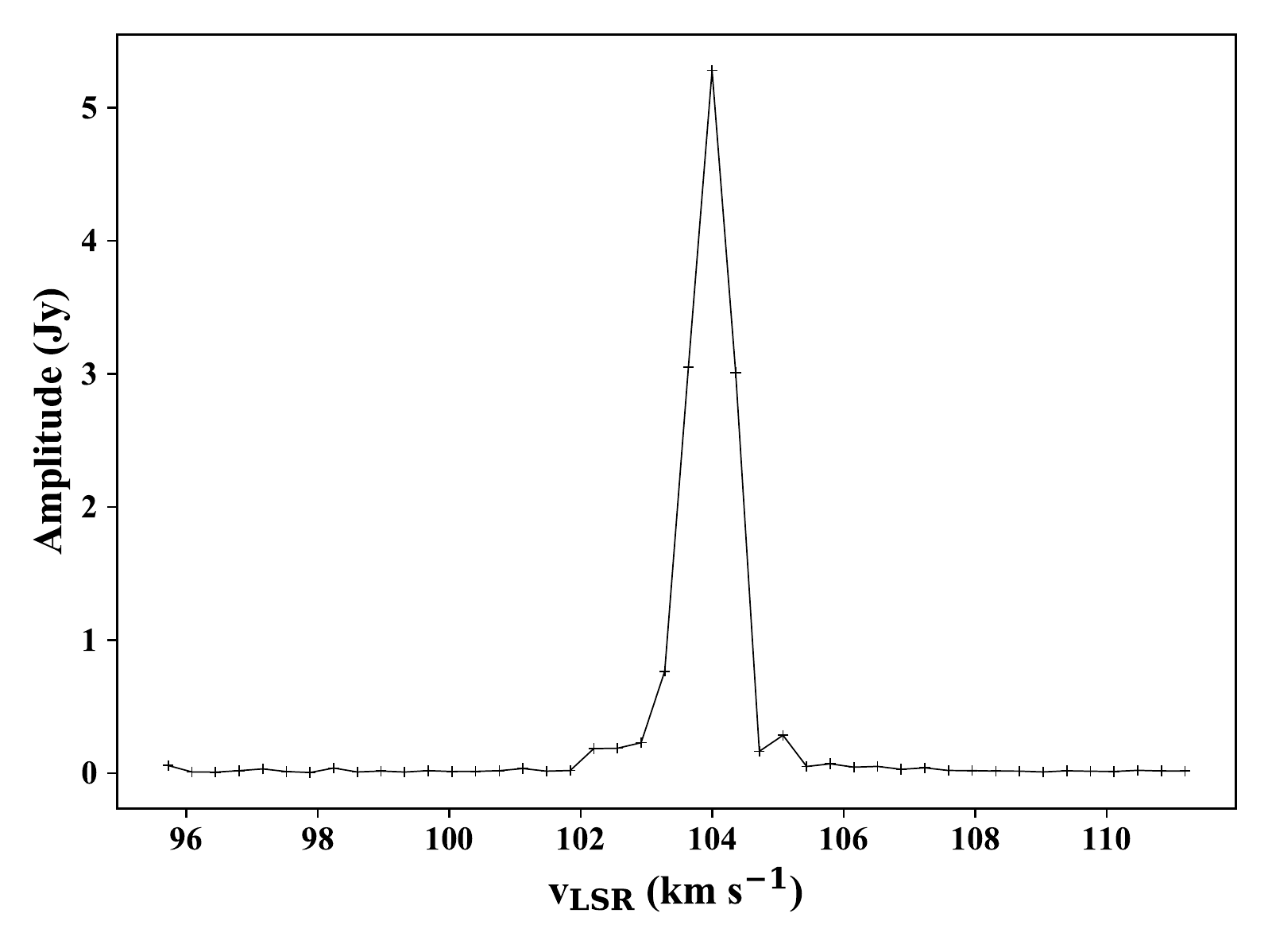}\label{G03309-Spectrum}}
\end{figure*}


\section{Parallax measurements}


\begin{figure*}[ht]
        \centering
        \caption{Parallax and proper motion data and fits for our 6.7 GHz methanol and 22 GHz water masers. The black symbols show the average of all quasar positions for 6.7 GHz masers. In each panel, the data of different maser spots are offset for clarity.
    First panel: Measured position of the maser on the sky. The positions  expected from the fit are shown as open circles.
    Second panel: East (upper part, solid lines) and north (lower part, dashed lines) position offsets vs. time.
    Third panel: As  second panel, but with proper motion removed, showing only the parallax effect for the maser underlined in the legend.}
       \subfloat[6.7 GHz methanol maser $-$ G028.14$-$00.00]{\includegraphics[width=0.8\textwidth]{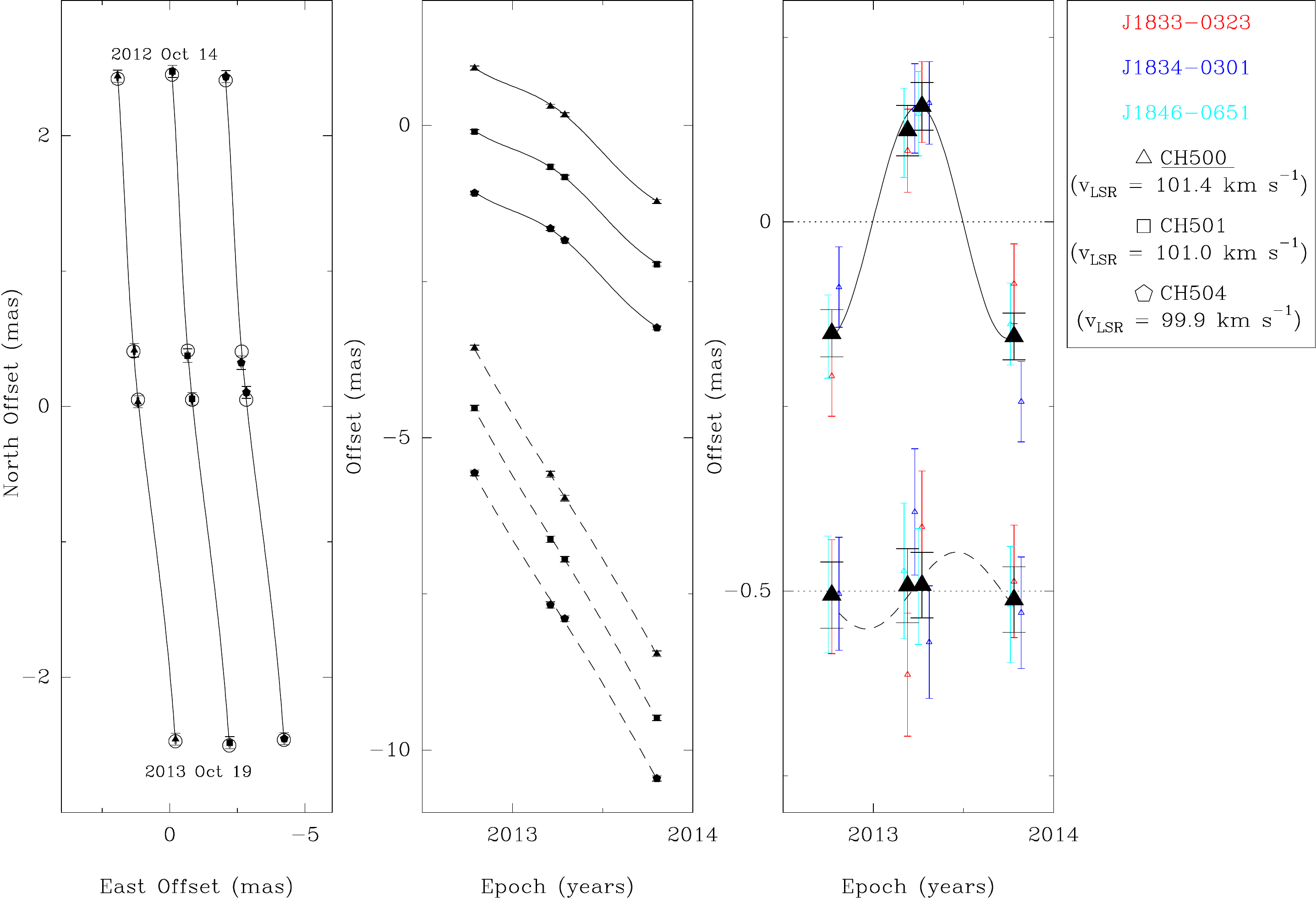}\label{G02814-Parallax}}\\
       \subfloat[6.7 GHz methanol maser $-$ G029.86$-$00.04]{\includegraphics[width=0.8\textwidth]{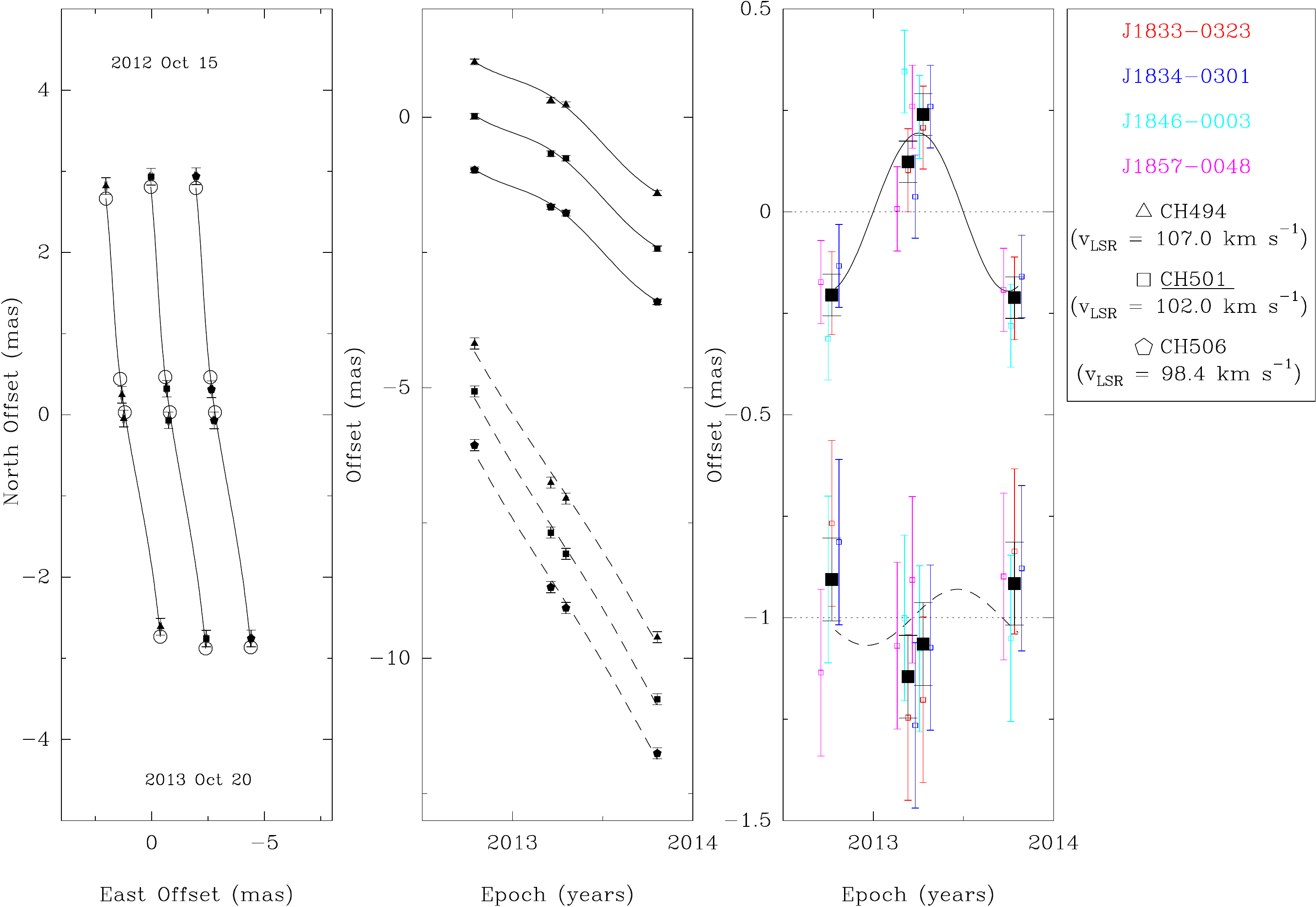}\label{G02986-Parallax}}
    \label{Parallax-Appendix}
\end{figure*}

\addtocounter{figure}{-1}

\begin{figure*}[ht]
        \centering
        \caption{Continued.}
                \addtocounter{subfigure}{2}
       \subfloat[6.7 GHz methanol maser $-$ G029.95$-$00.01]{\includegraphics[width=0.8\textwidth]{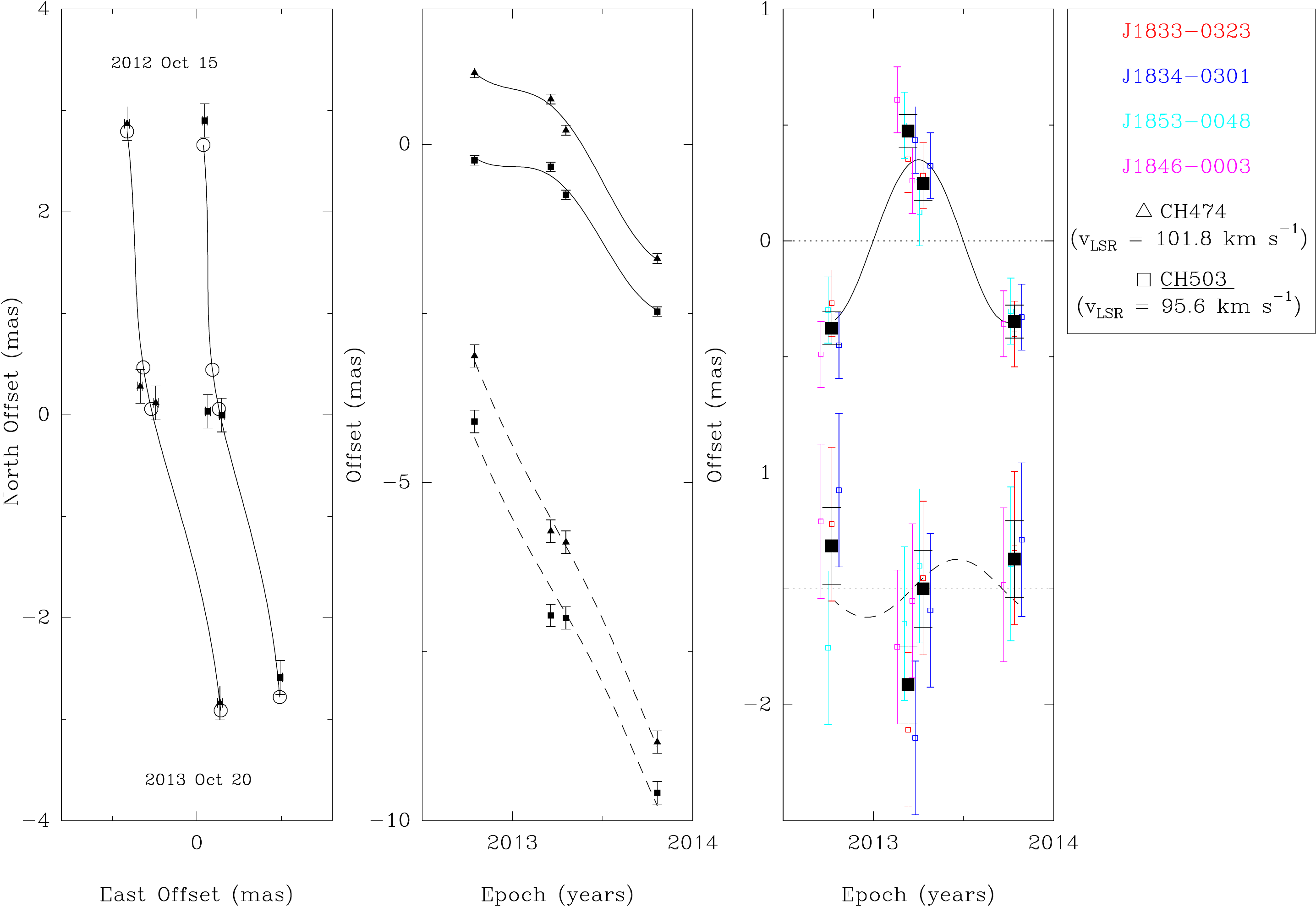}\label{G02995-Parallax}}\\
       \subfloat[22 GHz water maser $-$ G029.98+00.10]{\includegraphics[width=0.8\textwidth]{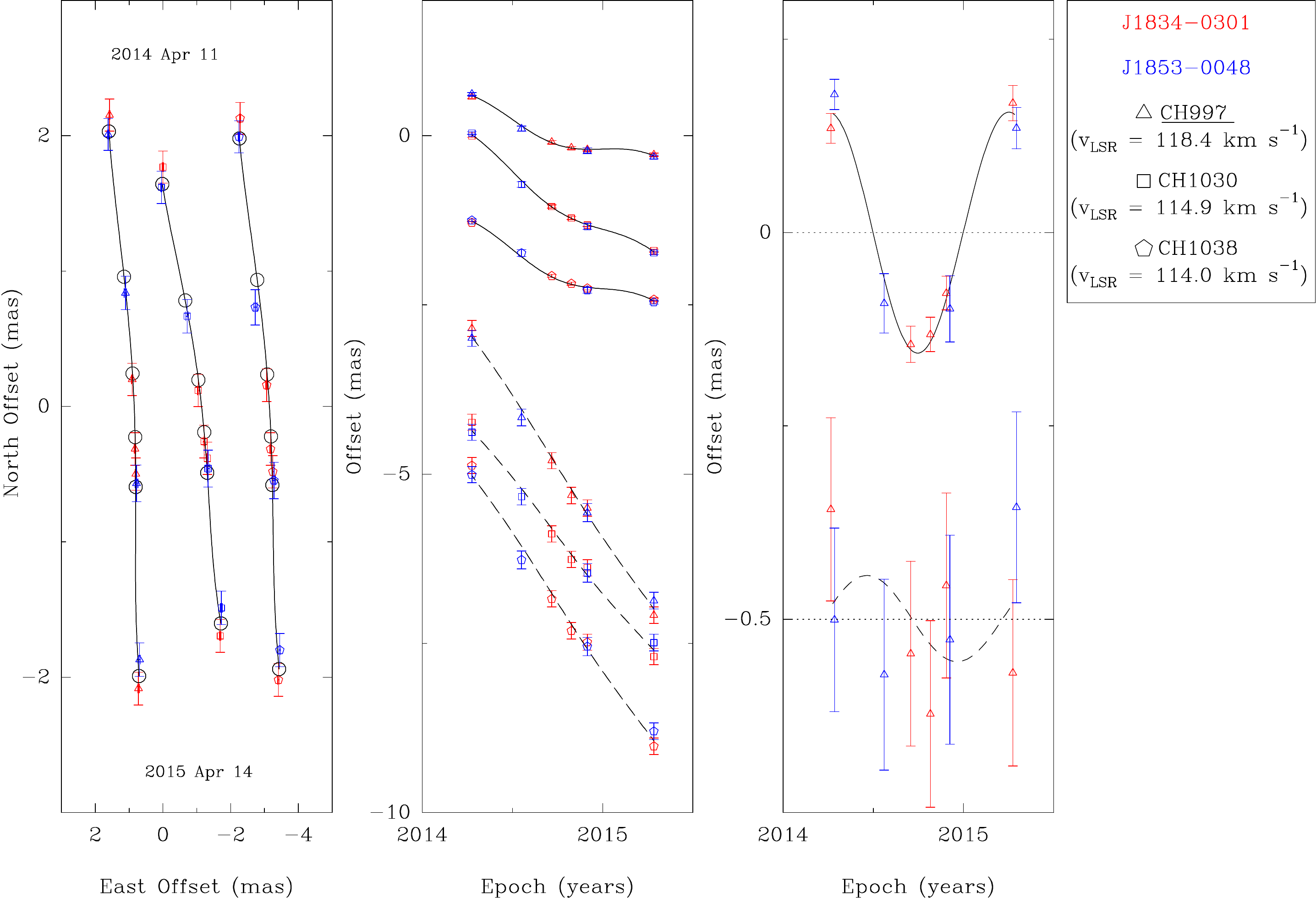}\label{G02998-Parallax}}
\end{figure*}

\addtocounter{figure}{-1}

\begin{figure*}[ht]
        \centering
        \caption{Continued.}
                \addtocounter{subfigure}{4}
       \subfloat[6.7 GHz methanol maser $-$ G030.22$-$00.18]{\includegraphics[width=0.8\textwidth]{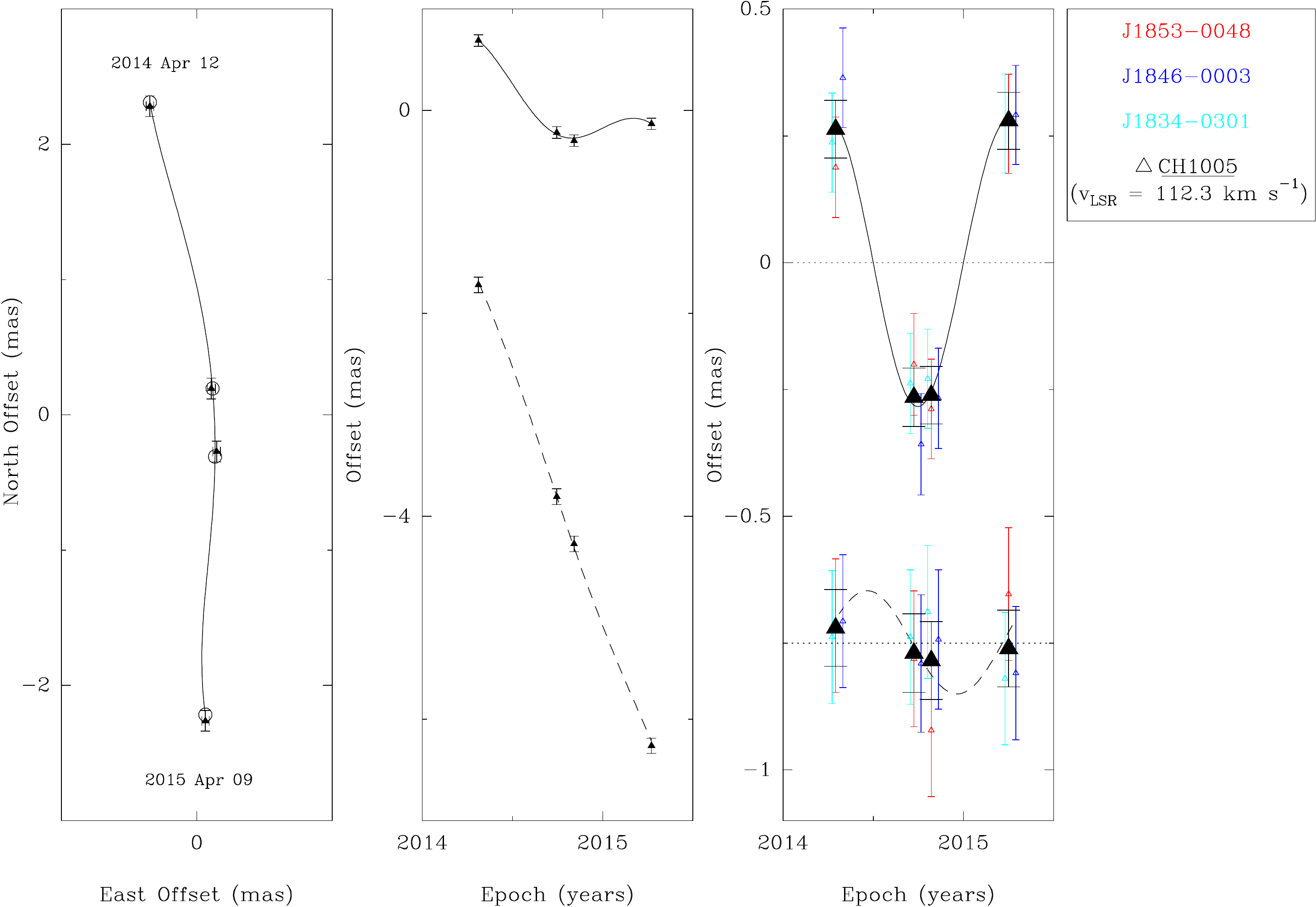}\label{G03022-Parallax}}\\
       \subfloat[6.7 GHz methanol maser $-$ G030.41$-$00.23]{\includegraphics[width=0.8\textwidth]{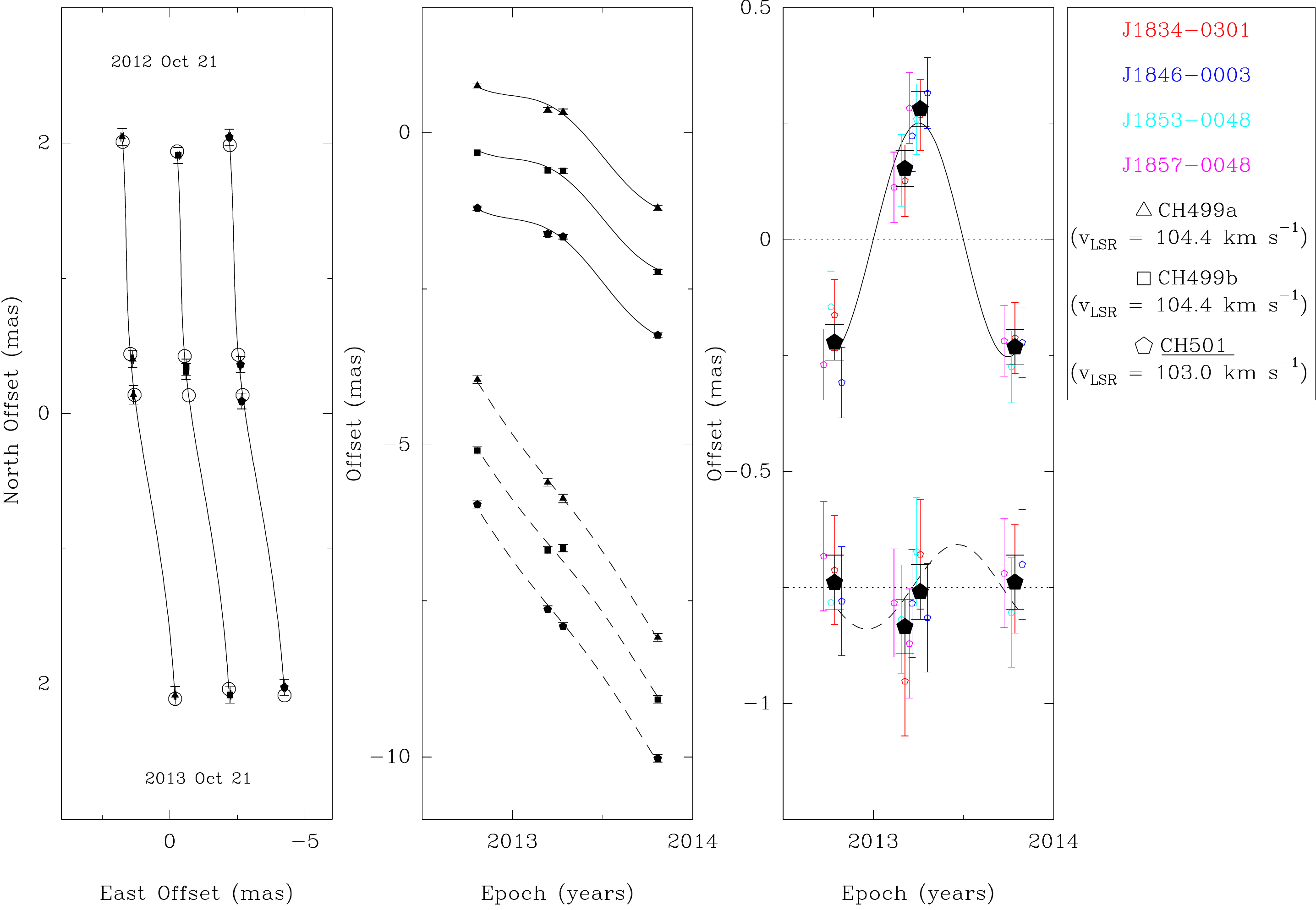}\label{G03041-Parallax}}
\end{figure*}

\addtocounter{figure}{-1}

\begin{figure*}[ht]
        \centering
        \caption{Continued.}
                \addtocounter{subfigure}{6}
       \subfloat[6.7 GHz methanol maser $-$ G030.70$-$00.06]{\includegraphics[width=0.8\textwidth]{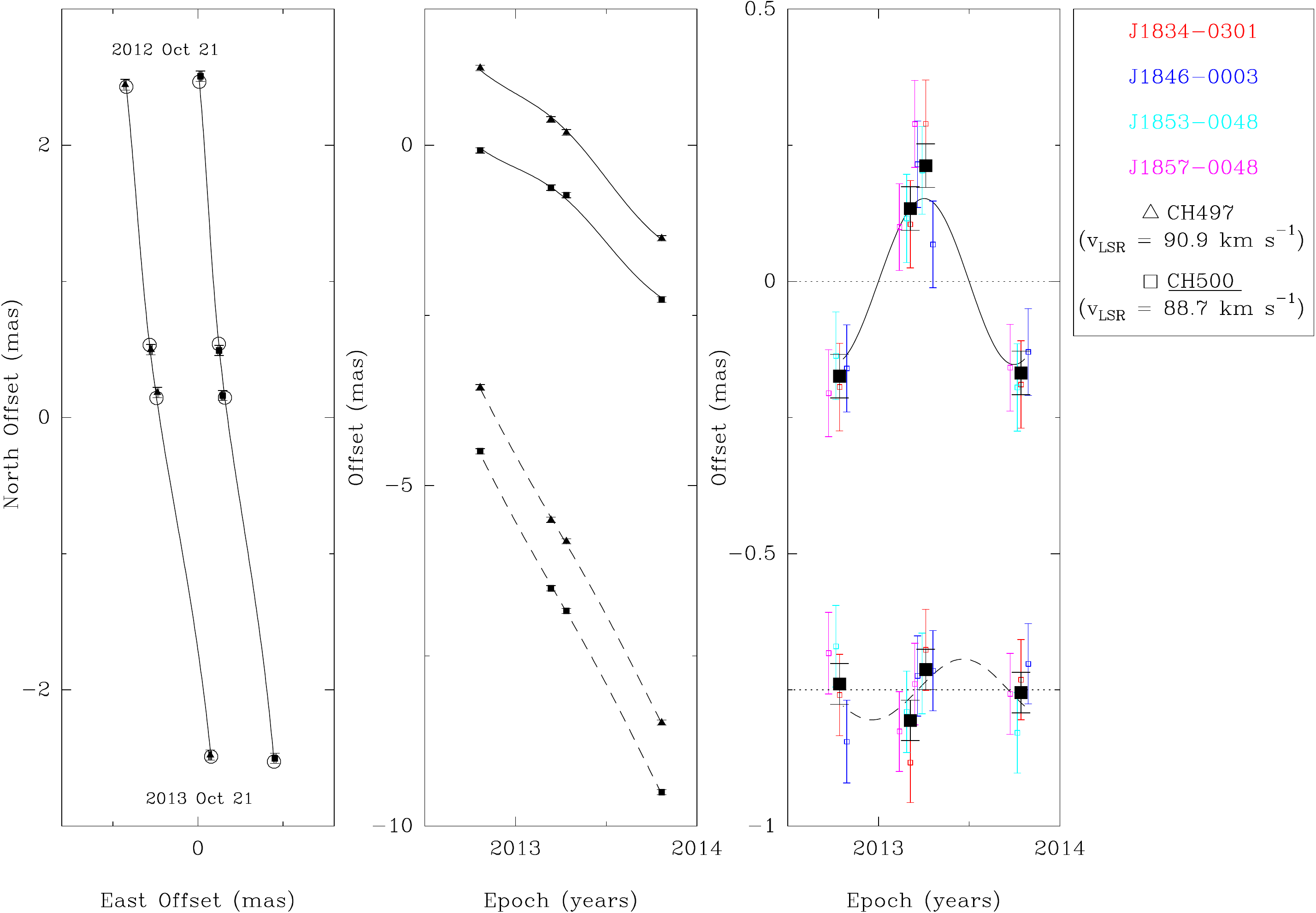}\label{G03070-Parallax}}\\
       \subfloat[6.7 GHz methanol maser $-$ G030.74$-$00.04]{\includegraphics[width=0.8\textwidth]{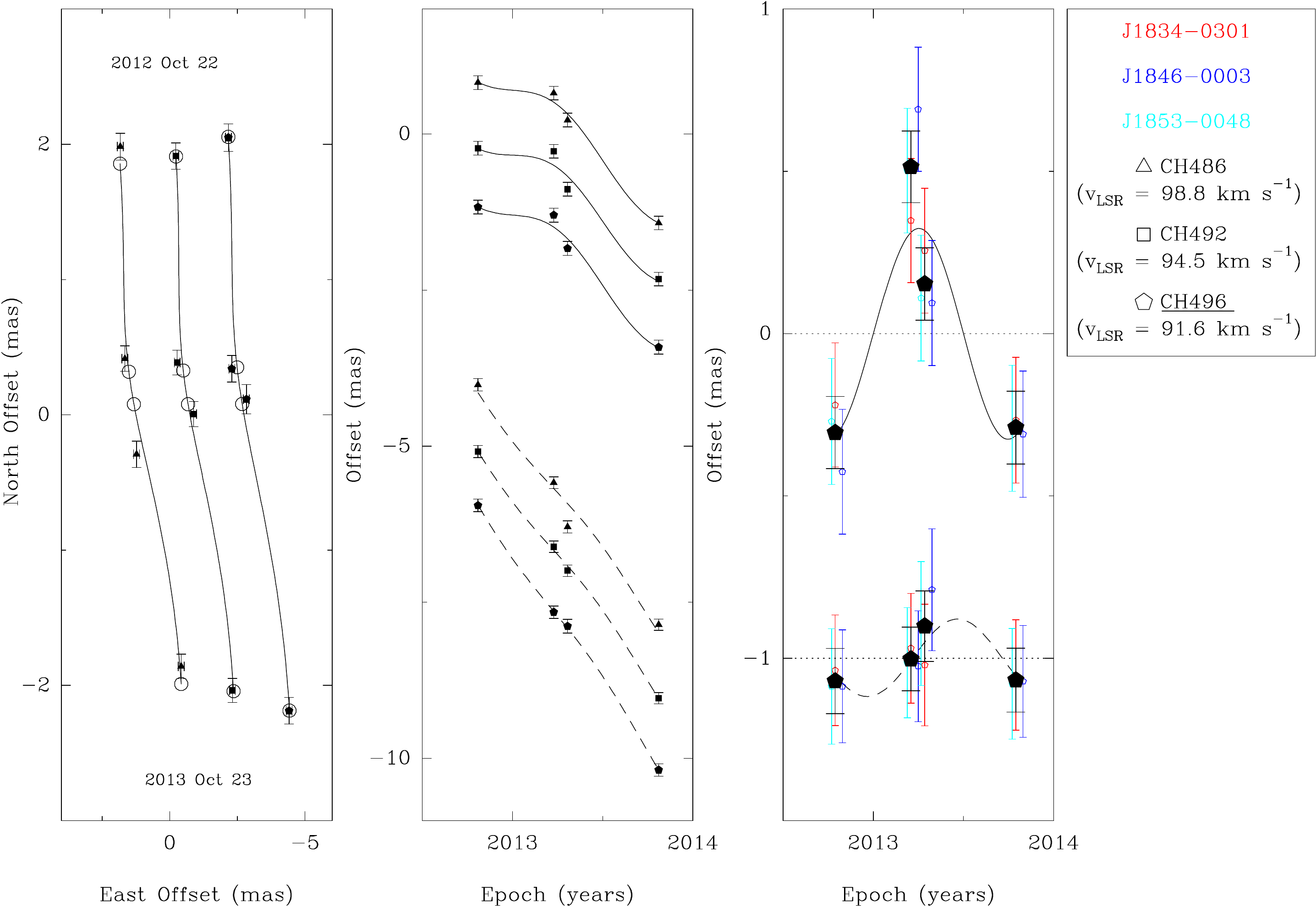}\label{G03074-Parallax}}
\end{figure*}

\addtocounter{figure}{-1}

\begin{figure*}[ht]
        \centering
        \caption{Continued.}
                \addtocounter{subfigure}{8}
       \subfloat[6.7 GHz methanol maser $-$ G030.78+00.20]{\includegraphics[width=0.8\textwidth]{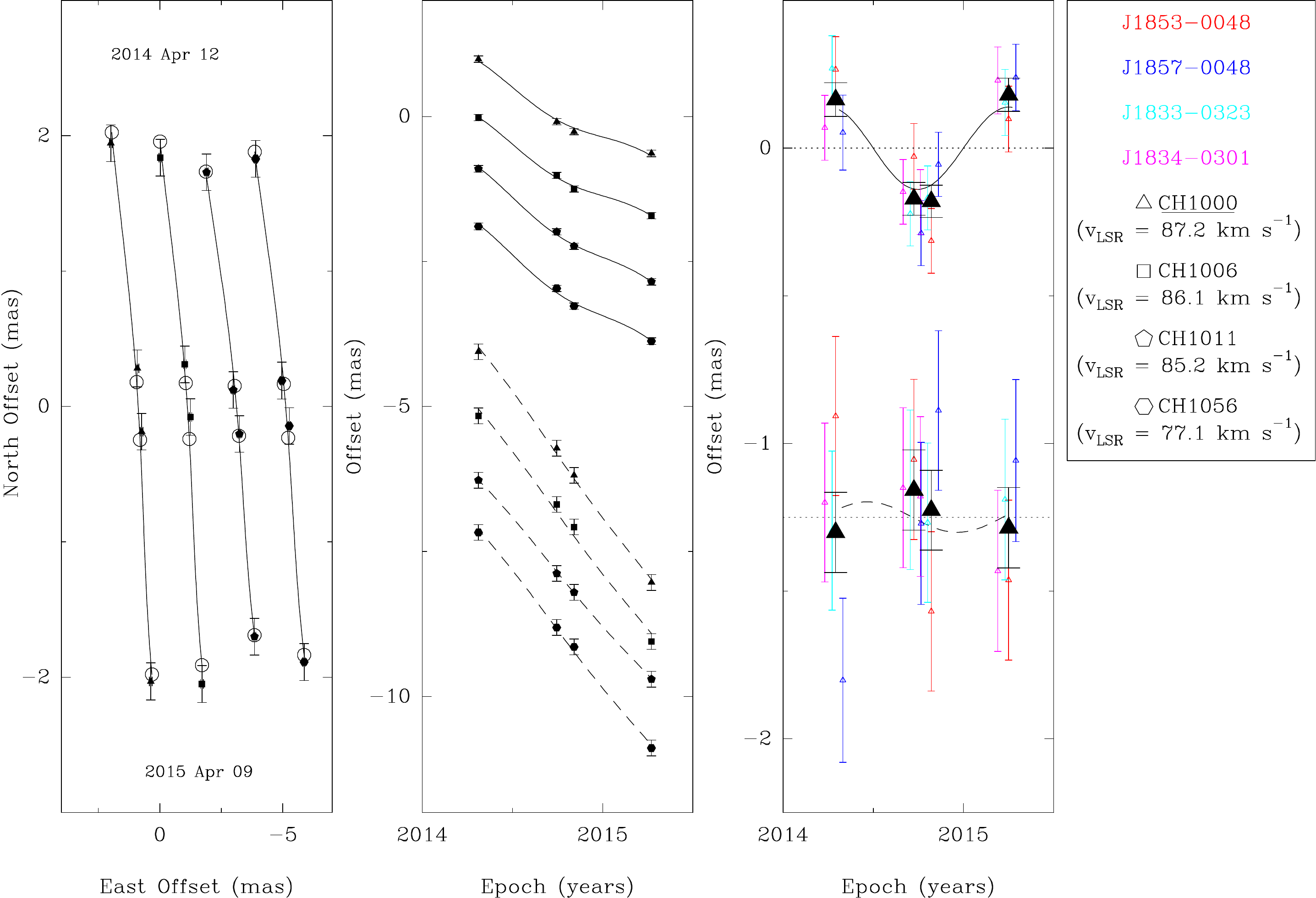}\label{G03078-Parallax}}\\
       \subfloat[6.7 GHz methanol maser $-$ G030.81$-$00.05]{\includegraphics[width=0.8\textwidth]{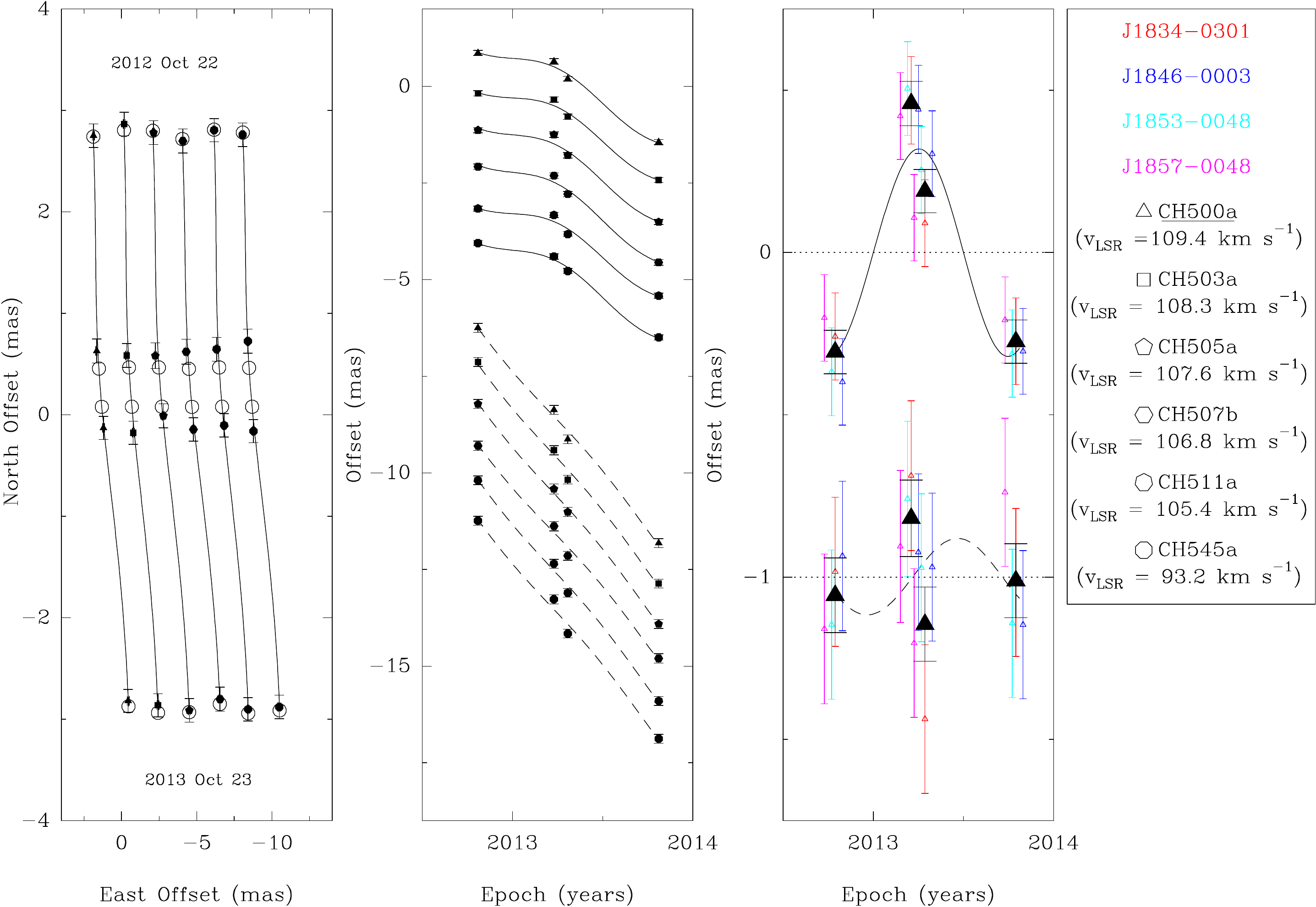}\label{G03081-Parallax}}
\end{figure*}

\addtocounter{figure}{-1}

\begin{figure*}[ht]
        \centering
        \caption{Continued.}
                \addtocounter{subfigure}{10}
       \subfloat[6.7 GHz methanol maser $-$ G030.97$-$00.14]{\includegraphics[width=0.8\textwidth]{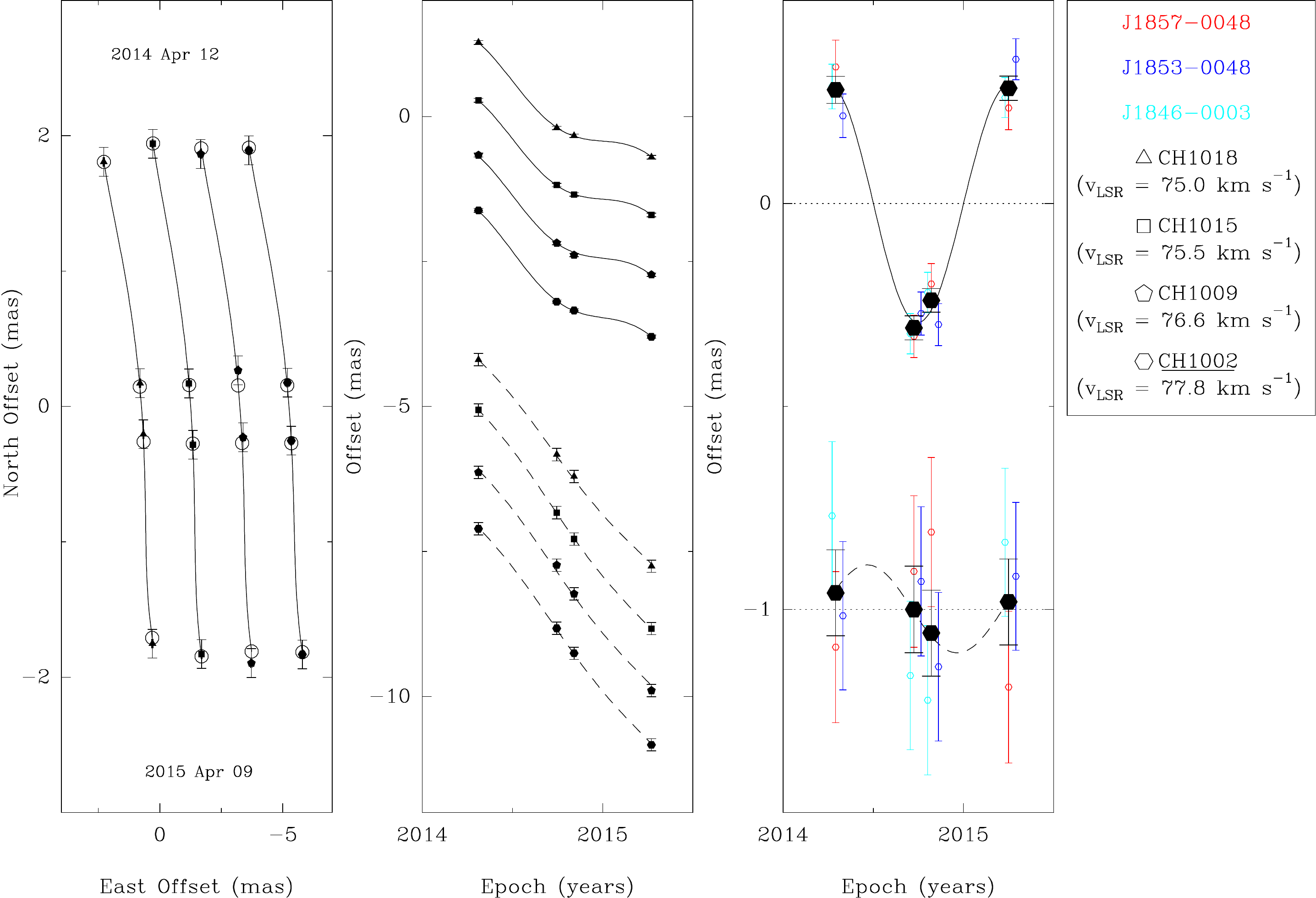}\label{G03097-Parallax}}\\
       \subfloat[22 GHz water maser $-$ G031.41+00.30]{\includegraphics[width=0.8\textwidth]{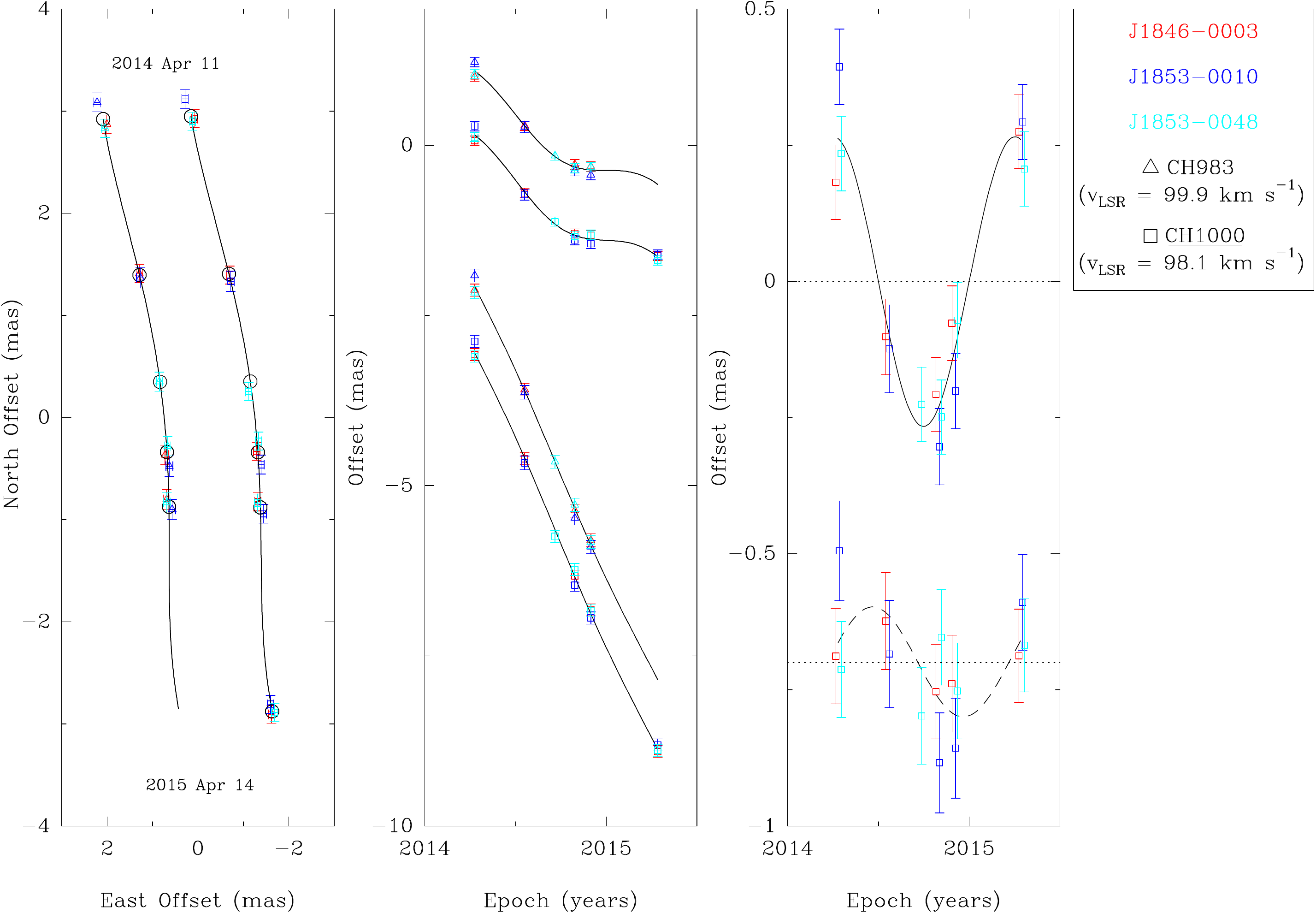}\label{G03141-Parallax}}
\end{figure*}

\addtocounter{figure}{-1}

\begin{figure*}[ht]
        \centering
        \caption{Continued.}
                \addtocounter{subfigure}{12}
       \subfloat[6.7 GHz methanol maser $-$ G032.04+00.05]{\includegraphics[width=0.8\textwidth]{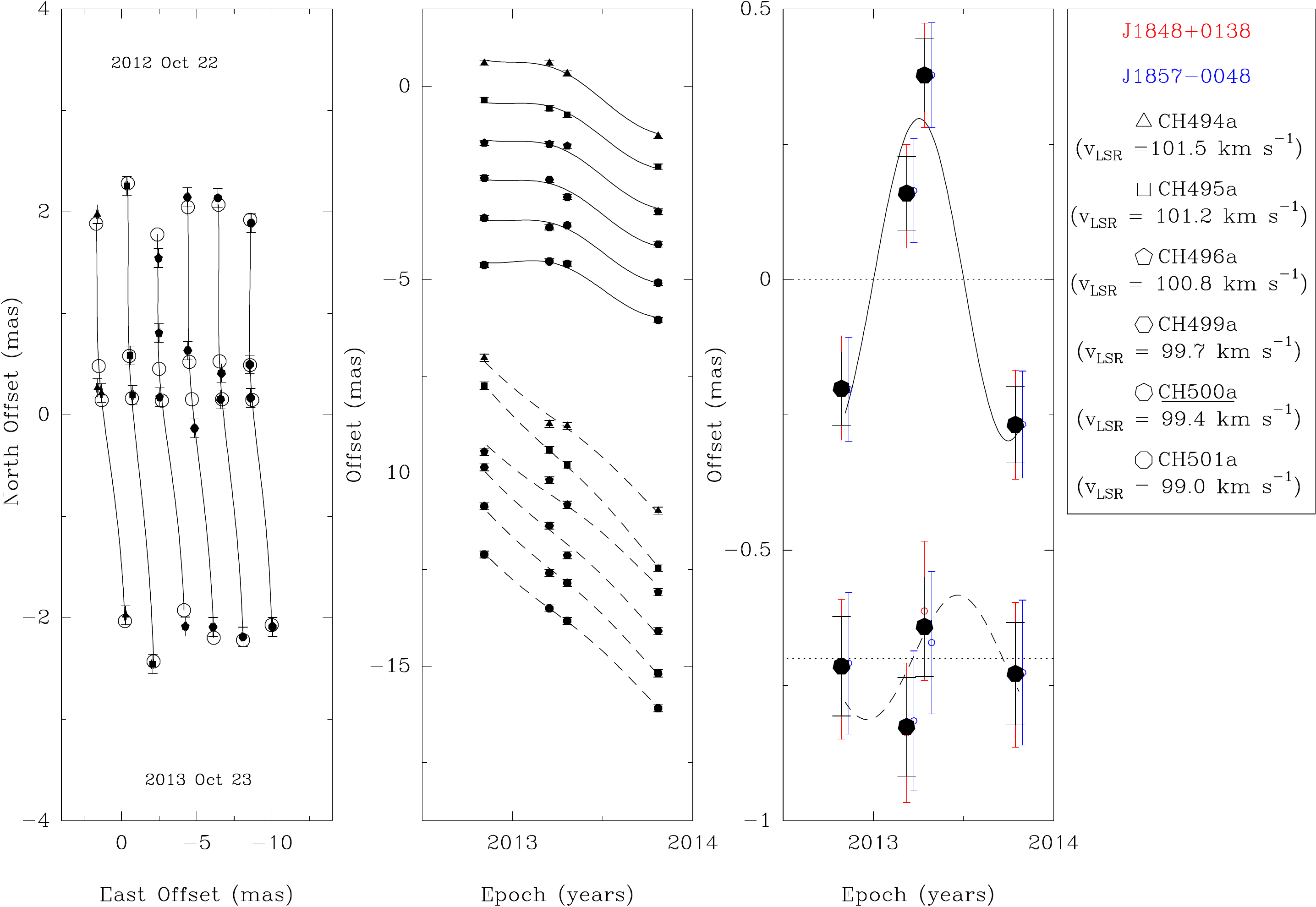}\label{G03204-Parallax}}\\
       \subfloat[6.7 GHz methanol maser $-$ G033.09$-$00.07]{\includegraphics[width=0.8\textwidth]{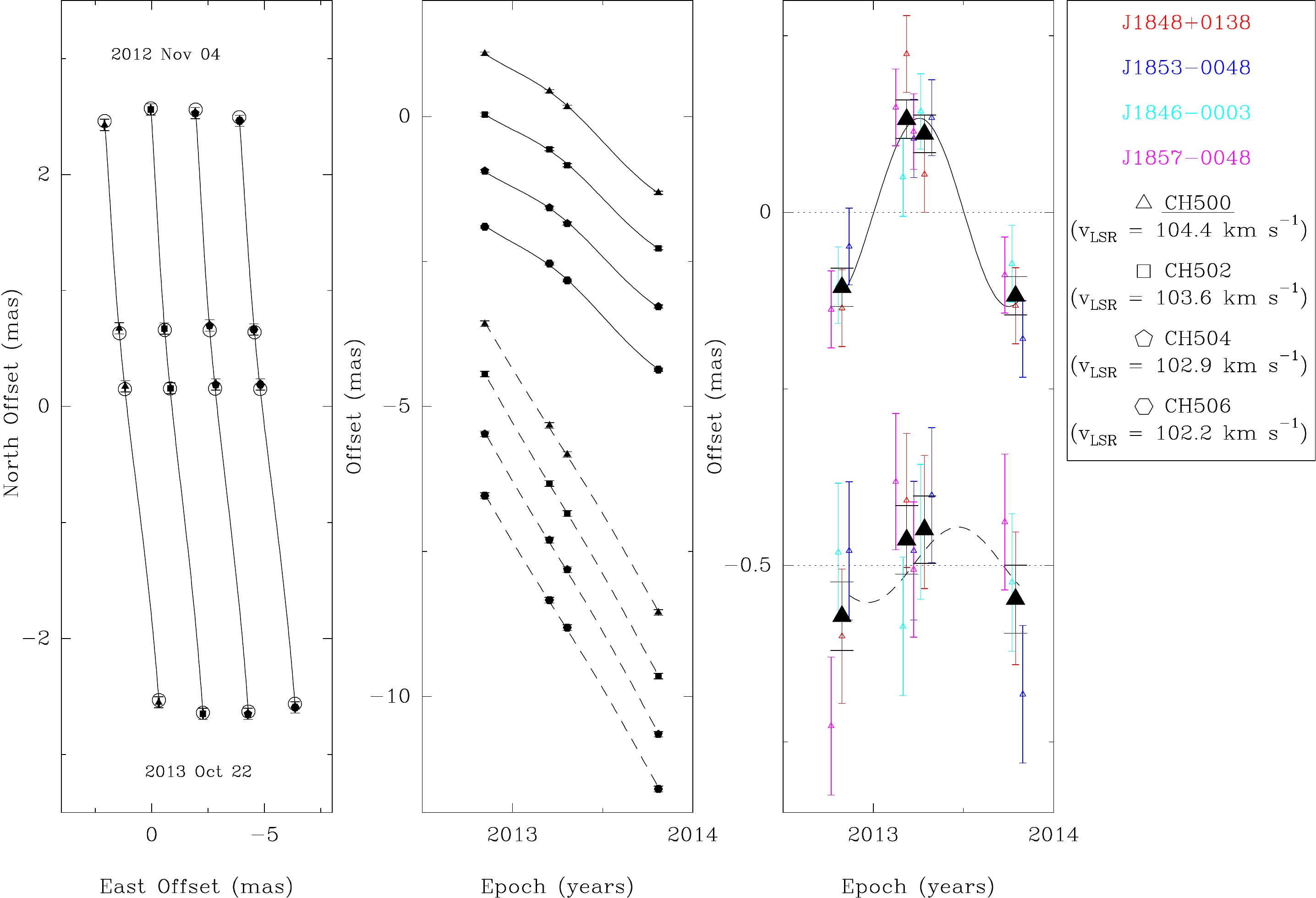}\label{G03309-Parallax}}
\end{figure*}



\section{Proper motion plots}
\label{PropMot-Measurements}

In some of the sources we give the position and proper motion of the central object. These were determined by averaging the positions and proper motions of all maser spots with respect 
to a reference maser spot and then adding the proper motion of the reference maser spot. The internal motion plots of these sources show 
the internal motion of all maser spots with respect to the central object, which is flagged with a star symbol in the plots.


 \begin{figure*}
        \centering     
\caption{Internal motion plot. The star symbol indicates the position of the central object.}
            \subfloat[22 GHz water maser $-$ G028.86+00.06]{\includegraphics[width=0.45\textwidth]{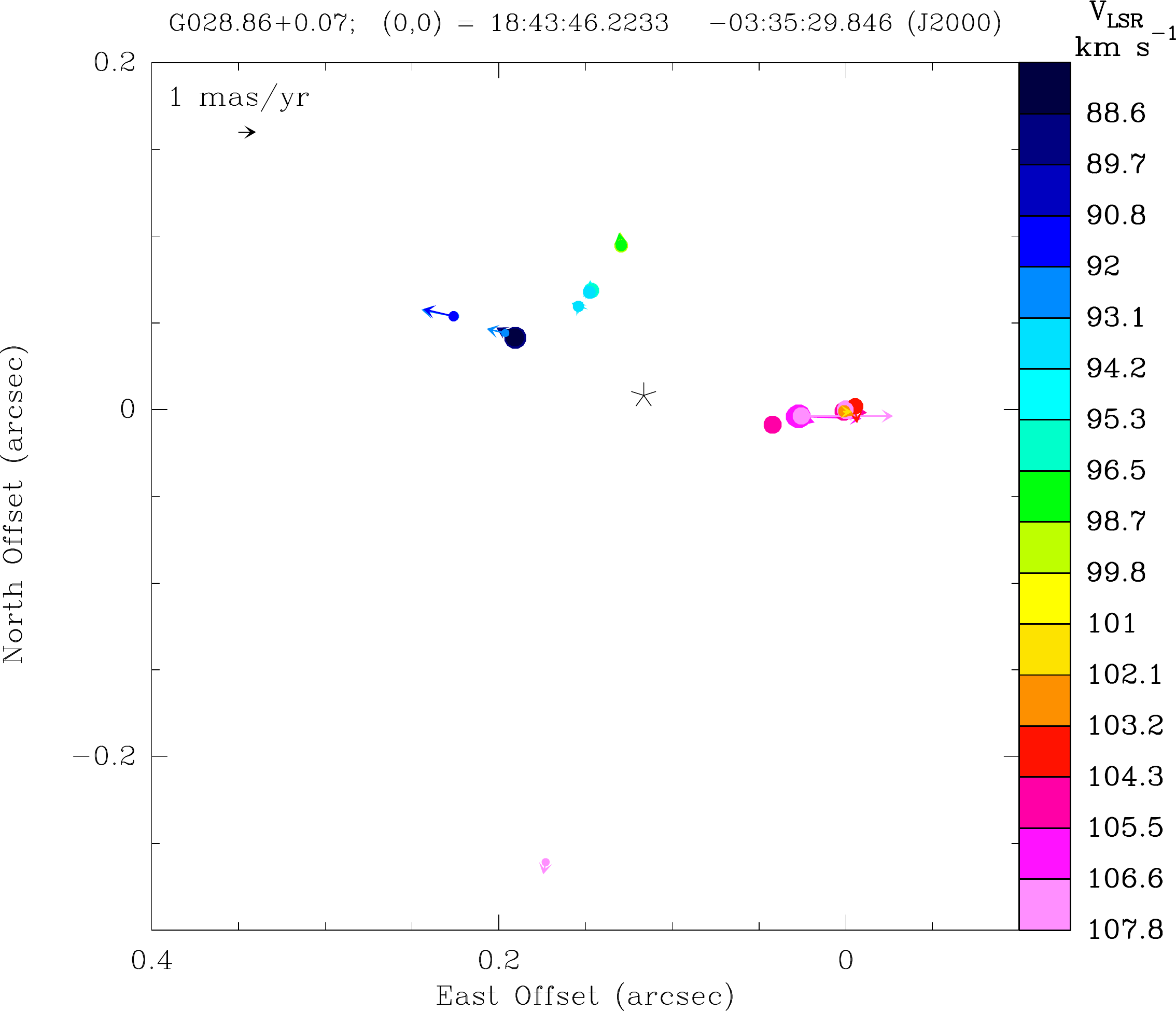}\label{G02886-ProperMotion}}
        \subfloat[6.7 GHz methanol maser $-$ G029.86$-$00.04]{\includegraphics[width=0.45\textwidth]{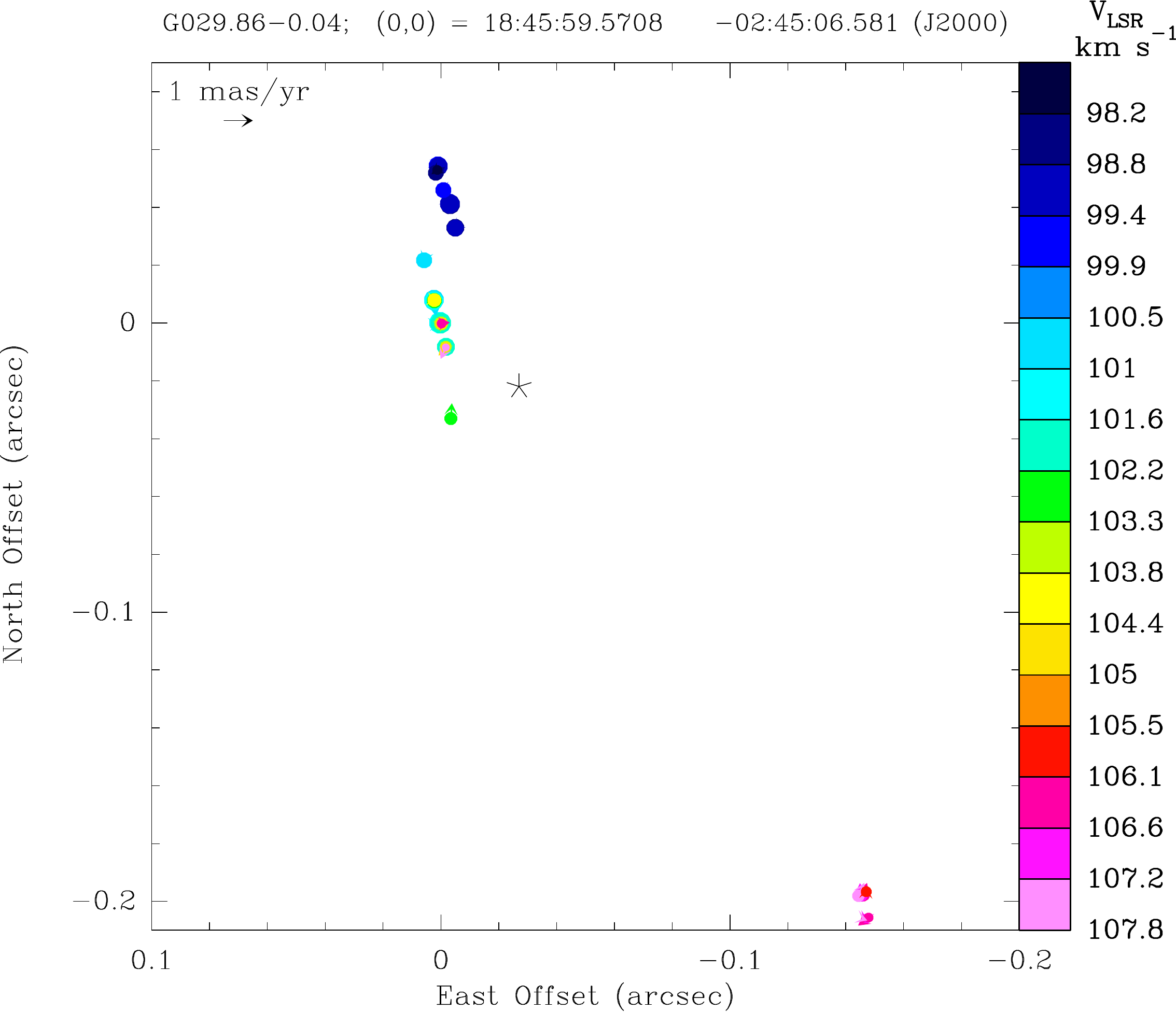}\label{G02986-ProperMotion}}

        \subfloat[6.7 GHz methanol maser $-$ G029.95$-$00.01]{\includegraphics[width=0.45\textwidth]{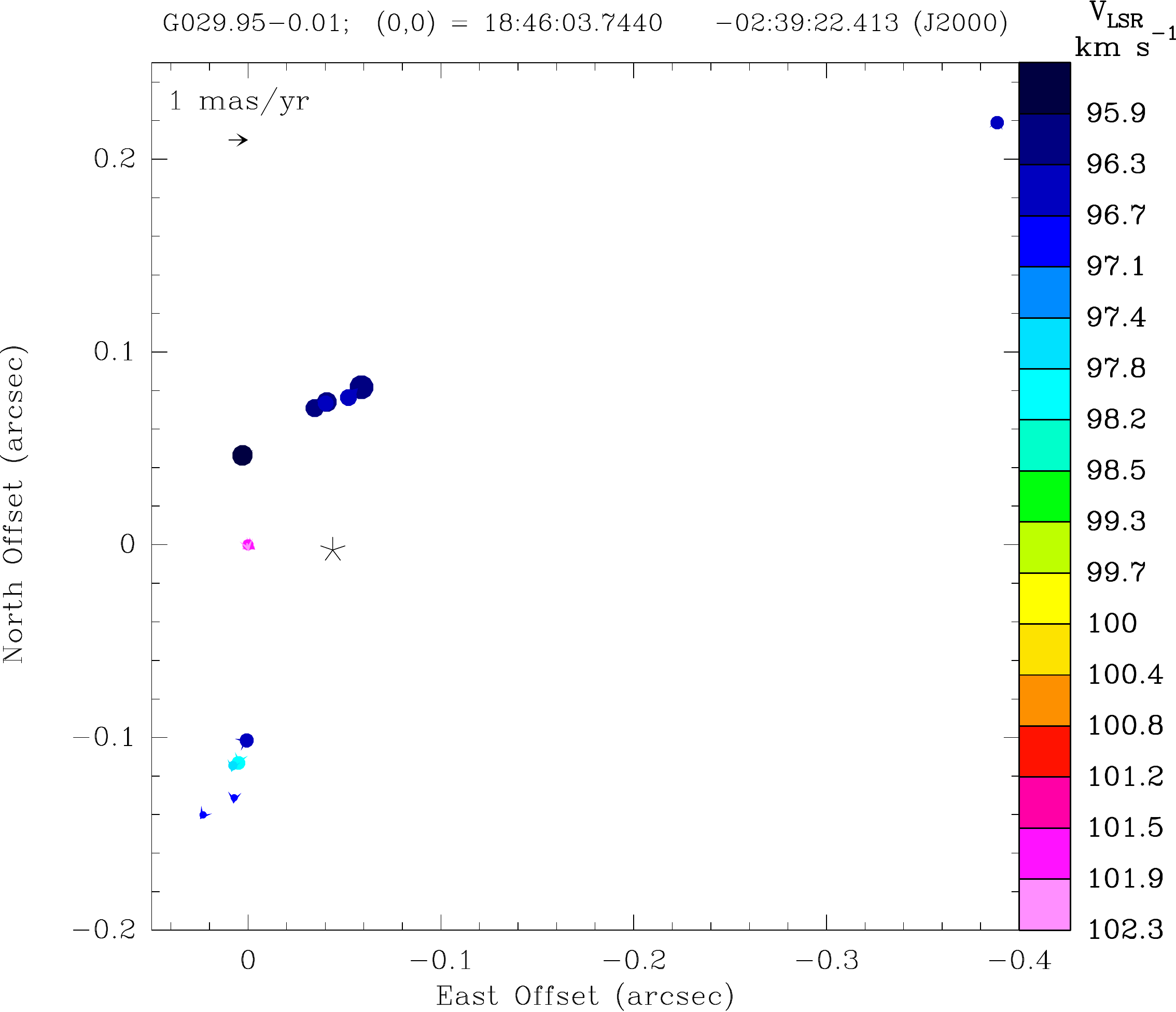}\label{G02995-ProperMotion}}
         \subfloat[22 GHz water maser $-$ G029.98+00.10]{\includegraphics[width=0.45\textwidth]{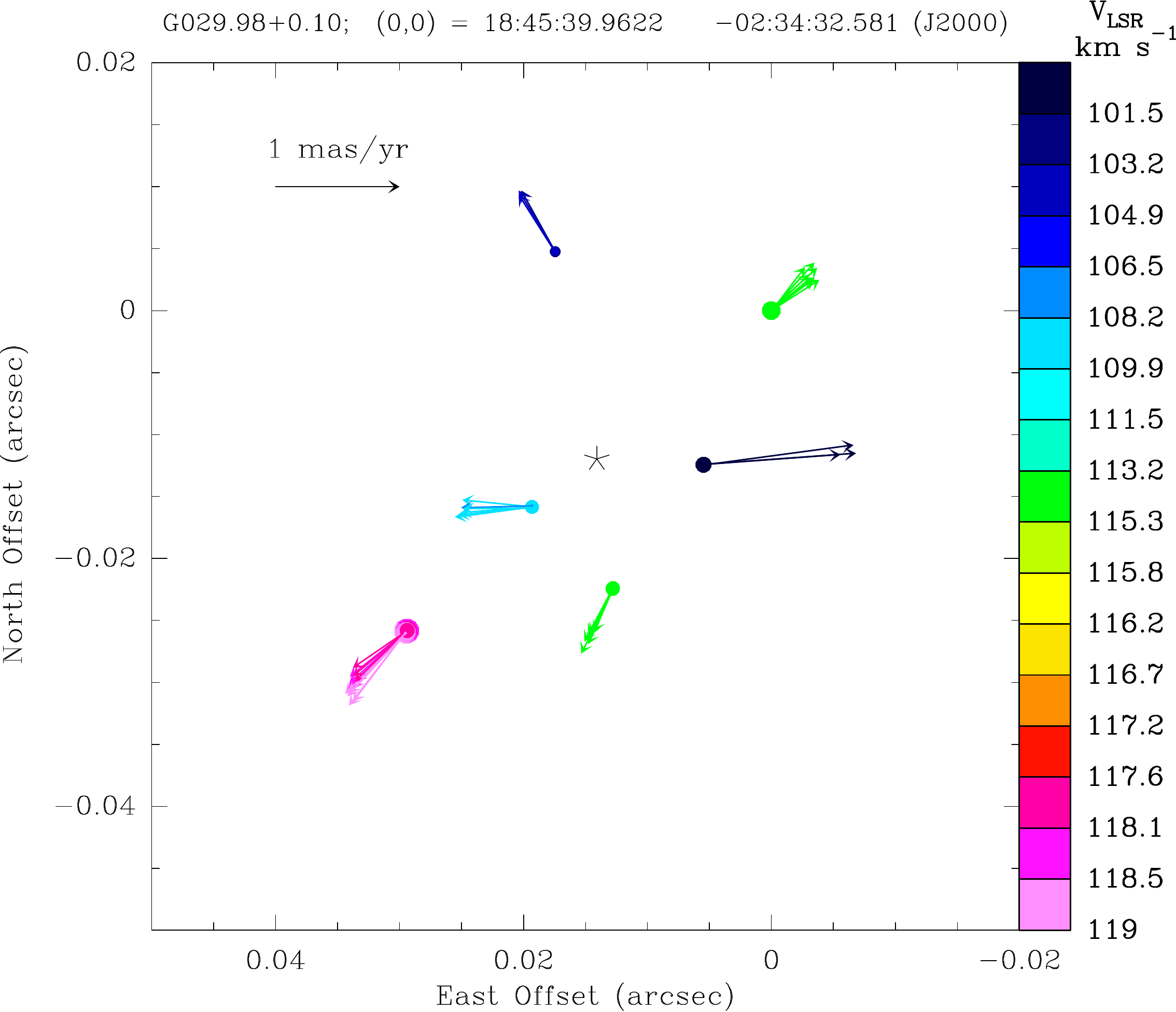}\label{G02998-ProperMotion}}
    \label{ProperMotion}
 \end{figure*}

\addtocounter{figure}{-1}

\begin{figure*}
\caption{Continued.}
        \centering      
       \subfloat[6.7 GHz methanol maser $-$ G030.41$-$00.23]{\includegraphics[width=0.45\textwidth]{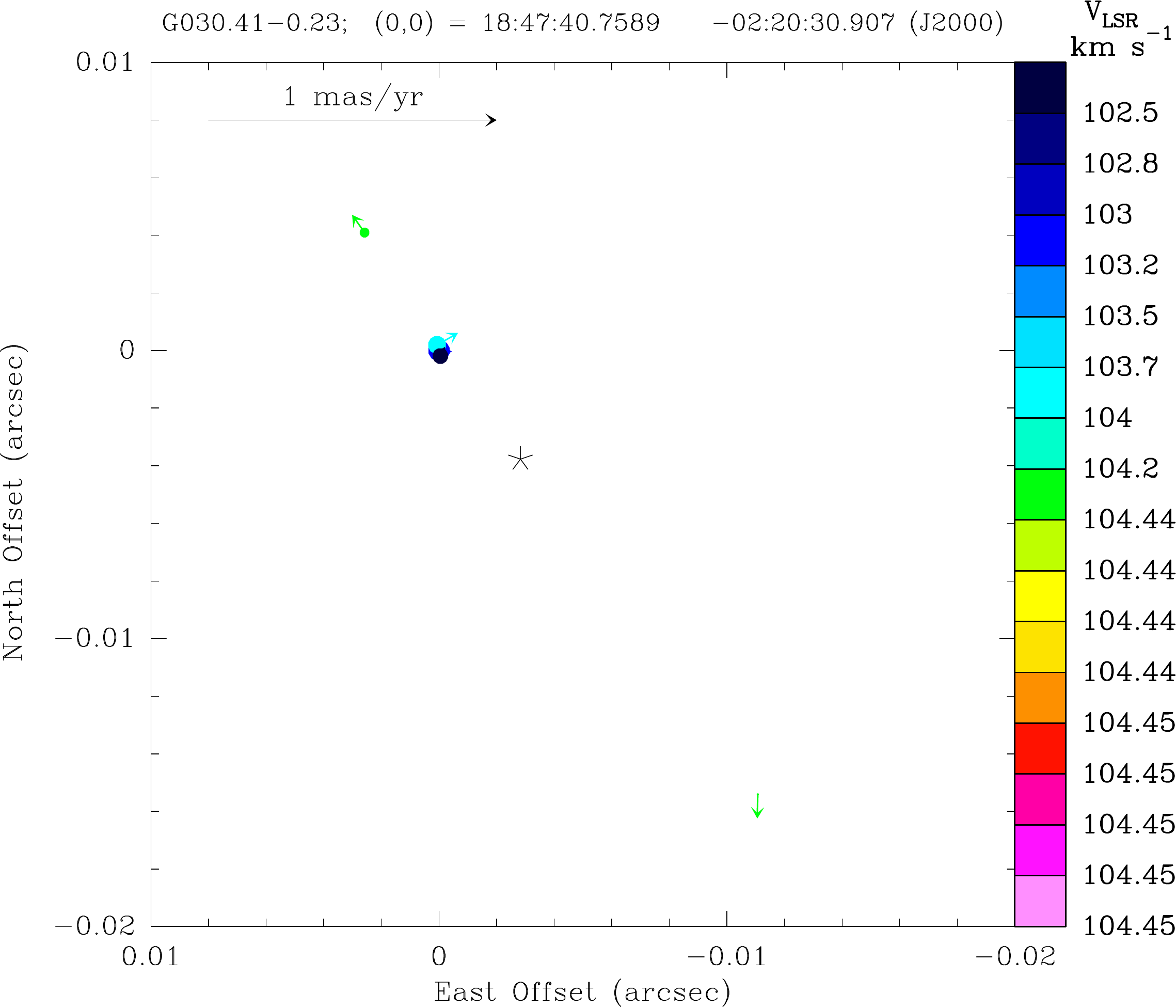}\label{G03041-ProperMotion}}
        \subfloat[6.7 GHz methanol maser $-$ G030.70$-$00.06]{\includegraphics[width=0.45\textwidth]{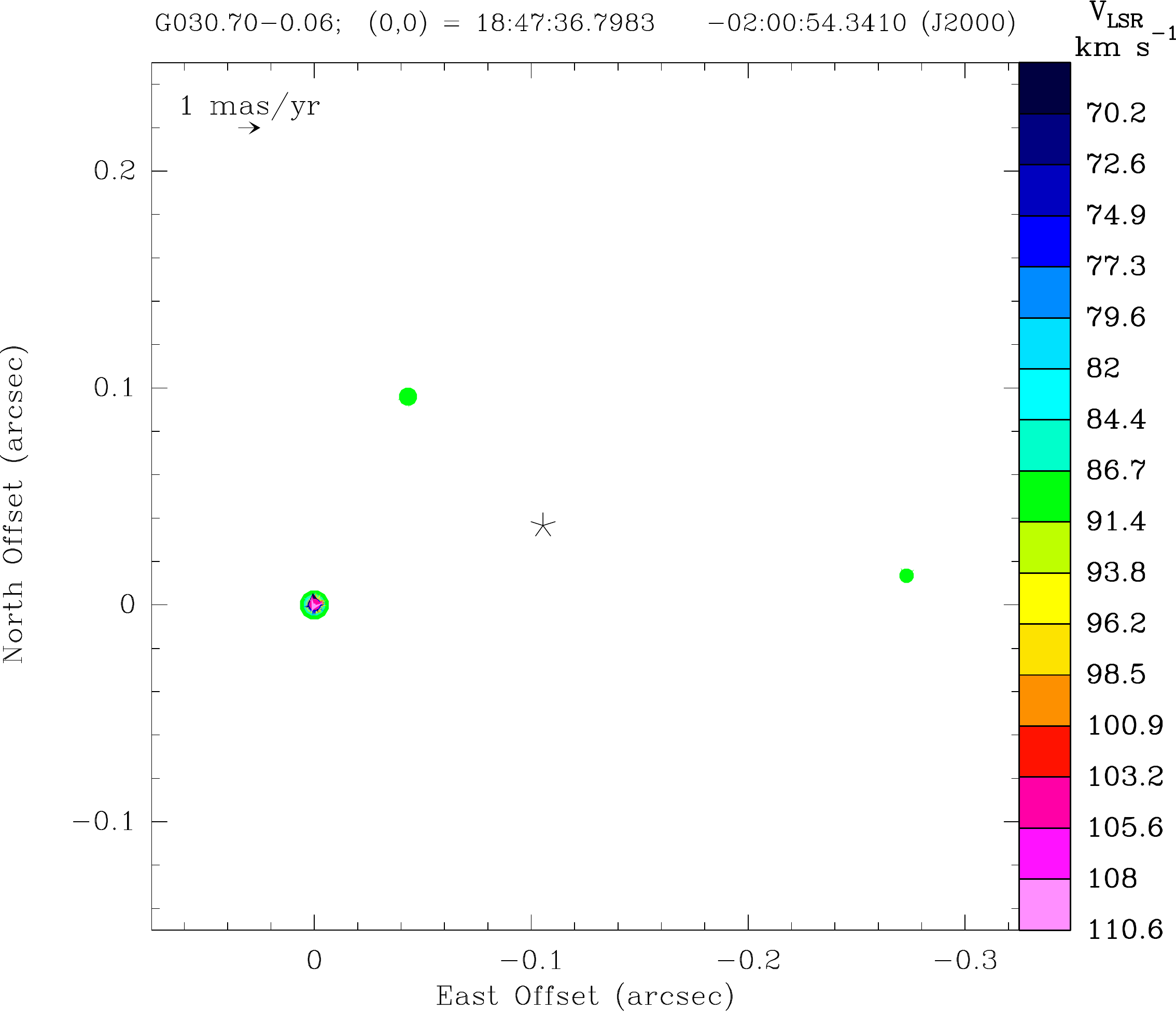}\label{G03070-ProperMotion}}

        \subfloat[6.7 GHz methanol maser $-$ G030.74$-$00.04]{\includegraphics[width=0.45\textwidth]{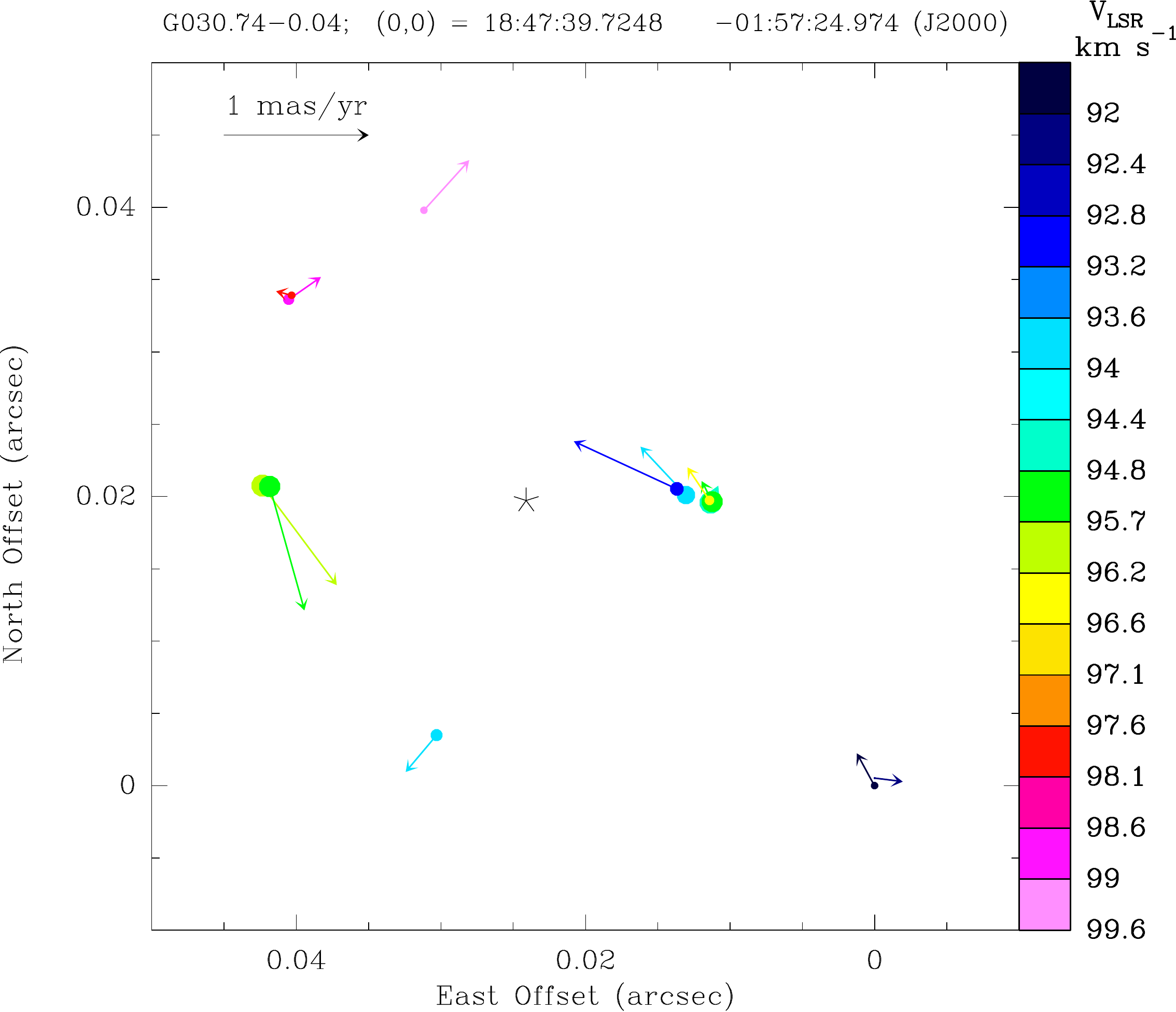}\label{G03074-ProperMotion}}
        \subfloat[22 GHz water maser $-$ G031.41+00.30]{\includegraphics[width=0.45\textwidth]{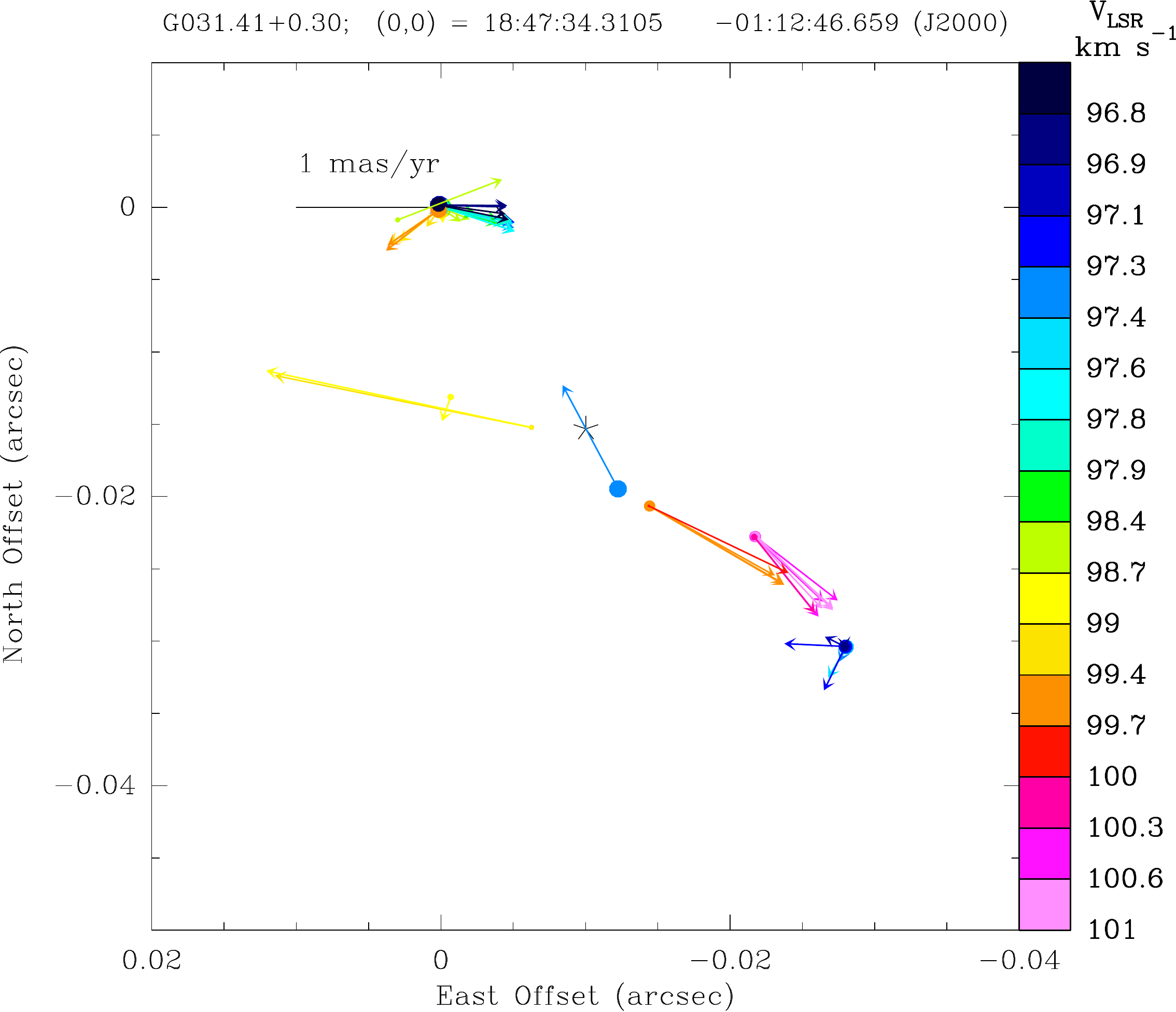}\label{G03141-ProperMotion}}
\end{figure*}


\section{Cloud characteristics}

\begin{figure*}
        \centering     
        \caption{Distribution of the bolometric luminosity (upper left), dust temperature (upper right), mass (lower left), and H$_{2}$ column density (lower right) of the host clouds of the high peculiar motion (red) and low peculiar motion (blue) sources (for sources with velocity uncertainty $\leq$ 20 km s$^{-1}$). The values of these parameters were taken from the ATLASGAL clump catalog of \citet{Urquhart2018}.}
    \subfloat{\includegraphics[width=0.45\textwidth]{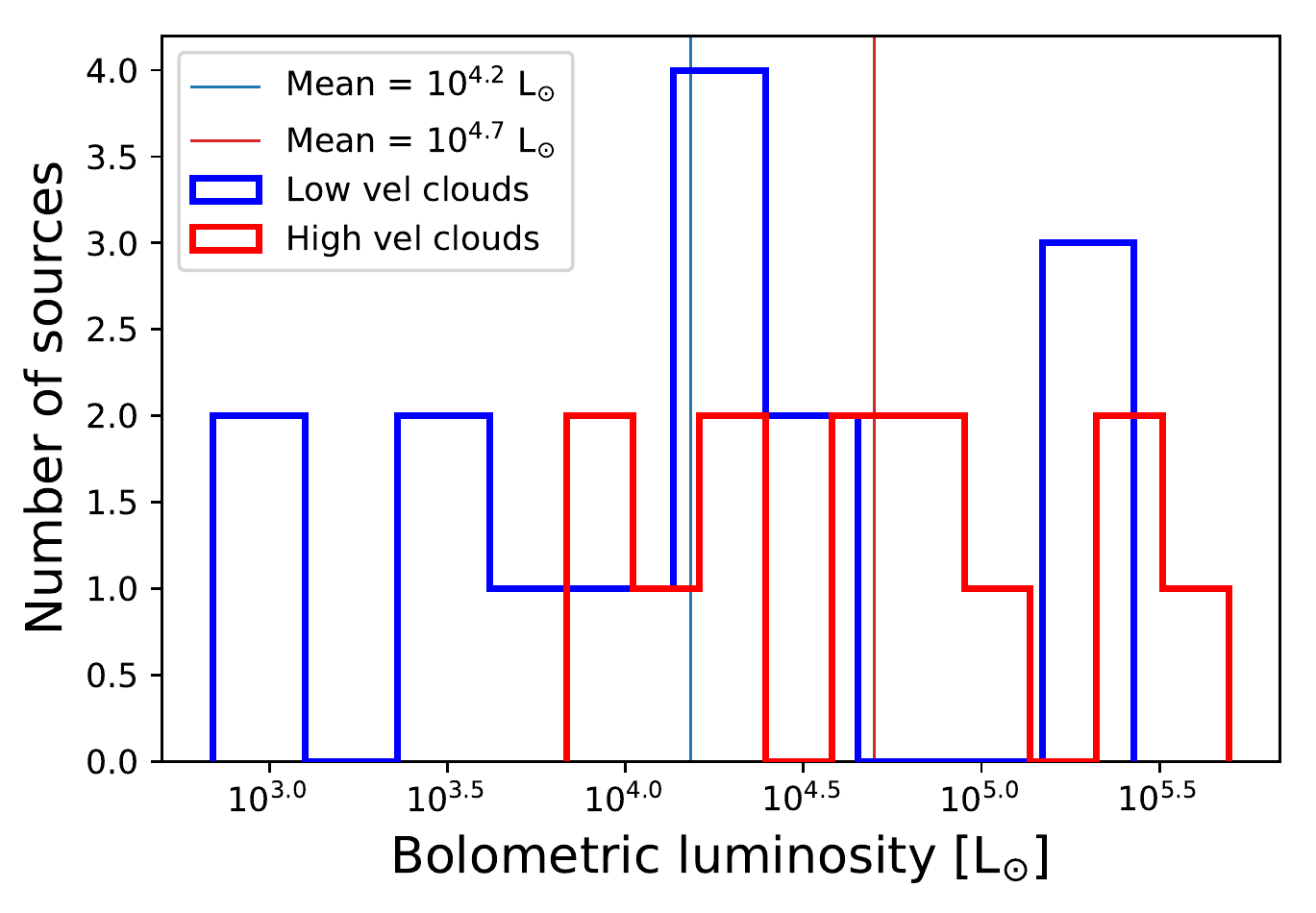}}
    \subfloat{\includegraphics[width=0.45\textwidth]{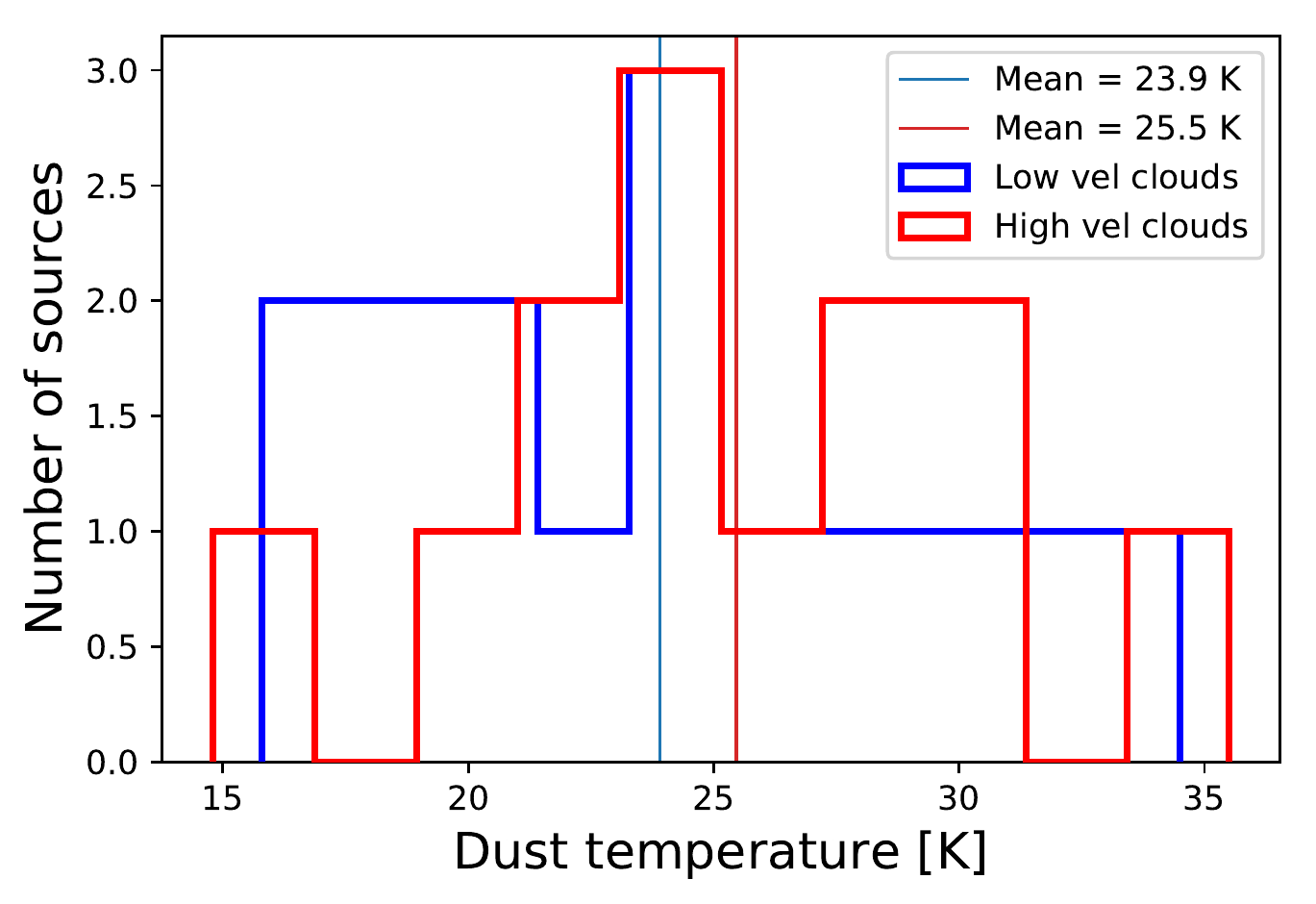}}\\

    \subfloat{\includegraphics[width=0.45\textwidth]{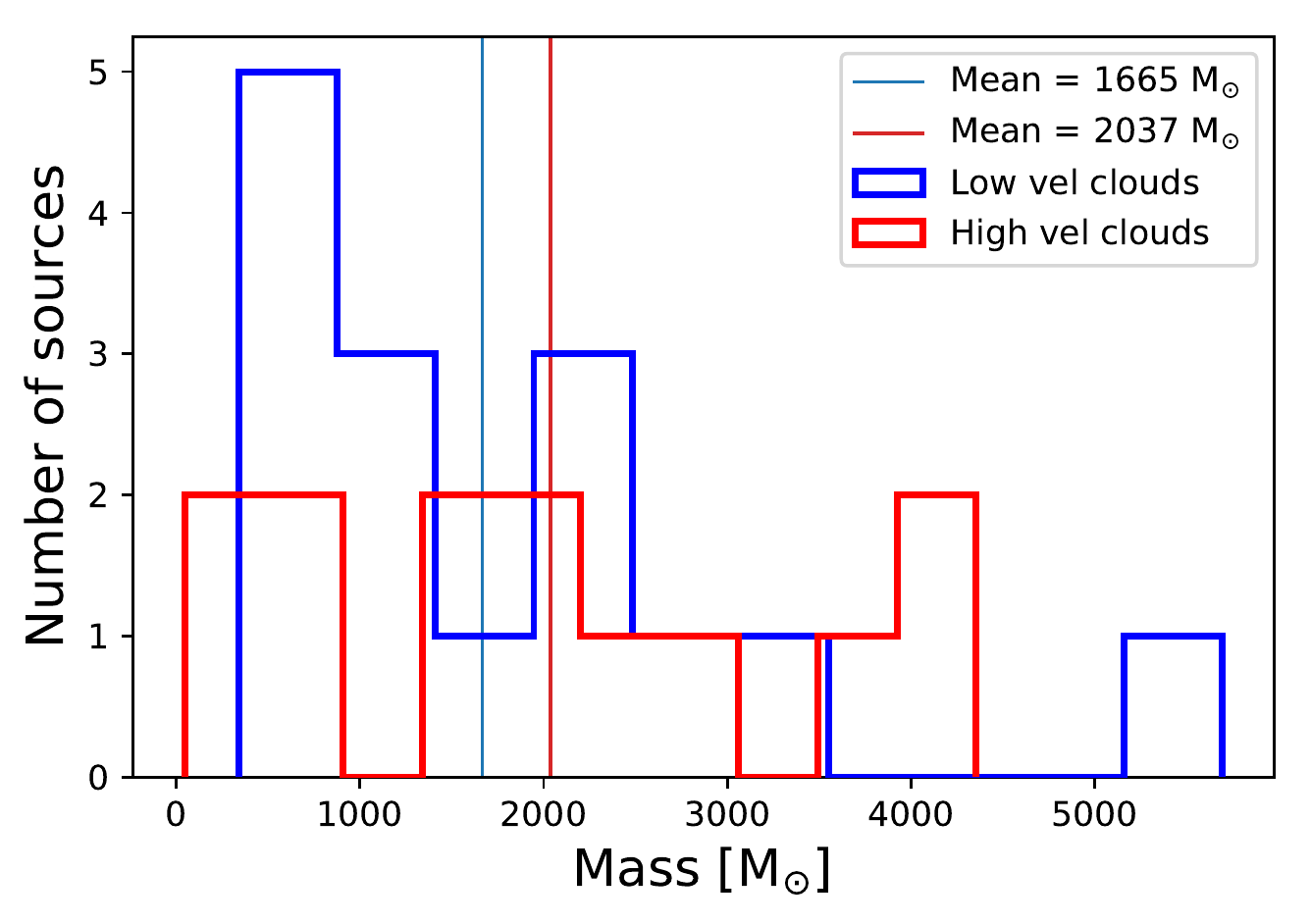}}
    \subfloat{\includegraphics[width=0.45\textwidth]{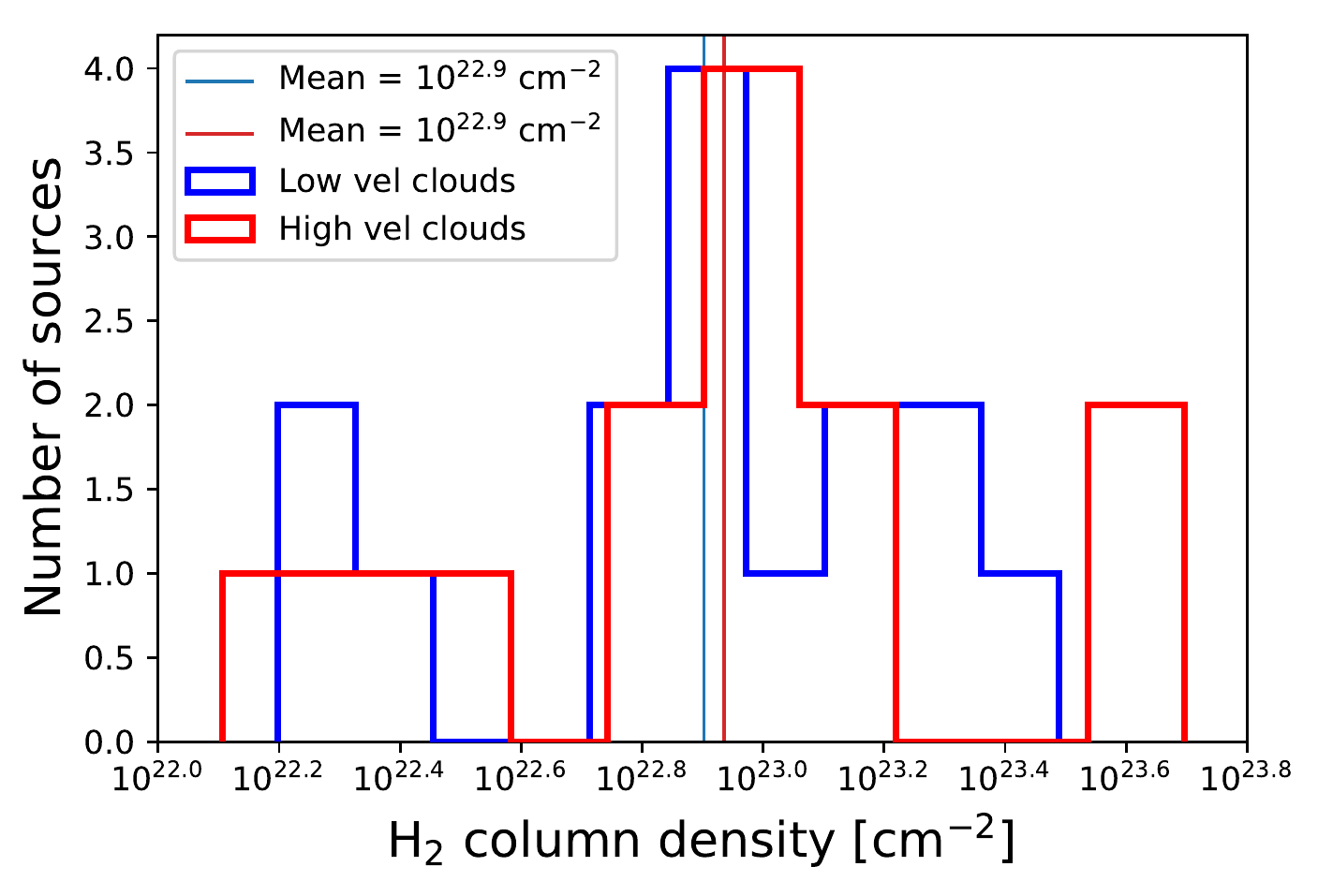}}
        \label{Scutum-Arm-Characteristica}
\end{figure*}

\clearpage
\onecolumn

\def\thetable{A.1}
\begin{center}
\renewcommand{\arraystretch}{1.2}
\begin{longtable}{rllcccc}
\caption{Observed coordinates, line-of-sight velocities, observing times,  frequency of the maser transition of our targets, and the VLBA program code of the corresponding observation . The J* sources are the background quasars that were observed as phase reference sources for the corresponding target.} \label{Sources-Coord} \\
\hline \hline Source & RA & Dec & v$_{LSR}$ & Observing Times & Frequency & VLBA \\
& [hh mm ss.ssss] & [dd mm ss.sss] & [km s$^{-1}$] & & [GHz] & Program Code \\ \hline
\endfirsthead
\hline \hline Source & RA & Dec & v$_{LSR}$ & Observing Times & Frequency & VLBA \\
& [hh mm ss.ssss] & [dd mm ss.sss] & [km s$^{-1}$] & & [GHz] & Program Code \\ \hline
\endhead
\hline
\endfoot
\hline
\endlastfoot
G023.25$-$00.24 & 18 34 31.2392 & $-$08 42 47.501 & 65  & Oct 2012 $-$ Oct 2013 & 6.7 & BR149F \\
J1825$-$0737    & 18 25 37.6096 & $-$07 37 30.013 & & & & \\
J1846$-$0651    & 18 46 06.3002 & $-$06 51 27.747 & & & & \\
G028.14$-$00.00 & 18 42 42.5916 & $-$04 15 35.163 & 101 & Oct 2012 $-$ Oct 2013 & 6.7 & BR149H\\
J1833$-$0323    & 18 33 23.9055 & $-$03 23 31.466 & & & \\
J1834$-$0301    & 18 34 14.0746 & $-$03 01 19.627 & & & \\
J1827$-$0405    & 18 27 45.0404 & $-$04 05 44.580 & & & \\
J1846$-$0651    & 18 46 06.3002 & $-$06 51 27.747 & & & \\
G028.86+00.06   & 18 43 46.2259 & $-$03 35 29.563 & 106 & Feb 2011 $-$ Apr 2012 & 22  & BR145J\\
J1833$-$0323    & 18 33 23.9055 & $-$03 23 31.466 & & & \\
J1846$-$0651    & 18 46 06.3003 & $-$06 51 27.746 & & & \\
J1857$-$0048    & 18 57 51.3586 & $-$00 48 21.950 & & & \\
G029.86$-$00.04  & 18 45 59.5734 & $-$02 45 05.528 & 102 & Oct 2012 $-$ Oct 2013  &  6.7 & BR149I  \\
J1834$-$0301    & 18 34 14.0746 & $-$03 01 19.627 & & & \\
J1833$-$0323    & 18 33 23.9055 & $-$03 23 31.466 & & & \\
J1857$-$0048    & 18 57 51.3586 & $-$00 48 21.950 & & & \\
J1846$-$0003    & 18 46 03.7826 & $-$00 03 38.283 & & & \\
G029.95$-$00.01  & 18 46 03.7431 & $-$02 39 20.999 & 96 & Oct 2012 $-$ Oct 2013 & 6.7 &  BR149I \\
J1834$-$0301    & 18 34 14.0746 & $-$03 01 19.627 & & & & \\
J1853$-$0048    & 18 53 41.9892 & $-$00 48 54.331 & & & & \\
J1833$-$0323    & 18 33 23.9055 & $-$03 23 31.466 & & & & \\
J1846$-$0003    & 18 46 03.7826 & $-$00 03 38.283 & & & & \\
G029.98+00.10   & 18 45 39.9637 & $-$02 34 32.624 & 118 & Nov 2013 $-$ Apr 2015 & 22  & BR198N\\
J1834$-$0301    & 18 34 14.0746 & $-$03 01 19.627 & & & \\
J1853$-$0048    & 18 53 41.9892 & $-$00 48 54.330 & & & \\  
J1846$-$0003    & 18 46 03.7850 & $-$00 03 38.280 & & & \\   
J1857$-$0048    & 18 57 51.3586 & $-$00 48 21.950 & & & \\    
G030.22$-$00.18  & 18 47 08.2900 & $-$02 29 29.400 & 113 & Apr 2014 $-$ Apr 2015 & 6.7 & BR198M  \\
J1853$-$0048    & 18 53 41.9892 & $-$00 48 54.330 & & & & \\  
J1857$-$0048    & 18 57 51.3586 & $-$00 48 21.950 & & & & \\  
J1846$-$0003    & 18 46 03.7850 & $-$00 03 38.280 & & & & \\  
J1834$-$0301    & 18 34 14.0746 & $-$03 01 19.627 & & & & \\ 
G030.41$-$00.23  & 18 47 40.7610 & $-$02 20 30.254 & 103 & Oct 2012 $-$ Oct 2013  & 6.7 & BR149J  \\
J1853$-$0048    & 18 53 41.9892 & $-$00 48 54.331 & & & & \\ 
J1834$-$0301    & 18 34 14.0746 & $-$03 01 19.627 & & & & \\ 
J1857$-$0048    & 18 57 51.3586 & $-$00 48 21.950 & & & & \\ 
J1846$-$0003    & 18 46 03.7826 & $-$00 03 38.283 & & & & \\ 
G030.70$-$00.06 & 18 47 36.8036 & $-$02 00 55.021 & 88  & Oct 2012 $-$ Oct 2013 & 6.7 & BR149J\\
J1853$-$0048    & 18 53 41.9892 & $-$00 48 54.331 & & & \\ 
J1834$-$0301    & 18 34 14.0746 & $-$03 01 19.627 & & & \\
J1857$-$0048    & 18 57 51.3586 & $-$00 48 21.950 & & & \\ 
J1846$-$0003    & 18 46 03.7826 & $-$00 03 38.283 & & & \\ 
G030.74$-$00.04  & 18 47 37.7289 & $-$01 57 54.699 & 88 & Oct 2012 $-$ Oct 2013  &  6.7 & BR149K  \\
J1853$-$0048    & 18 53 41.9892 & $-$00 48 54.331 & & & & \\ 
J1834$-$0301    & 18 34 14.0746 & $-$03 01 19.627 & & & &\\ 
J1857$-$0048    & 18 57 51.3586 & $-$00 48 21.950 & & & & \\ 
J1846$-$0003    & 18 46 03.7826 & $-$00 03 38.283 & & & & \\ 
G030.78+00.20   & 18 46 48.0800 & $-$01 48 54.000 & 87 & Apr 2014 $-$ Apr 2015 & 6.7 & BR198M\\
J1853$-$0048    & 18 53 41.9892 & $-$00 48 54.330  & & & \\   
J1857$-$0048    & 18 57 51.3586 & $-$00 48 21.950 & & & \\   
J1833$-$0323    & 18 33 23.9055 & $-$03 23 31.466 & & & \\   
J1834$-$0301    & 18 34 14.0746 & $-$03 01 19.627 & & & \\    
G030.81$-$00.05  & 18 47 46.9764        & $-$01 54 26.676 & 109 & Oct 2012 $-$ Oct 2013 & 6.7 & BR149K \\
J1853$-$0048    & 18 53 41.9892 & $-$00 48 54.331 & & & & \\
J1834$-$0301    & 18 34 14.0746 & $-$03 01 19.627 & & & & \\
J1857$-$0048    & 18 57 51.3586 & $-$00 48 21.950 & & & & \\
J1846$-$0003    & 18 46 03.7826 & $-$00 03 38.283 & & & & \\
G030.97$-$00.14  & 18 48 22.0300 & $-$01 48 30.600 & 78 & Apr 2014 $-$ Apr 2015 & 6.7 & BR198M \\
J1857$-$0048    & 18 57 51.3586 & $-$00 48 21.950 & & & & \\
J1853$-$0048    & 18 53 41.9892 & $-$00 48 54.330  & & & & \\
J1846$-$0003    & 18 46 03.7850 & $-$00 03 38.280   & & & & \\
G031.41$+$00.30  & 18 47 34.3124 & $-$01 12 47.000 & 98 & Nov 2013 $-$ Apr 2015 & 22 & BR198N  \\
J1853$-$0048    & 18 53 41.9892 & $-$00 48 54.330  & & & & \\
J1846$-$0003    & 18 46 03.7850 & $-$00 03 38.280   & & & & \\
J1853$-$0010    & 18 53 10.2692 & $-$00 10 50.740  & & & & \\
J1857$-$0048    & 18 57 51.3586 & $-$00 48 21.950 & & & & \\
G032.04$+$00.05 & 18 49 36.5732 & $-$00 45 42.650 & 99 & Nov 2012 $-$ Oct 2013 & 6.7 & BR149L\\
J1853$-$0048    & 18 53 41.9892 & $-$00 48 54.331 & & & & \\
J1857$-$0048    & 18 57 51.3586 & $-$00 48 21.950 & & & & \\
J1846$-$0003    & 18 46 03.7826 & $-$00 03 38.283 & & & & \\
J1848$+$0138    & 18 48 21.8104 & $+$01 38 26.632 & & & & \\
G033.09$-$00.07 & 18 51 59.4085 & $+$00 06 34.262   & 104 & Nov 2012 $-$ Oct 2013 & 6.7 & BR149L\\
J1853$-$0048    & 18 53 41.9892 & $-$00 48 54.331 & & & \\ 
J1857$-$0048    & 18 57 51.3586 & $-$00 48 21.950 & & & \\ 
J1846$-$0003    & 18 46 03.7826 & $-$00 03 38.283 & & & \\ 
J1848$+$0138    & 18 48 21.8104 & $+$01 38 26.632 & & & \\ \hline \hline
\end{longtable}
\end{center}

\clearpage

\def\thetable{C.1}
\begin{center}
\renewcommand{\arraystretch}{1.2}
\begin{longtable}{rcccc}
\caption{Velocity range of the maser emission, phase reference sources, and maser channels used for parallax fitting.} \label{Sources-Info} \\
\hline \hline Source & v$_{Maser}$ & Phase reference & Maser & Figs.\\
 & [km s$^{-1}$] & sources & channels & \\ \hline
\endfirsthead
\hline \hline Source & v$_{Maser}$ & Phase reference & Maser & Figs.\\
 & [km s$^{-1}$] & sources & channels & \\ \hline
\hline
\endhead
\hline
\endfoot
\hline
\endlastfoot
G023.25$-$00.24 & 55$-$70 & J1825$-$0737 & 500 & \ref{G02325-Spectrum}, \ref{G02325-Parallax} \\
                                  &         & J1846$-$0651 & 501 & \\
                                  &         &              & 502 & \\
                                  &         &              & 503a & \\
                                  &         &              & 503b & \\
                                  &         &              & 514 & \\
G028.14$-$00.00 & 98$-$105 & J1833$-$0323 & 500 & \ref{G02814-Spectrum}, \ref{G02814-Parallax} \\
                                  &         & J1834$-$0301 & 501 & \\
                                  &         & J1846$-$0651 & 504 & \\
G028.86+00.06     & 85$-$110& J1833$-$0323 & 126 & \ref{G02886-Spectrum}, \ref{G02886-Parallax}, \ref{G02886-ProperMotion} \\
                                  &         & J1857$-$0048 & 136 & \\
                                  &         & J1846$-$0651 & 155 & \\
                                  &         &              & 171 & \\
G029.86$-$00.04  & 95$-$107 & J1833$-$0323 & 494 & \ref{G02986-Spectrum}, \ref{G02986-Parallax}, \ref{G02986-ProperMotion} \\
                                          &         & J1857$-$0048 & 501 & \\
                                          &         & J1846$-$0003 & 506 & \\
                                          &         & J1834$-$0301 &     & \\
G029.95$-$00.01  & 93$-$108 & J1833$-$0323 & 474 & \ref{G02995-Spectrum}, \ref{G02995-Parallax}, \ref{G02995-ProperMotion} \\
                                  &         & J1853$-$0048 & 503 & \\
                                  &         & J1846$-$0003 &     & \\
                                  &         & J1834$-$0301 &     & \\
G029.98+00.10  & 98$-$120 & J1834$-$0301 & 997 & \ref{G02998-Spectrum}, \ref{G02998-Parallax}, \ref{G02998-ProperMotion} \\
                                 &         & J1853$-$0048 & 1030 & \\
                                 &         &              & 1038 & \\
G030.22$-$00.18  & 110$-$115 & J1834$-$0301 & 1005 & \ref{G03022-Spectrum}, \ref{G03022-Parallax} \\
                                 &         & J1853$-$0048 &  & \\
                                 &         & J1846$-$0003 &  & \\
G030.41$-$00.23  & 102$-$106 & J1834$-$0301 & 499a & \ref{G03041-Spectrum}, \ref{G03041-Parallax}, \ref{G03041-ProperMotion} \\
                                 &         & J1853$-$0048 & 499b & \\
                                 &         & J1846$-$0003 & 501  & \\
                                 &         & J1857$-$0048 &     & \\
G030.70$-$00.06  & 85$-$93 & J1834$-$0301 & 497 & \ref{G03070-Spectrum}, \ref{G03070-Parallax}, \ref{G03070-ProperMotion} \\
                                 &         & J1853$-$0048 & 500 & \\
                                 &         & J1846$-$0003 &     & \\
                                 &         & J1857$-$0048 &     & \\
G030.74$-$00.04  & 85$-$110 & J1834$-$0301 & 486 & \ref{G03074-Spectrum}, \ref{G03074-Parallax}, \ref{G03074-ProperMotion} \\
                                 &         & J1853$-$0048 & 492 & \\
                                 &         & J1846$-$0003 & 496 & \\
G030.78+00.20    & 75$-$91 & J1834$-$0301 & 1000 & \ref{G03078-Spectrum}, \ref{G03078-Parallax} \\
                                 &         & J1853$-$0048 & 1006 & \\
                                 &         & J1833$-$0323 & 1011 & \\
                                 &         & J1857$-$0048 & 1056 & \\
G030.81$-$00.05  & 87$-$112& J1834$-$0301 & 498 & \ref{G03081-Spectrum}, \ref{G03081-Parallax} \\
                                 &         & J1853$-$0048 & 499 & \\
                                 &         & J1846$-$0003 & 500 & \\
                                 &         & J1857$-$0048 & 501 & \\
                                 &         &              & 502 & \\
                                 &         &              & 503 & \\
                                 &         &              & 504 & \\
                                 &         &              & 505 & \\
                                 &         &              & 507 & \\
                                 &         &              & 509 & \\
                                 &         &              & 511 & \\
                                 &         &              & 512 & \\
                                 &         &              & 544 & \\
                                 &         &              & 545 & \\
G030.97$-$00.14  & 72$-$81 & J1846$-$0003 & 1002 & \ref{G03097-Spectrum}, \ref{G03097-Parallax} \\
                                  &         & J1853$-$0048 & 1009 & \\
                                  &         & J1857$-$0048 & 1015 & \\
                                  &         &              & 1018 & \\
G031.41$+$00.30  & 93$-$107 & J1846$-$0003 & 983 & \ref{G03141-Spectrum}, \ref{G03141-Parallax}, \ref{G03141-ProperMotion} \\
                                 &         & J1853$-$0010 & 1000 & \\
                                 &         & J1853$-$0048 &      & \\
                                 &         &              &      & \\
G032.04$+$00.05  & 90$-$103& J1848$+$0138 & 494a & \ref{G03204-Spectrum}, \ref{G03204-Parallax} \\
                                 &         & J1857$-$0048 & 495a & \\
                                 &         &              & 496a & \\
                                 &         &              & 499a & \\
                                 &         &              & 500a & \\
                                 &         &              & 501a & \\
                                 &         &              & 504 & \\
                                 &         &              & 505 & \\
                                 &         &              & 507 & \\
                                 &         &              & 509 & \\
                                 &         &              & 511 & \\
                                 &         &              & 512 & \\
                                 &         &              & 544 & \\
                                 &         &              & 545 & \\
G033.09$-$00.07  & 101$-$107& J1848$+$0138 & 500 & \ref{G03309-Spectrum}, \ref{G03309-Parallax} \\
                                          &         & J1857$-$0048 & 502 & \\
                                          &         &              & 504 & \\
                                          &         &              & 506 & \\ \hline \hline
\end{longtable}
\end{center}

\end{document}